\documentclass[ProofPDF]{./Styles/Jnls-Template-V3}
\ifdefined\XeTeXversion
  \AtBeginDvi{}
\fi
\newif\ifanonymous
\anonymousfalse      
\usepackage{fancyvrb}
\usepackage{adjustbox}
\usepackage{fvextra}

\usepackage{xcolor}
\usepackage{array}
\usepackage{booktabs}
\usepackage[most]{tcolorbox}

\usepackage{booktabs}
\usepackage{needspace}
\definecolor{refblue}{RGB}{67,86,138}
\newcommand{\tabref}[1]{\textcolor{refblue}{Table~\ref{#1}}}
\newcommand{\figref}[1]{\textcolor{refblue}{Figure~\ref{#1}}}
\newcommand{\exref}[1]{\textcolor{refblue}{Example~\ref{#1}}}
\newcommand{\appref}[1]{\textcolor{refblue}{Appendix~\ref{#1}}}
\newcommand{\gain}[1]{+#1}
\newcounter{munocodeblock}
\renewcommand{\themunocodeblock}{\arabic{munocodeblock}}
\makeatletter
\newcommand{\munocodeblockcaption}[2]{%
  \par\nobreak\smallskip
  \refstepcounter{munocodeblock}%
  \label{#2}%
  \noindent\begin{minipage}{\columnwidth}
    \@makecaptiontable{Example \themunocodeblock}{#1}%
  \end{minipage}
  \par\nobreak\smallskip
  \par\addvspace{1.25\baselineskip}%
}
\makeatother

\usepackage{graphicx}
\usepackage{color}
\usepackage[framemethod=TikZ]{mdframed}

\newtcbox{\munocode}{
    on line,
    colback=gray!16,
    colframe=gray!75,
    coltext=black,
    boxrule=0.5pt,
    arc=1pt,
    boxsep=0pt,
    left=2pt,
    right=2pt,
    top=1pt,
    bottom=1pt,
    fontupper=\sffamily\small
}

\usepackage{xcolor}
\usepackage{mdframed}

\usepackage[most]{tcolorbox}
\usepackage{fvextra}

\newenvironment{musicbox}[1]{%
  \par\addvspace{1.25\baselineskip}%
  \begin{mdframed}[
    backgroundcolor=gray!4,
    linecolor=gray!35,
    linewidth=0.8pt,
    roundcorner=3pt,
    innerleftmargin=5mm,
    innerrightmargin=5mm,
    innertopmargin=3mm,
    innerbottommargin=4mm,
    skipabove=0pt,
    skipbelow=4pt]
  \normalfont
}{%
  \end{mdframed}
}

\newcommand{\MAESTROCapsPieceCount}{148}
\newcommand{\MAESTROCapsQAPerPiece}{210}
\newcommand{\MAESTROCapsQACount}{31,080}
\newcommand{\MAESTROCapsAudioHours}{10.38}

\begin{document}

\jname{Transactions of the International Society for Music Information Retrieval }
\jvol{00}
\jissue{00}
\jmonth{season}
\jyear{xxx}
\jid{doi}
\artid{xx.xxxx/xxxx.xx}
\filename{main}
\def\MSFirstPage{1}
\def\MSLastPage{52}
\graphicspath{{./figures/}}
\title{Unifying Score and Performance for Fine-Grained Music Understanding in Audio-Language Models}
\articletype{DATASET ARTICLE}

\ifanonymous
    \author[1]{Anonymous Author(s)}
    \howtociteauthor[1]{Anonymous}
    \rhauthor[1]{Anonymous}
\else
    \author[1]{Milan Liessens Dujardin}
    \howtociteauthor[1]{M. Liessens Dujardin}
    \rhauthor[1]{Milan}

    \author[2]{Song-Ze Yu}
    \howtociteauthor[2]{S.-Z. Yu}
    \rhauthor[2]{Song-Ze Yu}

    \author[3]{Kevin Miao}
    \howtociteauthor[3]{K. Miao}
    \rhauthor[3]{Kevin Miao}

    \aff[1]{Bryel Labs and UC Berkeley (\email{milan@bryel.ai})}
    \aff[2]{UC Berkeley (\email{vaclis@berkeley.edu})}
    \aff[3]{Bryel Labs (\email{kevin@bryel.ai})}
    \afflnote{*Author affiliations can be found in the back matter of this article}
\fi

\begin{history}
\submitted{\submittedday{xxx}\submittedmonth{xxx}\submittedyear{xxx}}
\accepted{\acceptedday{xxx}\acceptedmonth{xxx}\acceptedyear{xxx}}
\published{\publishedday{xxx}\publishedmonth{xxx}\publishedyear{xxx}}
\end{history}

\begin{abstract}[ABSTRACT]
Large audio language models (LALMs) have shown promising progress in broad music-understanding tasks such as tagging, retrieval, and captioning. Music understanding that requires finer hearing over both the content and how it is realized within a performance through dynamics, phrasing, articulation, time, and other performance techniques, however, remains at an earlier stage. Existing audio-language model (ALM) training pipelines typically rely on coarse, weakly grounded captions and therefore provide little support for learning these subtle nuances in music, limiting their ability to serve real-world applications in education or artistic practice.

We therefore introduce \textbf{MuNo-SP} (Music Notation unifying Score and Performance), a text-based representation that jointly encodes score content and performance information. Building on MuNo-SP, we develop an \textbf{automatic training-data generation pipeline} that uses aligned scores and performances to produce long-form auditory analyses and musically informed question--answer pairs. We use this pipeline to construct \textbf{MAESTROCaps}, a classical piano dataset comprising \MAESTROCapsPieceCount{} long-form performance analyses and \MAESTROCapsQACount{} question--answer pairs derived from \MAESTROCapsPieceCount{} aligned score--performance pairs. In a human evaluation, MuNo-SP analyses were preferred by majority vote over MIDI-only analyses for eight of nine excerpts. MuNo-SP also performed strongly on a benchmark of score--performance understanding, suggesting that integrating score and performance information enables more reliable and musically informative LALM supervision than a MIDI-only baseline.
\end{abstract}


\maketitle

\section{Introduction}\label{sec:introduction}

A musician may identify from a performance how individual voices interplay,
how motifs are developed, and how harmonic nuances color a melody; recognize
fine-grained phrase boundaries, climaxes, and transitions; and interpret how
timing, dynamics, and articulation shape the performance. Musician-level music understanding therefore requires understanding both
\textit{what} is played and \textit{how} it is realized in sound, with a
precise notion of time and close attention to detail.

Despite progress on broad music-understanding tasks such as tagging, retrieval,
and captioning
(\cite{ghosh2025music,ghosh2026audioflamingonext,youGaMMAJointGlobalTemporal2026}),
large audio language models (LALMs) do not yet reliably demonstrate these more
sophisticated capabilities (\cite{liessensdujardin2026muspbench, dujardinPitchBenchMeasuringPitch2026}). Such capabilities could support learners and
musicians in interpreting, studying, improvising, and composing music, as well
as enable fine-grained control in music generation.

One route to this level of \textbf{music
intelligence} is supervision that captures musical content as well as
its realization, grounded precisely in time. A growing body of work uses large
language models to generate music captions, but existing training datasets
(\cite{salganikMusicSemSemanticallyRich2026b,melechovskyMidiCapsLargescaleMIDI2024,doh2023lp}) typically
contain coarse and/or weakly time-grounded descriptions without event-level alignment. Consequently, they often
omit nuances essential to deeper music understanding, including tempo
variation, articulation, expressive gestures, and phrasing.

In many musical traditions, particularly Western classical music, scores encode both
musical content and intent. As the renowned piano pedagogue Heinrich
Neuhaus \cite{neuhaus1973art} argues, performers form an artistic image of a
work through the lens of their musical knowledge, experience, and familiarity
with performance practices, among other influences, and realize that image through
tone and time. The score thus serves both as a discrete representation of
audible musical content and as the contextual foundation for the performer's
musical imagination, making it a valuable resource for advancing
understanding in audio-language models.

Creating fine-grained, musically grounded training data at scale requires
bringing score and performance together in a unified representation. We
introduce \textbf{MuNo-SP} (Music Notation unifying Score and Performance), a
text-based representation that integrates detailed score content with
temporally aligned performance information readily interpretable by language
models. Building on MuNo-SP, we develop an automatic data-generation pipeline
that produces long-form auditory analyses and question--answer pairs. Grounding
generation in the score encourages analyses that reflect how musicians reason
about musical content, while alignment with the performance conditions these
analyses on how that content is realized in sound and time.

\vspace{10pt}

Our contributions are the following:
\vspace{-10pt}
\begin{enumerate}
    \item We introduce \textbf{MuNo-SP}, a music representation that unifies aligned scores and performance information within a language-model-accessible format.
    
    \item We present an \textbf{automatic training-data generation pipeline} that converts aligned scores and performance MIDI into auditory analyses and temporally grounded question--answer pairs.
    
    \item We construct and release \textbf{MAESTROCaps}, comprising \MAESTROCapsPieceCount{} long-form analyses and \MAESTROCapsQACount{} question--answer pairs across \MAESTROCapsPieceCount{} aligned score--performance pairs.

    \item We release an updated set of \textbf{\MAESTROCapsPieceCount{} ASAP-derived scores}, together with the available (n)ASAP alignment data, after correcting over 2,000 errors.
    \vspace{-10pt}
\end{enumerate}

The MAESTROCaps dataset and code are available at
\href{https://huggingface.co/datasets/bryel-labs/MAESTROCaps}{huggingface.co/datasets/bryel-labs/MAESTROCaps}.

\section{Related Work}

\subsection{Music Understanding}

Scores and performances are two modalities through which musicians commonly
communicate musical intent. Accordingly, researchers have investigated
model-based music understanding from both symbolic notation and audio and have
developed benchmarks that probe increasingly detailed forms of musical
knowledge and reasoning across both modalities.

\paragraph{From Symbolic Scores}
A growing body of work evaluates whether large language models can reason about symbolic music representations and music-theoretical concepts. MusicTheoryBench (\cite{yuan2024chatmusician}) and ZIQI-Eval (\cite{li2024ziqieval}) assess broad music knowledge and reasoning, including tasks that use ABC notation, whereas SSMR-Bench (\cite{wang2025ssmrbench}) uses synthesized sheet-music questions to evaluate symbolic music reading and understanding. ABC-Eval~(\cite{wang2025abceval}) evaluates LLMs on tasks expressed in ABC notation, including bar sequencing, error detection, and emotion classification. MSU-Bench~(\cite{dai2025msubench}) targets score-level music understanding across four levels of analysis spanning timing, pitch, harmony, and form. MuSP-Bench (\cite{liessensdujardin2026muspbench}) extends this line of work by assessing understanding of scores, performances, and their interaction across multiple music representations, and including questions concerning interpretation and performance technique. Collectively, these benchmarks show substantial error rates on advanced score-understanding tasks, indicating that current systems remain unreliable in score understanding.

\paragraph{From Audio}
Recent audio-language and multimodal models, including Audio Flamingo Next~(\cite{ghosh2026audioflamingonext}), GaMMA~(\cite{youGaMMAJointGlobalTemporal2026}), Gemini 3.1 Pro~(\cite{gemini31pro}), Qwen3.5-Omni~(\cite{qwenteam2026qwen35omnitechnicalreport}), and GPT-4o Audio~(\cite{openai2024gpt4oaudio}), have expanded the ability of multimodal systems to reason directly from audio. Benchmarks such as MMAR (\cite{maMMARChallengingBenchmark2025}) and MMAU-Pro (\cite{kumarMMAUProChallengingComprehensive2025}) evaluate capabilities ranging from acoustic-event recognition and speech recognition to musical knowledge and higher-level reasoning. MusTBench (\cite{kwonMusTBENCHBenchmarkingAdvancing2026}) assesses temporal understanding in models. Despite this, current models remain unreliable for fine-grained musical understanding from audio, particularly when questions require precise identification of specific events in music and interpretation of them within a broader context. Evidence of these limitations can be found in the low performance of frontier models on benchmarks targeting capabilities including fine-grained form identification and pitch hearing, such as MuSP-Bench and PitchBench (\cite{dujardinPitchBenchMeasuringPitch2026}).

\subsection{Symbolic Music Representations}

Symbolic music representations have largely developed around two different purposes. The first is to serve as an interface between humans, human and computer, or computers: enabling composers, performers, and software systems to write, exchange, render, edit, or control musical information. The second is to provide machine-oriented representations that make symbolic music easier for computational models to understand or generate. These two objectives lead to substantially different representational choices.

\paragraph{Representations as human--human, human--computer and computer--computer interfaces.}
Widely used formats include MusicXML (\cite{good2001musicxml,good2001virtualscore}), ABC notation (\cite{walshaw2011abc}), MIDI (\cite{midi1996,dannenberg2006velocity}), and Open Sound Control (OSC) (\cite{Wright1997OpenSA}). 

MusicXML is a de facto standard interchange format for notation software and encodes rich score semantics, including dynamics, articulations, slurs, multi-part scores, and other directions. Its primary purpose is the interoperability of visual score rendering and notation software: much of its markup describes layout details, like page positioning, stem directions, and beaming, which are often irrelevant to musical reasoning. Its verbosity makes it impractical for limited LLM context windows without preprocessing.

ABC notation was designed for compact transcription of folk and traditional melodies. It efficiently represents pitch, duration, key, and meter in plain text, making it lightweight and easy to tokenize for transcription. However, models continue to perform poorly on several ABC-based reasoning tasks, as shown in advanced ABC-reading benchmarks like MuSP-Bench.

MIDI captures performance-level events or properties such as onsets, offsets, channels, and velocity. Originally designed as a machine-readable protocol for controlling electronic instruments and reproducing performances, it is well suited for playback in software or Digital Audio Workstations (DAWs), as well as expressive timing analysis. However, it omits explicit notational structure: phrase boundaries, harmonic function, the distinction between principal and accompanying voices, leaving many cues that performers derive directly from the score implicit. In this sense, MIDI captures what was performed from the instrument's point of view, obscuring the reasoning behind the performance.

Existing representations thus involve a tradeoff: ABC is compact but hard to interpret, MusicXML is expressive but verbose, and MIDI is performance-rich but inherently ambiguous and only a proxy for sound in time. None of these representations bring together score and performance in a compact, semantically meaningful, and time-grounded way.

\paragraph{Representations for machine understanding and generation.}
A separate line of work develops symbolic representations specifically for sequence models. These representations are primarily designed to make symbolic music more suitable for machine learning, specifically sequence modeling and generation. MIDI-like tokenizations (\cite{oore_midilike_2018}) represent music through flattened NOTE-ON, NOTE-OFF, VELOCITY, and TIME-SHIFT events, providing a simple event-based interface for autoregressive modeling. REMI (\cite{huang_remi_2020}) introduces explicit bar and within-bar position tokens to expose metrical structure. Subsequent representations such as TSD (\cite{fradet-etal-2023-byte}), Structured (\cite{pia2021hadjeres}), CPWord (\cite{cpword2021}), and Octuple (\cite{zeng2021musicbert}) differ in how they encode temporal information and organize note attributes into tokens, leading to different tradeoffs in sequence length and modeling efficiency (\cite{fradet2023miditokpythonpackagemidi}). 

However, these representations are mostly derived from MIDI-like event streams and optimize for musical event serialization and therefore inherit the semantic limitations of MIDI discussed above, making musical intent implicit and leaving out the relationship between written score instructions and their realization in performance.

\subsection{Automated Captioning}

Researchers have addressed the scarcity of human-written music
descriptions by having large language models (LLMs) verbalize tags,
metadata, and musical attributes. These approaches can produce captions at a
scale that manual annotation cannot readily match, but the resulting language is
generally bounded by the information available to the caption-generation
pipeline and the reliability of audio feature extraction systems used.

LP-MusicCaps (\cite{doh2023lp}) contains approximately
2.2 million pseudo-captions for 0.5 million audio clips. Its three subsets are
derived from MusicCaps (10-second audio fragments), MagnaTagATune (30-second audio fragments), and the ECALS subset of the Million Song
Dataset. For each clip, GPT-3.5 generates captions from clip-level tags using
four instructions: writing, summarization, paraphrasing, and attribute
prediction. LP-MusicCaps
excludes identifying metadata such as artist and album names to reduce
hallucinations involving known musical entities. It nevertheless remains
conditioned on clip-level tags, and therefore not tied to concrete musical events. 

MusicInstruct (\cite{dengMusiLingoBridgingMusic2024}) extends this paradigm from
captioning to instruction following. It contains question--answer pairs
generated by ChatGPT from MusicCaps captions, including specific questions
about attributes such as tempo, mood, genre, and instrumentation. 

MIDICaps (\cite{melechovskyMidiCapsLargescaleMIDI2024}) instead generates captions for over 168k files from the Lakh MIDI Dataset. Key, time signature, tempo, and instrumentation are extracted directly from MIDI, while
genre, mood, and chord features are estimated from synthesized audio using Essentia and Chordino. Claude 3
Opus then verbalizes these features. This produces a
large resource for applications such as text-to-MIDI generation, but captions do not preserve the
sequence, timing, or local realization of the underlying musical events within one audio fragment.

\cite{bukeyRethinkingMusicCaptioning2026a} perform captioning based on musical-content prediction outcomes. Their system predicts structured attributes such as genre and mood from audio and supplies that
data to an LLM at inference time. Its descriptive scope, however, remains constrained by the predicted metadata
schema and does not provide event-level temporal grounding.

MusicBench (\cite{melechovskyMustangoControllableTexttoMusic2024}) likewise consists of 53K music-caption pairs containing musical attributes like chords, key, and beats on 10-second clips.

FUTGA (\cite{wuFutgaFinegrainedMusic2024}) proposes a model that synthesizes captions with transition boundaries for complete songs, thereby augmenting MusicCaps and the Song Describer datasets. SonicVerse (\cite{chopraSonicVerseMultiTaskLearning2025a}) more directly
incorporates audio by jointly learning caption generation and auxiliary
musical-feature prediction. For long recordings, it divides the audio into
fixed-length segments (10 s), generates a caption for each segment, and prompts GPT-4
to combine the chronologically ordered captions into a continuous description
with timestamps. This provides temporal localization at the coarseness level of the analysis window. 

\cite{zhouUnlockingStrongSupervision2026} propose a tagging system to generate audio pre-training data at scale, but their system builds on existing captioners in doing so. Music-to-Text Synaesthesia (\cite{kuangMusictoTextSynaesthesiaGenerating2023a}) collects 1,955 pairs of
Western classical recordings and rich yet composition-level or movement-level descriptions from EARSENSE, a chamber-music
database. 

Other work prioritizes the reliability of human-authored descriptions. The Song
Describer Dataset (\cite{mancoSongDescriberDataset2023}) contains approximately
1,100 crowdsourced descriptions of 706 recordings from MTG-Jamendo (\cite{bogdanov2019mtg}). These
captions provide high-quality, but coarse, data and reduce dependence on
automatically predicted attributes, but the expense of human annotation limits
the dataset's scale. 

~\tabref{tab:caption_examples} illustrates the granularity of representative
captions. Most emphasize global properties such as genre, mood,
instrumentation, tempo, harmony, or the character of a work. Recent systems
introduce structural boundaries or coarse timestamped windows, but existing
resources do not provide dense, human-verified descriptions that include precisely identified
musical events, the evolution of melodies and textures,
or the realization of phrasing, articulation, dynamics, and timing in a
particular performance across entire works. Fine-grained music understanding requires
descriptions that are musically precise, explicitly grounded in
time, and expressed at the level used by musicians.

\begin{table*}[t]
\centering
\scriptsize
\setlength{\tabcolsep}{5pt}
\renewcommand{\arraystretch}{1.05}
\begin{tabular}{@{}p{0.17\textwidth}p{0.78\textwidth}@{}}
\toprule
\textbf{Dataset} & \textbf{Example caption} \\
\midrule

LP-MusicCaps &
This gentle guitar song featuring a piano, percussion, and a soothing
melody is perfect for relaxation with a slow tempo. \\
\midrule

MusicInstruct &
Q: Could you provide a detailed description of
the church choir song?
A: The song is a prime example of a traditional
church choir performance, one recorded in an era
past. The arrangement sees the choir members joining their voices in an expressive harmony, imbued
with a devout and contemplative essence. The pace
of the song is moderated, neither too slow nor too
fast, which enhances its immersive nature. Accompanying the choir is a church organ that amplifies
the sacred ambiance without overshadowing the [...] \\
\midrule

MIDICaps &
A melodic and happy rock and pop song featuring a string ensemble,
piano, clean electric guitar, slap bass, and drums. The song is in the
key of F major with a 4/4 time signature and a tempo of 120 BPM. The
chord progression alternates between B$\flat$ and F throughout the song,
creating a motivational and energetic corporate vibe. \\
\midrule

MusicBench &
This mellow instrumental track showcases a dominant electric guitar
that opens with a descending riff, followed by arpeggiated chords,
hammer-ons, and a slide. The percussion section keeps it simple with rim
shots and a common time count, while the bass adds a single note on the
first beat of every bar. Minimalist piano chords round out the song while
leaving space for the guitar to shine. There are no vocals, making it
perfect for a coffee shop or some chill background music. The key is in
E major, with a chord progression that centers around that key and a
straightforward 4/4 time signature. \\
\midrule

FUTGA &
The song has a mellow and calming mood. The theme is serene and
contemplative. The tempo is moderate. The melody is simple and
memorable. The instruments used in the song include a bansuri, tabla,
string instruments, and a piano.
The music begins with a slow, sustained
note from the bansuri. The tabla joins in with a rhythmic pattern. The
string instruments join in with a simple melody. The tempo increases
slightly. The bansuri plays a more prominent role in the melody.
[\ldots] \\
\midrule

SonicVerse &
``Bohemian Rhapsody'' is a dynamic and genre-blending piece that takes
listeners on a musical journey. The song begins with a gospel-inspired
section, featuring a choir of female vocalists singing harmoniously in a
major key, accompanied by a keyboard playing the melody. The tempo is
brisk at 171.0 beats per minute, creating an uplifting and spiritual
atmosphere. Around the 60-second mark, the song transitions into a
slower tempo, introducing a male vocalist singing a soft melody
accompanied by a piano and a cello playing a single note. [\ldots] \\
\midrule

\cite{zhouUnlockingStrongSupervision2026} &
The audio begins with a lively atmosphere in a large, reverberant indoor
space, likely a gymnasium or auditorium. A male voice, energetic and
encouraging, shouts, ``Alright. Good,'' as a crowd of mixed ages reacts
with cheers and applause. A bright, synthesized melody reminiscent of
the ``Mega Man'' video game series plays in the background, accompanied
by a male vocalist singing, ``I'm a hero,'' and ``It's a hero's life.''
The music and crowd noise overlap, creating a festive, communal
ambiance. Shortly after [\ldots] \\
\midrule

Music-to-Text Synaesthesia &
Perhaps the work's center of gravity is the slow movement,
\textit{Andante espressivo}. The motif is given a peaceful, and
extraordinarily beautiful treatment. The stormy middle section makes a
very strong impression. \\
\midrule

Song Describer Dataset &
A folky song with a warm, organic and cosy sound, sung by a single male
voice and accompanied by a number of acoustic guitars. \\
\bottomrule
\end{tabular}

\caption{Representative examples illustrating the granularity of
existing music captions. Most emphasize global properties such as genre,
mood, instrumentation, tempo, harmony, or the general character of a
work. Recent systems introduce structural boundaries or coarse
timestamped windows, but existing resources do not provide dense descriptions of precisely identified musical events across
entire works. The examples are reproduced from the cited papers or their
official dataset releases, with typographic normalization and editorial
truncation indicated by [\ldots]. In general, these are the first examples within their respective containers, or the most relevant ones. The FUTGA example combines the global and segment-level descriptions presented together in its source figure.}

\label{tab:caption_examples}
\end{table*}

\subsection{Music Performance Analysis}

Music Performance Analysis (MPA) examines how performers interpret and realize a musical score in sound. Existing work commonly uses statistical analysis as well as machine-learning methods to identify performance patterns and relationships between notated and performed attributes.
For instance, in jazz performance, \cite{inproceedings} use statistical tests to investigate relationships between dynamics and features such as note duration, pitch, and metrical position, finding that higher and longer notes tend to be played more loudly. \cite{chenDualContrastiveSparseAutoencoders} use a dual-contrastive sparse autoencoder to extract interpretable features from residual-stream activations of a frozen music transformer; these features characterize both composition- and performance-related attributes. 

Research on classical music has examined the relationship between notation and performance. \cite{Kosta03052016} study the correspondence between notated dynamics and performed loudness in Chopin mazurkas. Using models like decision trees, support vector machines, and neural networks, they predict performed loudness from dynamic markings and vice versa. Prediction works comparatively well across performances of the same piece but poorly when models are trained on a pianist's performances of different pieces, suggesting that piece-specific score context contributes more strongly than individual style.

Several datasets provide expert annotations of musical performances. PercePiano (\cite{parkPianoPerformanceEvaluation2024}), for example, contains expert assessments along 19 perceptual dimensions, including pedaling (\textit{clean--blurred}), timing (\textit{stable--unstable}), and emotion (\textit{pleasant--sad}), for piano performances of classical repertoire. \cite{zhangHowDoesTeacher2024} analyze NeuroPiano, a dataset that contains annotations by professional pianists of audio recordings in which students perform six technical exercises, including scales, octaves, chords, and arpeggios.  These resources nevertheless address restricted contexts: NeuroPiano focuses on short technical exercises, while PercePiano provides ratings for excerpts of at most 16 bars. Neither dataset supplies long-form, context-sensitive expert explanations covering complete performances. 

\section{Unifying Score and Performance}
\vspace {-10pt}
In this section, we outline the design criteria for MuNo-SP, its syntax, and the evaluation methods used.

MuNo-SP unifies notated musical content, including pitch, rhythm, dynamics,
articulation, phrasing, and written voices, with aligned performance
information, covering timing, duration, velocity, and pedal activity. We design
the representation according to three criteria:
\vspace {-10pt}
\begin{itemize}
    \item \textbf{Efficient:} Compact relative to expressive formats such as
    MusicXML.
    \item \textbf{Interpretable:} Organized through a musical hierarchy and readable by LLMs.
    \item \textbf{Expressive:} Able to represent detailed score and performance information.
\end{itemize}


\pagebreak
\vspace*{-1.7\baselineskip}
\subsection{Syntax}

\textbf{MuNo-SP} is a line-oriented, bar-ordered, beat-organized representation that unifies a
written score with an aligned performance. Its score-only variant,
\textbf{MuNo-S}, retains the same structure but without performance
information. Each file begins with an \munocode{INSTRUMENTS} declaration and
follows a hierarchical \munocode{BAR} $\rightarrow$ \munocode{BEAT}
$\rightarrow$ \munocode{instrument} $\rightarrow$ \munocode{voice} organization. The instrument layer is omitted for solo
music, and \munocode{v<N>} voice wrappers are retained only where independent or
overlapping written voices must be distinguished.

Each bar and beat header in MuNo-SP records its performed onset, total duration, and rounded mean note velocity, as in
\munocode{BAR 3: 00:09.69+04.27 v59}. Onsets use \munocode{MM:SS.cc}; durations use \munocode{SS.cc} or \munocode{MM:SS.cc}. Header
velocities average the performed notes in that scope, counting chord members
separately.

Note lines contain pitch, written duration, MIDI velocity, performed onset, and sounding duration, for example
\munocode{C\#4 qtr v64 00:01.25+00.48}. Chords use comma-separated pitches with correspondingly ordered velocities. Rests are written
as \munocode{rest 8th}; their timing is implicit. Tuplets, dots,
articulations, ornaments, ties, slurs, fermatas, arpeggios, grace notes, and
cues remain attached to their written events.

Dynamics, tempo indications, expression markings, pedal events, and
other markings are nested within the beat where they occur. MIDI sustain transitions appear as timestamped
\munocode{pedal-start} and \munocode{pedal-stop} lines and are assigned by
proximity to consecutive beat onsets.

\begin{figure*}[!b]
    \centering
    \includegraphics[width=\textwidth]{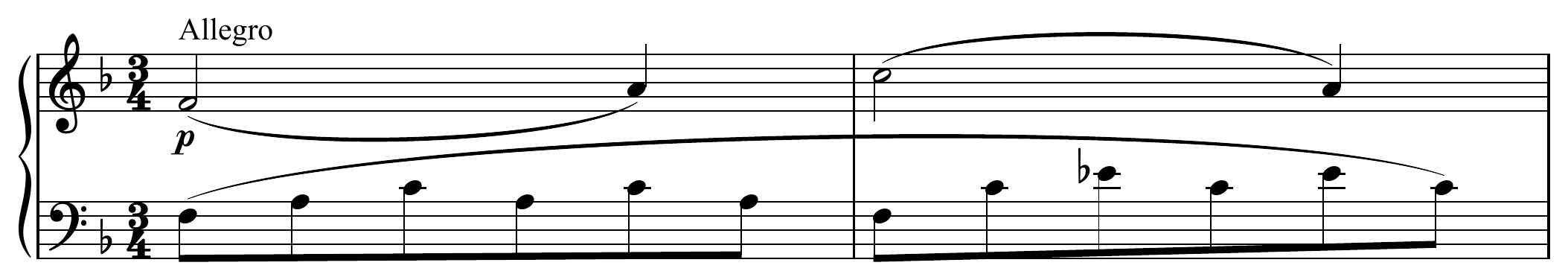}
    \caption{Bars 1--2 of Mozart's \textit{Piano Sonata No.~12 in F major,
    K.~332/I}, showing the opening melody above its broken-chord accompaniment
    as represented in~\exref{munoblock:mozart-example}.}
    \label{fig:mozart-k332-opening}
\end{figure*}

Written repeats are expanded as explicit \munocode{REPEAT BAR N} blocks when
repeated in the performance. Each performed occurrence retains its own timing,
duration, velocity, markings, and pedal data. These conventions are illustrated
in~\exref{munoblock:mozart-example} using a short
score--performance excerpt from Recording~1. Recording identifiers for all
main-text examples are listed in~\appref{app:main-text-recordings}.~\figref{fig:mozart-k332-opening} shows its corresponding score.

\begin{musicbox}{The first two bars of Mozart's \textit{Piano Sonata No.~12 in F major, K.~332/I} in MuNo-SP representation}

\begin{Verbatim}[
fontfamily=\rmdefault,
breaklines=true,
breakanywhere=true,
samepage=false
]
INSTRUMENTS piano

BAR 1: 00:01.05+01.16 v44
TIME SIGNATURE 3/4 -- 3 beats per bar
  BEAT 1: 00:01.05+00.45 v41
    pedal-start v127 00:00.00
    ALLEGRO role:local-expression
    piano (p)
    v1:
      F3 8th v38 00:01.07+01.14 slur:1:start
      A3 8th v37 00:01.31+00.37
    v4:
      F4 half v47 00:01.05+01.23 slur:4:start
  BEAT 2: 00:01.50+00.37 v41
    v1:
      C4 8th v37 00:01.50+00.39
      A3 8th v44 00:01.68+00.35
    v4:
      ---
  BEAT 3: 00:01.87+00.34 v49
    v1:
      C4 8th v34 00:01.89+00.38
      A3 8th v55 00:02.04+00.24
    v4:
      A4 qtr v58 00:01.84+00.43 slur:4:end

BAR 2: 00:02.21+01.12 v55
  BEAT 1: 00:02.21+00.39 v53
    pedal-stop v51 00:02.27
    pedal-start v68 00:02.39
    v1:
      F3 8th v39 00:02.22+01.11
      C4 8th v51 00:02.41+00.36
    v4:
      C5 half v69 00:02.21+01.17 slur:4:start
  BEAT 2: 00:02.60+00.37 v57
    v1:
      Eb4 8th v55 00:02.60+00.38
      C4 8th v58 00:02.76+00.39
    v4:
      ---
  BEAT 3: 00:02.97+00.36 v56
    v1:
      Eb4 8th v47 00:02.98+00.39
      C4 8th v53 00:03.15+00.22 slur:1:end
    v4:
      A4 qtr v68 00:02.96+00.41 slur:4:end
\end{Verbatim}

\end{musicbox}
\munocodeblockcaption{MuNo-SP representation of bars 1--2 of Mozart's
\textit{Piano Sonata No.~12 in F major, K.~332/I}.}
{munoblock:mozart-example}


\providecommand{\result}[2]{}
\renewcommand{\result}[2]{#1\,{\scriptsize\textcolor{black!45}{[#2]}}}
\providecommand{\notapp}{}
\renewcommand{\notapp}{\textcolor{black!45}{\textnormal{N/A}}}
\providecommand{\notrun}{}
\renewcommand{\notrun}{\textcolor{black!45}{\textnormal{---}}}
\providecommand{\partialrun}{}
\renewcommand{\partialrun}{\textcolor{black!45}{\textnormal{partial}}}

\providecommand{\comparisonrule}{}
\renewcommand{\comparisonrule}{\specialrule{0.045em}{1.2pt}{0.6pt}}
\providecommand{\sectionrule}{}
\renewcommand{\sectionrule}{\specialrule{0.18em}{2.2pt}{1.4pt}}

\begin{table*}[b]
\centering
\begingroup
\fontsize{6.5pt}{7.2pt}\selectfont
\setlength{\tabcolsep}{3pt}
\renewcommand{\arraystretch}{1.05}

\begin{tabular*}{\textwidth}{@{\extracolsep{\fill}}llrccc@{}}
\toprule
\textbf{Subset} &
\textbf{Input} &
\textbf{\#Q} &
\shortstack{\textbf{GPT-5.6 Sol}} &
\shortstack{\textbf{Muse Spark 1.2}} &
\shortstack{\textbf{Qwen3.6-Plus}} \\
\specialrule{0.10em}{0pt}{0.7pt}

\multirow{4}{*}{Score only (S)}
& ABC
& 180
& 51.1
& 31.7
& 11.1 \\

& Image
& 180
& 27.2
& 15.0
& 13.3 \\

\cmidrule(lr){2-6}

& MuNo-S
& 180
& \textbf{83.3}
& \textbf{65.6}
& \textbf{20.6} \\

& MuNo-SP
& 180
& 82.8
& 62.8
& 15.6 \\

\comparisonrule

\multirow{3}{*}{Performance only (P)}
& MIDI-as-text (integer)
& 106
& 32.1
& 26.4
& 3.8 \\

& MIDI-as-text (ABC)
& 106
& 59.4
& 30.2
& 10.4 \\

\cmidrule(lr){2-6}

& MuNo-SP
& 106
& \textbf{76.4}
& \textbf{57.5}
& \textbf{22.6} \\

\comparisonrule

\multirow{2}{*}{\shortstack[l]{Score and\\performance (SP)}}
& ABC+MIDI
& 72
& 45.8
& 30.6
& 6.9 \\

\cmidrule(lr){2-6}

& MuNo-SP
& 72
& \textbf{80.6}
& \textbf{52.8}
& \textbf{20.8} \\

\comparisonrule

\multirow{6}{*}{\shortstack[l]{Score or performance\\(S/P)}}
& ABC
& 84
& 63.1
& 53.6
& 34.5 \\

& MIDI-as-text (integer)
& 84
& 42.9
& 51.2
& 11.9 \\

& MIDI-as-text (ABC)
& 84
& 65.5
& 46.4
& 25.0 \\

& Image
& 84
& 47.6
& 57.1
& 35.7 \\

\cmidrule(lr){2-6}

& MuNo-S
& 84
& 78.6
& \textbf{67.9}
& 40.5 \\

& MuNo-SP
& 84
& \textbf{81.0}
& 66.7
& \textbf{42.9} \\

\sectionrule

\multirow{4}{*}{\shortstack[l]{Score-available\\comparison}}
& ABC
& 264
& 54.9
& 38.6
& 18.6 \\

& Image
& 264
& 33.7
& 28.4
& 20.5 \\

& MuNo-S
& 264
& 81.8
& \textbf{66.3}
& \textbf{26.9} \\

& MuNo-SP
& 264
& \textbf{82.2}
& 64.0
& 24.2 \\

\comparisonrule

\multirow{3}{*}{\shortstack[l]{Performance-text\\comparison}}
& MIDI-as-text (integer)
& 190
& 36.8
& 37.4
& 7.4 \\

& MIDI-as-text (ABC)
& 190
& 62.1
& 37.4
& 16.8 \\

& MuNo-SP
& 190
& \textbf{78.4}
& \textbf{61.6}
& \textbf{31.6} \\

\comparisonrule

\multirow{2}{*}{\shortstack[l]{Joint score--performance\\comparison}}
& ABC+MIDI
& 72
& 45.8
& 30.6
& 6.9 \\

& MuNo-SP
& 72
& \textbf{80.6}
& \textbf{52.8}
& \textbf{20.8} \\

\sectionrule

\multirow{4}{*}{Aggregate scores}
& Baseline mean
& 442
& 44.8
& 31.2
& 12.9 \\

& Baseline best
& 442
& 55.0
& 36.0
& 15.8 \\

\cmidrule(lr){2-6}

& MuNo mean
& 442
& 80.4
& 61.3
& 24.1 \\

& MuNo best
& 442
& \textbf{80.8}
& \textbf{62.0}
& \textbf{25.3} \\

\bottomrule
\end{tabular*}
\endgroup

\caption{Accuracy (\%) by question subset, input representation, and model.
MIDI-as-text (integer) replaces pitches with MIDI note numbers, whereas
MIDI-as-text (ABC) represents performance events using named pitches.
ABC+MIDI preserves the symbolic ABC score while aligning its written notes with
performance onset, offset, velocity, and pedal information.
The General (S/P) subset combines the general and tonality/style questions from
the benchmark. We exclude composer/work identification questions since the
training pipeline will have access to this information. Matched-coverage
comparisons use identical question sets. Aggregate scores are weighted by the
number of questions in each subset. Mean scores first average the available
representations within each subset and then pool the subset scores over all
442 questions; best scores select the strongest representation within each
subset before pooling over the same questions. Bold indicates the best input
representation for each model within each comparison group. Thin rules separate
individual comparisons, while heavier rules separate the main table sections.}
\label{tab:main-results}
\end{table*}

\subsection{Evaluation of Score and Performance Understanding}

Having defined the syntax of MuNo-SP, we next evaluate how effectively it
communicates score content, performance realization, and their relationship to language models. We use MuSP-Bench, a benchmark assessing both score
and performance understanding. Comparison with alternative representations on
this benchmark provides evidence of how effectively MuNo-SP makes its encoded
musical information accessible to language models.

MuSP-Bench comprises 490 open-ended questions spanning 24 classical works.
Questions are categorized by the evidence required: score only (S),
performance only (P), joint score and performance (S\&P), or either score or
performance independently (S/P). They cover pitch, temporal organization,
performance realization, melody, harmony, orchestration, structure,
interpretation, and musical context, with reasoning horizons ranging from
individual events to whole-piece analysis. Each question specifies a constrained answer format, such as a pitch, chord,
timestamp, measure range, numerical value, or brief musical description.

We assess MuNo across three model families using OpenAI's GPT-5.6 Sol (\cite{openAIGPT56SystemCard2026}),
Meta's Muse Spark~1.2 (\cite{metaMuseSpark2026}), and Alibaba's Qwen3.6-Plus (\cite{teamQwen36Plus2026}). We evaluate MuNo-S and
MuNo-SP, the score and performance-enriched blocks of MuNo-SP, where applicable
to each question subset. Each model receives the complete work-level
representation, the benchmark question, identical response instructions, and
a concise representation-specific syntax description.

We compare MuNo against four baseline families: ABC notation, score images,
MIDI performance events serialized as text, and ABC+MIDI. The MIDI baseline
is evaluated with both named-pitch and integer-pitch encodings. ABC+MIDI
preserves the ABC score and aligns written notes with performed onset, offset,
velocity, and pedal information through parallel comment rows. Baseline
results are taken from \cite{liessensdujardin2026muspbench} and are compared
with MuNo only on matched question sets. Our primary representation comparison
excludes 48 composer- and work-identification questions, for which the
training pipeline already supplies identifying metadata, leaving 442
questions.

\subsection{Results}

\subsubsection{MuNo enables greater understanding}

Across all three model families and every question subset, the strongest MuNo
representation outperforms the conventional baselines of ABC, MIDI-as-text,
score images, and ABC+MIDI, as shown in~\tabref{tab:main-results}. For joint
score-and-performance (S\&P) questions, MuNo-P outperforms ABC+MIDI for all
three models. These results suggest that MuNo presents musically relevant
structure in a form that is accessible to language models.

This advantage also holds in the aggregate results, which are weighted over
the same 442 questions. Selecting the strongest representation within each
subset, GPT-5.6 Sol achieves 80.8\% with MuNo, compared with 55.0\% for the
conventional baselines. The corresponding results are 62.0\% versus 36.0\%
for Muse Spark~1.2 and 25.3\% versus 15.8\% for Qwen3.6-Plus.

\subsubsection{GPT-5.6 Sol leads}

GPT-5.6 Sol is the strongest model evaluated with MuNo. On score-only
questions, it reaches 83.3\% with MuNo-S and 82.8\% with MuNo-SP, compared with
51.1\% using ABC. Across the combined set of
264 questions for which score-based inputs are available, MuNo-S and MuNo-SP
achieve 81.8\% and 82.2\%, respectively, compared with 54.9\% for ABC and
33.7\% for score images.

On performance-only questions, MuNo-SP reaches 76.4\%, an improvement of 17.0 percentage points over MIDI-as-text with notes written with ABC pitch notation. Across the matched set
of 190 performance-text-comparable questions, MuNo-SP similarly achieves
78.4\%, compared with 62.1\% for MIDI-as-text (ABC).

The largest gain occurs for joint score-and-performance questions. MuNo-SP
achieves 80.6\%, compared with 45.8\% for ABC+MIDI, an improvement of
34.8 percentage points. For the General (S/P) subset, MuNo-SP reaches 81.0\%
and MuNo-S reaches 78.6\%, while the strongest conventional baseline,
MIDI-as-text, reaches 65.5\%.

The results on MuSP-Bench make GPT-5.6 Sol the strongest candidate for the caption-generation model: it combines high score-reading accuracy with substantial gains in performance and joint score--performance reasoning. Because it does not achieve perfect accuracy, struggling, for example, to identify every occurrence of a fugue subject or analyze complex chord progressions, we further assess the quality of the resulting captions in~\tabref{sec:evaluation}.

\begin{table}[h]
\centering
\vspace{2pt}
\small
\setlength{\tabcolsep}{4pt}
\renewcommand{\arraystretch}{1.10}

\resizebox{\columnwidth}{!}{%
\begin{tabular}{@{}llrccccc@{}}

\toprule
\textbf{Evaluation set} &
\textbf{Horizon} &
\textbf{\#Q} &
\shortstack{\textbf{Best}\\\textbf{baseline}} &
\textbf{MuNo-S} &
\shortstack{\textbf{MuNo-S}\\\textbf{improvement}} &
\textbf{MuNo-SP} &
\shortstack{\textbf{MuNo-SP}\\\textbf{improvement}} \\
\midrule

\multirow{5}{*}{\shortstack[l]{Score-only}}
& Short
& 59
& 47.5\%
& 88.1\%
& \gain{40.7}
& \textbf{89.8\%}
& \textbf{\gain{42.4}} \\

& Mid
& 58
& 48.3\%
& \textbf{79.3\%}
& \textbf{\gain{31.0}}
& \textbf{79.3\%}
& \textbf{\gain{31.0}} \\

& Long
& 99
& 52.5\%
& \textbf{76.8\%}
& \textbf{\gain{24.2}}
& 74.7\%
& \gain{22.2} \\

& Any
& 48
& 77.1\%
& 87.5\%
& \gain{10.4}
& \textbf{91.7\%}
& \textbf{\gain{14.6}} \\

& \textbf{Overall}
& \textbf{264}
& \textbf{54.9\%}
& \textbf{81.8\%}
& \textbf{\gain{26.9}}
& \textbf{82.2\%}
& \textbf{\gain{27.3}} \\

\specialrule{1.2pt}{5pt}{5pt}

\multirow{5}{*}{\shortstack[l]{Performance-\\inclusive}}
& Short
& 94
& 48.9\%
& \notapp
& \notapp
& \textbf{88.3\%}
& \textbf{\gain{39.4}} \\

& Mid
& 123
& 48.0\%
& \notapp
& \notapp
& \textbf{78.9\%}
& \textbf{\gain{30.9}} \\

& Long
& 177
& 55.9\%
& \notapp
& \notapp
& \textbf{74.6\%}
& \textbf{\gain{18.6}} \\

& Any
& 48
& 77.1\%
& \notapp
& \notapp
& \textbf{91.7\%}
& \textbf{\gain{14.6}} \\

& \textbf{Overall}
& \textbf{442}
& \textbf{54.5\%}
& \notapp
& \notapp
& \textbf{80.5\%}
& \textbf{\gain{26.0}} \\

\bottomrule
\end{tabular}%
}

\caption{GPT-5.6 Sol accuracy by evidence horizon. Short, Mid, and Long
contain non-context questions, while Any contains the 48 tonality/style
questions, whose evidence may span the entire piece. Score-only includes
questions answerable from score information (S and S/P). Performance-inclusive
includes all evaluated modalities. The best baseline result uses ABC for S and S/P
questions, MIDI-as-text for P questions, and ABC+MIDI for SP questions.
Improvements are absolute percentage-point gains over the corresponding
baseline on identical questions.}
\label{tab:horizon-results}
\end{table}
~\tabref{tab:horizon-results} shows per-horizon results for questions requiring access to the score only, and those requiring access to performance, with or without score. Advantages are the largest for short-horizon questions.

\subsubsection{Token Efficiency}

We measure each text representation using OpenAI's
\munocode{GPT-5.x} tokenizer, used for GPT-5.6, and compare it against timed
MusicXML, an XML representation containing performance timing information for
each note. Averages use the same 24 evaluation pieces.

As shown in~\tabref{tab:token-efficiency}, all compact representations are
substantially smaller than timed MusicXML---the original score in which each note has an associated timestamp---with average reductions compared to timed XML ranging from
84.4\% to 97.6\%. MuNo-S averages 33,973 tokens (7.0\% of timed MusicXML),
while the more detailed MuNo-SP averages 75,779 tokens (15.6\%). The average
token-count gap between MuNo-S and ABC is 22,428 tokens, larger than the
12,463-token gap between MuNo-SP and ABC+MIDI. Although ABC+MIDI is more
compact, MuNo-SP offers an advantage in interpretability.
\providecommand{\munocode}[1]{\texttt{#1}}

\begin{table}[h]
\centering
\small
\setlength{\tabcolsep}{3pt}
\renewcommand{\arraystretch}{1.08}

\resizebox{\columnwidth}{!}{%
\begin{tabular}{@{}lrrrrl@{}}

\toprule
\textbf{Format} &
\textbf{Avg.\ tokens} &
\textbf{XML (\%)} &
\textbf{Reduction (\%)} &
\textbf{Max.\ tokens} &
\textbf{Longest example} \\
\midrule

MuNo-S
& 33,973
& 7.0
& 93.0
& 105,253
& Balakirev: \textit{Islamey} \\

MuNo-SP
& 75,779
& 15.6
& 84.4
& 228,257
& Balakirev: \textit{Islamey} \\

\midrule

ABC
& 11,545
& 2.4
& 97.6
& 42,341
& Scriabin: \textit{Piano Sonata No.\ 5} \\

MIDI-as-text
& 57,692
& 11.9
& 88.1
& 192,080
& Liszt: \textit{Ballade No.\ 2} \\

ABC+MIDI
& 63,316
& 13.1
& 86.9
& 211,651
& Scriabin: \textit{Piano Sonata No.\ 5} \\

\midrule

Timed MusicXML
& 484,246
& 100.0
& 0.0
& 1,456,624
& Scriabin: \textit{Piano Sonata No.\ 5} \\

\bottomrule
\end{tabular}%
}
\caption{Average OpenAI GPT-5.x tokenizer counts over the 24 evaluation
pieces. Percentages and reductions are relative to timed MusicXML. The final
columns give the longest example for each format.}
\label{tab:token-efficiency}
\end{table}

\section{Caption Generation Pipeline}

Our goal is to produce captions that leverage the rich symbolic information
contained in musical scores to support the training of audio-language models
(ALMs). Existing collections of classical music provide scores paired with
audio, performance MIDI, or both. Our pipeline converts these resources into
MuNo-SP representations from which fine-grained training data can be
generated.

\begin{figure*}[t]
    \centering
    \includegraphics[width=\textwidth]{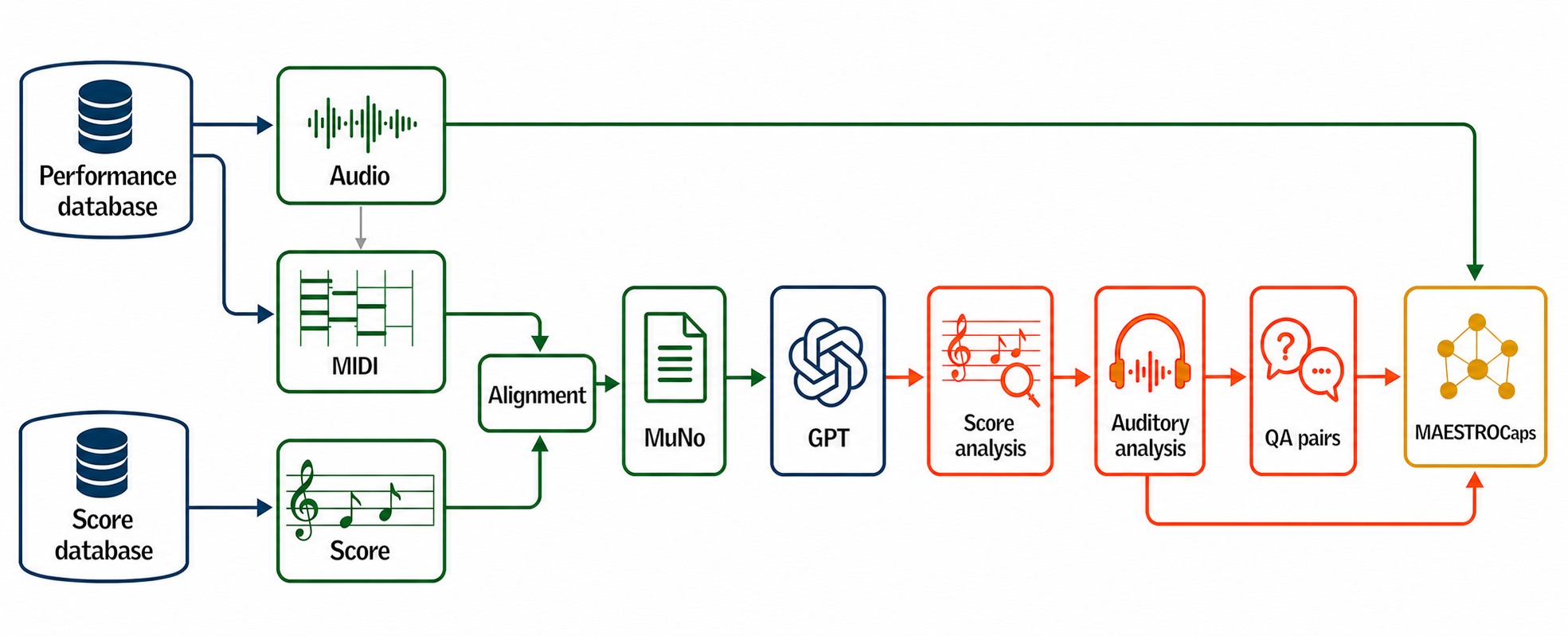}
    \caption{Overview of the data-generation pipeline. Performance audio and MIDI are paired with symbolic scores, and the resulting data are converted into the MuNo-S and MuNo-SP representations. Language models use these representations for score analysis, performance-informed auditory analysis, and question--answer generation. These outputs, together with the corresponding audio, are consolidated into MAESTROCaps.}
    \label{fig:pipeline}
\end{figure*}

The first part of the pipeline consists of four stages: data collection, score
cleaning, score--performance alignment, and conversion into MuNo-SP. The
second part generates auditory analyses from MuNo-SP and its score-only variant MuNo-S,
followed by question--answer pairs derived from those analyses. The full
pipeline is shown in ~\figref{fig:pipeline}.

\newpage

\vspace*{-2\baselineskip}

\subsection{Data Collection}
\paragraph{Piano Works}

We construct the piano portion of the corpus from four linked resources:
MAESTRO (\cite{hawthorneEnablingFactorizedPiano2019a}), ASAP
(\cite{asap-dataset}), the associated (n)ASAP alignment annotations
(\cite{Peter-2023}), and PianoCoRe
(\cite{borovikPianoCoReCombinedRefined}). ASAP aligns 222 MusicXML scores,
obtained from the online MuseScore repository, with 1,068 performances. Of
these, over 500 originate from MAESTRO, which provides synchronized audio and
performance MIDI for Western classical piano music. 
As a result, each pair provides three complementary views of the same performance: the
written notation and performance instructions encoded in the score;
the notes, velocities, timings, and sustain-pedal events captured in the
performance MIDI; and the corresponding acoustic realization in the audio. Together, the ASAP scores
linked to MAESTRO cover 179 pieces or movements.

\paragraph{Orchestral Works}

As proof of extensibility beyond piano music, we run our pipeline on the Beethoven Symphony Expression Dataset
(BSED) (\cite{berendesBeethovenSymphonyExcerpt}), which contains 20 orchestral
excerpts paired with four recorded performances each. BSED provides MusicXML,
audio, and note-level score--performance annotations. Because its annotations
do not include velocities, we approximate them from the dynamic markings in
the score using fixed values and interpolation for gradual changes in loudness.

\subsection{Data Cleaning}

The MusicXML scores used by ASAP were obtained from public repositories and may contain transcription or engraving errors that persist in datasets derived from ASAP. We therefore clean the scores using evidence aggregated across aligned MIDI performances, treating consensus across performances as the primary source of evidence.

We supplement this evidence with symbolic scores collected from online sources for works by Bach, Beethoven, Chopin, Mozart, and Scriabin, as well as with public-domain editions from IMSLP. A complete list of the score collections and their coverage is provided in~\appref{app:score-sources}. Because neither symbolic scores nor automatic performance alignments are uniformly reliable, corrections are accepted through threshold-based consensus and subsequently reviewed manually.

We audit the 179 MAESTRO-linked ASAP score folders at the movement level. We filter out alternate score realizations and those available with fewer than 10 PianoCoRe recordings. 

\paragraph{Score Correction Algorithm}

We generate deletion candidates from score notes marked absent in the (n)ASAP
alignments and, where an external symbolic score is available, from differences
between that score and the ASAP score. For each candidate, we estimate a local
performance-time window from neighboring aligned score events. Pitches found
in that window provide insertion or replacement hypotheses; a same-pitch
predecessor provides a possible tie explanation. If we find more than 50 deletions, we cancel correction for that score.

We evaluate the hypotheses against the available (n)ASAP and PianoCoRe
alignments using explicit source-specific threshold routes. PianoCoRe evidence
is used only when at least ten alignments are valid and they cover at least half
of the available recording pool. A simultaneous written unison in another
voice blocks deletion, and a supported tie is handled before deletion. An
insertion or replacement is materialized only when its proposed pitch meets an
applicable performance- or reference-supported criterion.~\appref{app:score-correction-procedure} gives the complete decision rules.

For each of the \MAESTROCapsPieceCount{} catalog entries, the release includes the source-preserving,
annotated, and cleaned MusicXML variants together with separate records of all
assessed candidates and accepted changes. These records retain the location,
candidate source, evidence counts, threshold route, decision, and materialized
outcome.

\subsection{Score--Performance Alignment}

\paragraph{MAESTRO-Linked Performances}

PianoCoRe initializes score--performance correspondences with Parangonar and
subsequently improves their temporal accuracy through refinement,
interpolation, and synchronization. Its score-informed MIDI cleaning assumes
that the reference score is correct; when that score contains transcription
errors, the resulting MIDI may be altered to match those errors and therefore
no longer faithfully preserve the captured performance. We instead retain the
original, unmodified performance MIDI and pair it with our cleaned score. We
then propagate PianoCoRe's refined correspondences through the note-level
conversion traces between the original and cleaned scores and the original and edited MIDI.

\paragraph{BSED Performances}

For BSED, we use the dataset-supplied note-level score--performance annotations. These annotations directly associate written score events with their corresponding performed events.

\subsection{MuNo-SP Conversion}

We convert each aligned score--performance pair into MuNo-SP in two steps. First, we incorporate the aligned performance attributes into the MusicXML structure, producing a timing-enriched intermediate representation. We then convert this representation into MuNo-SP.

\subsection{Automated Captioning}

The pipeline consists of three stages. First, GPT-5.6 Sol segments the score
into sections, thematic presentations, and short phrases. Second, it generates
long-form, performance-informed auditory analyses from MuNo-SP. Third, another
LLM generates question--answer pairs from those analyses. Two auditory-analysis excerpts can be viewed in~\exref{munoblock:auditory-analysis-excerpts}.

\begin{figure}[tbp]
\begin{musicbox}{Excerpts of auditory analyses generated by our pipeline}

\begin{itemize}
    \item \textbf{Debussy, \textit{Reflets dans l'eau}.} At
    04:52.46--04:55.11, \munocode{Db1+Ab1+Db2}, \munocode{Eb3+Ab3}, Eb4,
    Bb4, and \munocode{Bb5+Bb6} form a sustained plateau with no continuing
    melodic path, the delayed upper B-flat entry at 04:53.21 providing the
    only new attack. At
    04:55.03--04:57.68, the same D-flat/A-flat bass and E-flat/B-flat upper
    structure remains almost motionless, creating unresolved added-sixth and
    ninth color whose tension is held through duration and very low intensity.
    At 04:57.28--04:58.61, the final sonority enters in layers as
    \munocode{Ab3+Ab4}, \munocode{F3+F4}, Db1, and \munocode{Ab6+Ab7},
    spanning the complete register without an independent melody. At
    05:02.08--05:06.71, only Db1 and \munocode{Ab6+Ab7} remain explicitly
    renewed, and the open D-flat--A-flat fifth with F still resonating from
    04:57.50--05:01.45 leaves a pure D-flat-major frame that fades without
    further motion.
    \item \textbf{Beethoven, \textit{Piano Sonata No.~21 in C major,
    Op.~53, ``Waldstein''}, I.} At 00:00.80--00:03.40, softly detached C-major
    pulsations built from \munocode{C2+G2} and \munocode{C3+E3} establish an
    even, dry rhythmic field before \munocode{D3+F\#3} introduces
    dominant-directed tension. At 00:03.59, the first melodic gesture
    \munocode{G3--B3--A3--G3} projects clearly over \munocode{B1+G2}, with
    the final G3 at 00:04.25 shortened into a distinct release. At 00:04.93,
    the second gesture \munocode{C\#6--D6--C6--B5--A5--G5} descends from a
    forcefully projected high D6, while the persistent \munocode{B1+G2}
    sonority keeps the harmony suspended over dominant-function support. From
    00:05.66--00:06.16, the clipped G5 and continuing low pulsation leave a
    deliberate gap, preserving tension without a local resolution.

\end{itemize}
\end{musicbox}
\munocodeblockcaption{Representative time-grounded auditory-analysis excerpts
generated by the captioning pipeline for Debussy (Recording~2) and Beethoven
(Recording~3). \munocode{+} indicates notes form a chord, \munocode{--} a
sequence.}
{munoblock:auditory-analysis-excerpts}
\end{figure}

\subsubsection{Stage 1: Structural Segmentation}

The first stage involves dividing the score, represented as MuNo-S, into large-scale \textit{sections}, thematic \textit{presentations and returns}, and \textit{phrases} that typically span one to four bars using GPT-5.6 Sol. Each segment is assigned inclusive start and end bars, a musical function, and supporting evidence. The prompt, provided in~\appref{app:segmentation-prompt}, emphasizes grounding decisions in musical evidence, including slurs, thematic recurrence, cadences, harmony, texture, and phrase-length patterns. An example is provided in~\appref{app:example-segmentation}.  We include this intermediate stage because empirical validation showed that downstream analyses are more coherent and reliably structured when the model is first given an explicit formal scaffold.

\subsubsection{Stage 2: Long-form Auditory Analysis}
\label{enu:dimensions-analysis}

\paragraph{Requirements}
The structural segments are embedded into the MuNo-SP representation; the resulting text is sent to the model, which we prompt to generate captions that are:
\vspace{-6pt}
\begin{itemize}
    \item \textbf{Time-grounded:} Events and observations are localized within the recording using timestamps formatted as \munocode{MM:SS.cc}.
    \item \textbf{Precise:} Claims identify concrete musical and performance evidence: notes, harmonies, voices, texture, and expressive features.
    \item \textbf{Musician-like:} The analysis uses musical language and leverages indications in MuNo-SP, including slurs, bar structure, and segmentation information, to produce its description in the same way as a human might listen to the music.
    \item \textbf{Informative:} The caption prioritizes salient melody, harmony, texture, dynamics, rhythm, tension, and performance decisions.
\end{itemize}

The auditory analysis is created based on the following seven dimensions: 

\begin{itemize}
    \item \textbf{Melody:} Identify the principal melodic line and important repetitions, variations, transformations, and countermelodies.
    \item \textbf{Harmony and Tension:} Identify progressions and explain how harmony, bass motion, chromaticism, dissonance, and cadential motion create tension and release.
    \item \textbf{Texture:} Describe the register, balance, and interaction of melody, inner voices, bass, and accompaniment.
    \item \textbf{Dynamics and Phrasing:} Describe how intensity, emphasis, articulation, resonance, and breathing points shape the phrase.
    \item \textbf{Structural Function:} Explain the musical role of the phrase within its enclosing theme and section.
    \item \textbf{Rhythm and Time:} Characterize the pulse, rhythmic patterns, pacing, and temporal development of the phrase.
    \item \textbf{Performance Decisions:} Identify the clearest interpretive choices involving articulation, rubato, pacing, balance, or pedaling.
\end{itemize}

Formal requirements include a dedicated discussion of each phrase and a
synthesis for each theme, section, and complete piece.

The prompt is provided in~\appref{app:auditory-analysis-prompt}. 

Paired whole-piece analyses of
the same Rachmaninoff recording from segmented MuNo-SP and segmented MIDI are
provided in~\appref{app:auditory-analysis-examples}.



\subsubsection{Stage 3: Conversion into QA pairs}

The third stage of our generation pipeline consists of translating the musical analyses into specific question-answer pairs. We prompt an LLM to generate questions that are standalone, musically grounded, audible (i.e. answerable from listening alone), unambiguous, and informative, but, most importantly, reflect human inquiry. We use Gemini 3.7 Flash for this stage based on our preference for the quality and variety of its outputs. The prompt is provided in~\appref{app:qa-prompt}.

\section{The MAESTROCaps Dataset}

We introduce \textbf{MAESTROCaps}, a dataset of \MAESTROCapsPieceCount{} classical piano
performances aligned one-to-one with \MAESTROCapsPieceCount{} finalized score pieces. Every piece has one structural segmentation, one MuNo-S representation, one MuNo-SP representation, and one long-form auditory analysis. Each recording is accompanied by \MAESTROCapsQAPerPiece{} question--answer pairs, yielding \MAESTROCapsQACount{} pairs in total.~\tabref{tab:maestrocaps-composers} shows a breakdown of the dataset by composer.

\begingroup
\newcommand{\tblpct}[1]{#1\%}

\begin{table}[h]
    \centering
    \begin{tabular}{lrrrr}
        \toprule
        & \multicolumn{2}{c}{\textbf{Pieces}}
        & \multicolumn{2}{c}{\textbf{Audio}} \\
        \cmidrule(lr){2-3}
        \cmidrule(lr){4-5}
        \textbf{Composer}
        & \textbf{Count}
        & \textbf{Share}
        & \textbf{Hours}
        & \textbf{Share} \\
        \midrule
        Bach         & 56 & \tblpct{37.8} & 2.07 & \tblpct{19.95} \\
        Beethoven    & 37 & \tblpct{25.0} & 3.65 & \tblpct{35.20} \\
        Chopin       & 21 & \tblpct{14.2} & 1.45 & \tblpct{13.99} \\
        Liszt        &  9 & \tblpct{6.1}  & 0.82 & \tblpct{7.87}  \\
        Haydn        &  6 & \tblpct{4.1}  & 0.50 & \tblpct{4.77}  \\
        Schubert     &  4 & \tblpct{2.7}  & 0.40 & \tblpct{3.85}  \\
        Mozart       &  3 & \tblpct{2.0}  & 0.35 & \tblpct{3.41}  \\
        Rachmaninoff &  3 & \tblpct{2.0}  & 0.17 & \tblpct{1.64}  \\
        Schumann     &  3 & \tblpct{2.0}  & 0.24 & \tblpct{2.28}  \\
        Debussy      &  2 & \tblpct{1.4}  & 0.15 & \tblpct{1.42}  \\
        Scriabin     &  2 & \tblpct{1.4}  & 0.36 & \tblpct{3.42}  \\
        Balakirev    &  1 & \tblpct{0.7}  & 0.14 & \tblpct{1.32}  \\
        Glinka       &  1 & \tblpct{0.7}  & 0.09 & \tblpct{0.88}  \\
        \midrule
        \textbf{Total}
        & \textbf{\MAESTROCapsPieceCount}
        & \textbf{\tblpct{100.0}}
        & \textbf{\MAESTROCapsAudioHours}
        & \textbf{\tblpct{100.00}} \\
        \bottomrule
    \end{tabular}
    \caption{Composer distribution of pieces and audio duration across the
    \MAESTROCapsPieceCount{} aligned score--performance pairs in MAESTROCaps.
    Audio duration is computed from the aligned performance timelines. Row-level
    durations and shares are rounded for display, whereas totals are calculated
    from the underlying unrounded durations; consequently, displayed component
    values may not sum exactly to the reported totals.}
    \label{tab:maestrocaps-composers}
\end{table}

\endgroup

We further categorize the auditory analyses according to the seven dimensions and report their coverage in~\tabref{tab:auditory-analysis-characteristics}. Our algorithm uses word matching and counts sequences and chords as part of melody and harmony respectively.
\newcommand{\NumAuditoryAnalyses}{\MAESTROCapsPieceCount}

\newcommand{\WordAverage}{6374}
\newcommand{\TimestampAverage}{382}
\newcommand{\PitchMentionAverage}{1429}
\newcommand{\PitchSequenceAverage}{197}
\newcommand{\PitchCollectionAverage}{149}
\newcommand{\ChordMentionAverage}{18}

\newcommand{\MelodyPercent}{20.61}
\newcommand{\HarmonyPercent}{27.76}
\newcommand{\TexturePercent}{7.85}
\newcommand{\DynamicsPercent}{12.07}
\newcommand{\FunctionPercent}{15.78}
\newcommand{\RhythmPercent}{6.80}
\newcommand{\PerformancePercent}{9.14}

\begin{table}[t]
    \centering
    \small
    \setlength{\tabcolsep}{4pt}
    \begin{tabular}{lr}
        \toprule
        \multicolumn{2}{l}{\textbf{General characteristics}} \\
        \midrule
        Characteristic & Mean per analysis \\
        \midrule
        Words
            & \WordAverage \\
        Timestamps
            & \TimestampAverage \\
        Scientific-pitch mentions
            & \PitchMentionAverage \\
        Pitch sequences (\munocode{F3-G3-Ab3})
            & \PitchSequenceAverage \\
        Explicit pitch collections (e.g., \munocode{A4+C5})
            & \PitchCollectionAverage \\
        Explicit chord-term mentions
            & \ChordMentionAverage \\
        \midrule
        \multicolumn{2}{l}{\textbf{Musical dimensions}} \\
        \midrule
        Dimension & Share of mentions (\%) \\
        \midrule
        Melody
            & \MelodyPercent\% \\
        Harmony and tension
            & \HarmonyPercent\% \\
        Texture
            & \TexturePercent\% \\
        Dynamics and phrase shaping
            & \DynamicsPercent\% \\
        Function
            & \FunctionPercent\% \\
        Rhythm and time
            & \RhythmPercent\% \\
        Performance decisions
            & \PerformancePercent\% \\
        \bottomrule
    \end{tabular}
    \caption{Qualitative characteristics detected across
    \NumAuditoryAnalyses{} generated auditory-analysis files. General
    characteristics are reported as rounded arithmetic means per file.
    Scientific-pitch mentions include pitches occurring inside dash-separated
    sequences and plus-separated collections. Musical-dimension percentages report each dimension's share
    of all dimension assignments. Assignments are non-exclusive.}
    \label{tab:auditory-analysis-characteristics}
\end{table}

\providecommand{\munosp}{MuNo-SP}
\providecommand{\geminiflash}{Gemini 3.7 Flash}

\begin{table*}[h]
    \centering
    \small
    \begin{tabular}{llrrrp{0.27\textwidth}}
        \toprule
        Judge and scope
            & Dimension
            & \munosp
            & MIDI
            & Tie
            & Majority-vote outcome \\
        \midrule

        Human evaluation
            & Correctness
            & 46.51\% & 27.91\% & 25.58\%
            & \munosp{} 7/9; no majority 2/9 \\
        (9-comparison set)
            & Musician-likeness
            & 53.49\% & 41.86\% & 4.65\%
            & \munosp{} 7/9; MIDI 2/9 \\
        {}
            & Insightfulness
            & 62.79\% & 30.23\% & 6.98\%
            & \munosp{} 7/9; MIDI 1/9; no majority 1/9 \\
        {}
            & Overall preference
            & 58.14\% & 39.53\% & 2.33\%
            & \munosp{} 8/9; MIDI 1/9 \\

        \midrule

        \geminiflash
            & Correctness
            & 92.59\% & 7.41\% & 0.00\%
            & \munosp{} 8/9; MIDI 1/9 \\
        (9-comparison set as above)
            & Musician-likeness
            & 92.59\% & 7.41\% & 0.00\%
            & \munosp{} 8/9; MIDI 1/9 \\
        {}
            & Insightfulness
            & 92.59\% & 7.41\% & 0.00\%
            & \munosp{} 8/9; MIDI 1/9 \\
        {}
            & Overall preference
            & 92.59\% & 7.41\% & 0.00\%
            & \munosp{} 8/9; MIDI 1/9 \\

        \midrule

        \geminiflash
            & Correctness
            & 83.95\% & 14.81\% & 1.23\%
            & \munosp{} 23/27; MIDI 4/27 \\
        (9-comparison set + 18 additional)
            & Musician-likeness
            & 83.95\% & 16.05\% & 0.00\%
            & \munosp{} 23/27; MIDI 4/27 \\
        {}
            & Insightfulness
            & 83.95\% & 16.05\% & 0.00\%
            & \munosp{} 23/27; MIDI 4/27 \\
        {}
            & Overall preference
            & 83.95\% & 16.05\% & 0.00\%
            & \munosp{} 23/27; MIDI 4/27 \\

        \bottomrule
    \end{tabular}
    \caption{Pairwise evaluation results for \munosp{} and MIDI-only
    descriptions. Human percentages give the share of responses assigned to
    each condition or to an exact tie. For the matching nine-comparison
    \geminiflash{} subset, each caption pair was evaluated three times, giving
    27 judgments. The complete \geminiflash{} evaluation similarly contains
    three independently sampled, order-randomized judgments for each of 27
    caption pairs, giving 81 judgments. Gemini item-level outcomes are obtained
    by majority vote across the three samples for each caption pair. Gemini
    received the complete recordings and the two blinded descriptions but no
    human judgments.}
    \label{tab:human-gemini-pairwise}
\end{table*}

\newpage
\section{Evaluation}
\label{sec:evaluation}

We assess the generated auditory analyses through complementary programmatic
validation, human evaluation, ALM-as-a-judge evaluation, and qualitative comparison.

\subsection{Human Evaluation}

We conducted a human evaluation with three participants who reported 1--3,
7--10, and 16+ years of piano training, respectively; their piano-performing
experience ranged from limited to professional. Two participants reported not
having absolute pitch, one reported having relative pitch.

\subsubsection{Caption quality derived from MuNo and MIDI}
We compare descriptions generated from MuNo-SP with a MIDI-only baseline inspired by
MIDICaps. Both approaches first segment the performance and then generate
temporally grounded auditory analyses using comparable instructions adapted to
the information available in each representation.

Participants choose the preferred description separately for correctness, musician-likeness, insightfulness, and overall preference. Here, musician-likeness concerns both the language used and its correspondence to the participant's perception of the music, while insightfulness concerns the useful musical understanding conveyed. Participants can also explain their judgments in accompanying notes. The nine caption pairs were drawn at random from the three assigned pieces. 

~\tabref{tab:human-gemini-pairwise} reports the resulting preference shares and the
piece-level majority outcomes. MuNo-SP received the largest response share on
all four dimensions. The difference was greatest for insightfulness, for which
MuNo-SP received 62.79\% of responses compared with 30.23\% for MIDI. MuNo-SP
was the majority choice for
eight of nine comparisons. The preference for MuNo-SP therefore occurs despite the
MIDI condition's slightly greater average sentence count, although the modest
length difference remains a potential nuisance variable.

\subsubsection{Auditory Analysis Quality}
The same three evaluators independently rate MuNo-SP analysis
excerpts drawn at random from their assigned recordings. These are complete
passages from the saved model-generated analyses.

Each excerpt receives five ratings on fully anchored five-point scales: temporal correctness, pitch correctness, high-level descriptive correctness, musical intelligence, and musician-likeness. The three correctness dimensions distinguish accurate event localization, accurate notes and pitch relationships, and accurate characterization of features such as texture, phrasing, dynamics, and musical development. Musical intelligence measures the useful musical understanding conveyed, while musician-likeness measures whether the terminology and phrasing resemble a musician's description and the listener's own perception. The interface supports optional playback of notated pitch examples.

Across the 36 ratings per dimension, mean scores were 4.78 for
temporal correctness, 4.89 for pitch correctness, 4.42 for high-level
descriptive correctness, 4.64 for musical intelligence, and 4.39 for
musician-likeness. Across all five dimensions, 83.33\%--97.22\% of ratings
were 4 or 5.

\subsection{ALM-as-a-Judge Evaluation}

To evaluate a larger set of caption pairs beyond those included in the human study, we conduct an ALM-as-a-judge evaluation as a complementary source of evidence. We use
Gemini 3.7 Flash (\cite{googledeepmind2026gemini37flash}) to evaluate 27 MIDI--MuNo-SP caption pairs across nine
pieces, including the nine pairs used in the human study. 
To mirror the human evaluation, we evaluate each pair in three independent blind trials, yielding 81 judgments in total. Each trial provides the model with the complete recording and the two descriptions, whose presentation order is randomized
independently. We use the model's high-reasoning setting and ask it to assess
the descriptions according to the same four criteria used by the human
evaluators: correctness, musician-likeness, insightfulness, and overall
preference. The model is instructed to disregard differences in description
length and temporal coverage and to explain its decisions. The list of pieces
and the evaluation prompt are provided in~\appref{app:llm-as-a-judge-list} and~\appref{app:llm-as-a-judge-prompt},
respectively. Results are reported in~\tabref{tab:human-gemini-pairwise}.

MuNo-SP is preferred overall in 68 of the 81 trials (84.0\%). After taking the
majority decision across the three trials for each caption pair, MuNo-SP is
preferred in 23 of the 27 comparisons (85.2\%). The model produces the same
overall decision in all three trials for 22 of the 27 caption pairs, indicating
a strong and generally consistent preference for MuNo-SP over the MIDI
baseline. Rationales include more accurate pitch tracking, better separation of contrapuntal voices and chordal textures, and more coherent harmonic and phrase-structural interpretations.

Within the nine comparisons shared with the human study, majority voting across
the three model trials favors MuNo-SP in eight cases and MIDI in one. Moreover, Gemini and the human majority agree on all 9/9 majority outcomes, providing evidence for the preference patterns reported above.

\subsection{Case Study: MIDI and MuNo Comparison}

\begin{figure*}[!t]
    \centering
    \includegraphics[width=\textwidth]{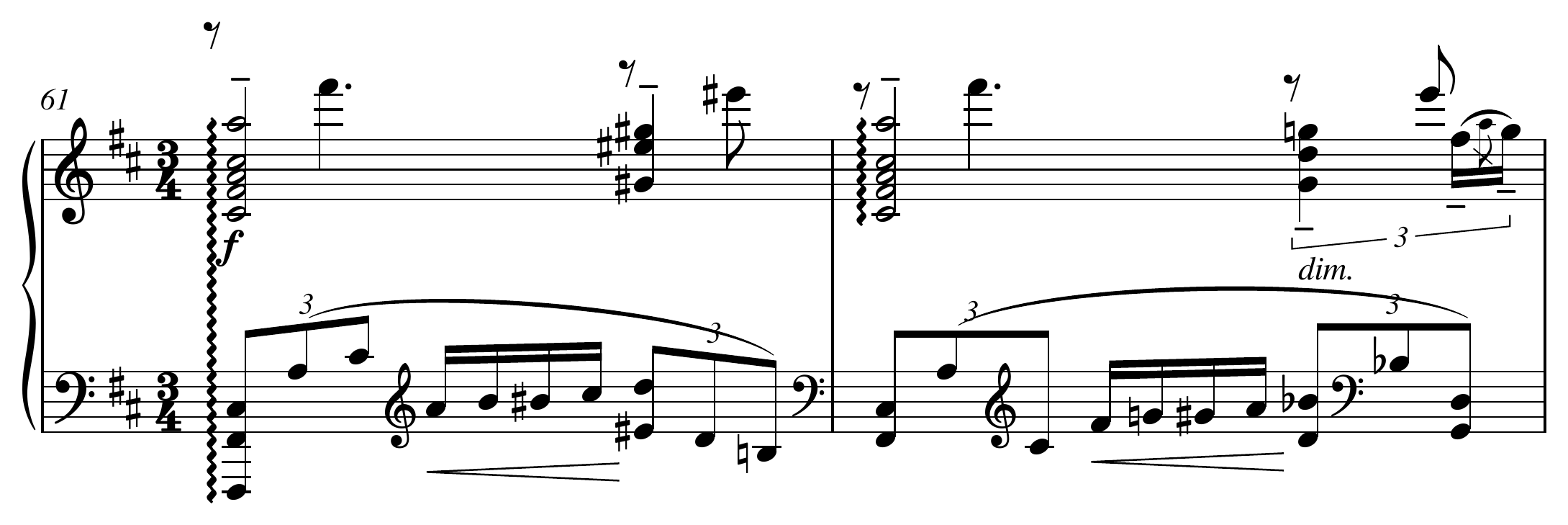}
    \caption{Bars 61--62 of Rachmaninoff's \textit{Prelude in D major,
    Op.~23, No.~4}, showing the concurrent main melody, upper-voice response,
    inner contrapuntal motion, and arpeggiated harmonic structure discussed in the MIDI-only and MuNo-SP
    descriptions.}
    \label{fig:rachmaninoff-midi-muno-voices}
\end{figure*}

MuNo makes explicit use of information contained in the score that is only
implicit in performance MIDI. We illustrate this difference using bars 61--62
of Rachmaninoff's \textit{Prelude in D major, Op.~23, No.~4}. We provide the
same piece to the same model, changing only the input representation. The output is shown in \textcolor{refblue}{Example~3}; the original score itself is shown in ~\figref{fig:rachmaninoff-midi-muno-voices}.

\begin{figure*}[h]
    \begin{musicbox}{MIDI-only and MuNo-SP descriptions of the same passage}
    
    \textbf{MIDI-only input.} At 03:12.26--03:14.11, the principal path
    \munocode{C\#4--F\#4--A4--C\#5--A5}--\munocode{F\#6}--\munocode{C\#4--A4--B4--C5--C\#5} descends and rebounds within a broad
    C\#-minor-related chordal span. At 03:14.25--03:14.74,
    \munocode{F4+G\#5+G\#4+C\#5} over D4 moves to \munocode{F\#4+F6} above B3,
    creating a compressed altered dominant sonority whose semitone
    \munocode{F4--F\#4} resolves downward in register rather than tonally. At
    03:15.10--03:17.34,
    \munocode{F\#2+C\#3+F\#4+A4+C\#5+A5} over A3 and F\#6 gradually descends through
    \munocode{C\#4--F\#4--G4--G\#4--A4}, with broader spacing and reduced upper
    weight releasing the secondary peak into an F\#-minor-centered resonance.
    [\ldots] At 03:17.21--03:20.46, the principal path
    \munocode{D4--G5--F\#5--E5--G5--G4--E4--F4--F\#4} recalls the lyrical
    contour in fragments while high G5 remains the principal projected strand.
    
    \medskip
    
    \textbf{MuNo-SP input.} At 03:12.44--03:14.74, the principal chordal line
    \munocode{A5--G\#5} reaches the return's forte apex, while
    \munocode{F\#6--E\#6} forms a separate high response and the inner ascent
    \munocode{A4--B4--B\#4--C\#5} intensifies the arrival. At
    03:12.20--03:14.31, \munocode{F\#1+F\#2+C\#3} and
    \munocode{C\#4+F\#4+A4+C\#5+A5} establish F\# minor, after which
    \munocode{G\#4+E\#5+G\#5} and the E\# pitches create dominant pressure back
    toward F\#. At 03:15.38--03:18.48, A5 is restated above F\# minor and then
    relaxes through \munocode{G5--F\#5--G5}, with the grace-like
    \munocode{A5} at 03:18.40 briefly heightening the G5 return over \munocode{G2+D3}.
    
    \end{musicbox}
    \begingroup
\setlength{\columnwidth}{\textwidth}
\munocodeblockcaption{
MIDI-only and MuNo-SP descriptions generated for the
same passage from Rachmaninoff's \textit{Prelude in D major, Op.~23, No.~4}
(Recording~4).
\texttt{+} indicates notes form a chord, \texttt{--} a sequence.
}
{
helloworld}
\endgroup
\label{munoblock:midi-muno-comparison}
\end{figure*}

The score shows that the passage layers four distinct components: the main
melody, an upper-voice response, inner contrapuntal motion, and an arpeggiated accompaniment. The
MIDI-only description instead concatenates arpeggiated chord tones, melodic
notes, upper-voice events, and inner motion. For example, it joins the arpeggio
\munocode{C\#4--F\#4--A4--C\#5}, the melodic \munocode{A5}, the separate
upper-voice \munocode{F\#6}, and the
inner line \munocode{C\#4--A4--B4--C5--C\#5}. MuNo-SP preserves the score's voice
and notation structure, allowing the model to distinguish the main
\munocode{A5--G\#5} line, the higher \munocode{F\#6--E\#6} response, and the
inner \munocode{A4--B4--B\#4--C\#5} ascent. It also relates the bass and the
\munocode{E\#} pitches to dominant pressure toward F-sharp minor, whereas the
MIDI-only description identifies C-sharp minor. Moreover, MuNo-SP correctly identifies the
\munocode{A5} at 03:18.40 as a grace note.

Additionally, the MIDI-only description often uses enharmonically plausible
pitch names while grouping them in ways that imply a different conception of
the music. Examples include \munocode{F4+G\#5+G\#4+C\#5} and
\munocode{F\#4--A4--C\#5--A5}. The MIDI-only description also treats the countermelody as beginning on
\munocode{C\#4}, presumably because this note directly follows the melodic
\munocode{A5} and \munocode{F\#6}. Although this assignment is arguable because of the passage's
subtle performance-level ambiguity, a careful listener might hear
\munocode{C\#4} as falling outside the foregrounded inner motion and still within the arpeggiated accompaniment. It is played
more softly and is perceptually masked by the recent forte attack of
\munocode{F\#6}, which directs the listener's attention toward the upper voice. The organization of the musical material is naturally encoded in the score through multiple voices and a crescendo that starts on \munocode{A4} leading towards the third beat, thereby contextualizing the performance data.

The MIDI-only description also makes several claims that are not supported by
the passage. The collection
\munocode{F\#2+C\#3+F\#4+A4+C\#5+A5} is grounded by \munocode{F\#2}, not
\munocode{A3}; there is no identifiable \munocode{F\#4+F6} sonority above
\munocode{B3}; and the purported path
\munocode{D4--G5--F\#5--E5--G5}, followed by
\munocode{G4--E4--F4--F\#4}, cannot be identified as a coherent voice in
either the score or the recording. This passage illustrates why MIDI alone is not suited for reliably conveying the intricate structure and layering
of a piece, as MIDI supervision can conflate simultaneous
musical layers; MuNo-SP, on the other hand, makes their notated functions directly
available to the model.

\subsection{Programmatic Validation}

Because no reference auditory analyses are available, we check objectively
verifiable claims against MuNo-SP using a rule-based validator. The validator checks whether all timestamps originate from actual events and whether all notes, chords, sequences, and harmonies are supported by the music.

The validator was applied to all \NumAuditoryAnalyses{} analyses; its per-analysis reports retain every supported and unsupported claim together with the corresponding MuNo-SP evidence.

Across the whole corpus, the validator found that 63 out of 56,600 timestamp occurrences---or 12 out of 56,600 when allowing a tolerance of \ensuremath{\pm}100 ms---did not match an onset or offset in the piece. Across the entire piece, 10 out of 6,943 named harmonies, 14 out of 17,527 pitch sequences, 208 out of 13,033 simultaneous pitch groups or chords, and none of the 17,143 single pitches were unsupported by the music. Closer analysis revealed that the 208 unsupported simultaneous pitch groups were partly caused by the model combining a bass note with a chord even when they were played separately, as in Schubert's Moment musical No. 3. The overall low hallucination rates indicate a high degree of correspondence between the generated analyses and the musical evidence.

\section{Discussion}
Our results demonstrate that representation design substantially influences language models' ability to reason about music. Compared with alternative formats, MuNo-S and MuNo-SP provide models with a richer understanding of score and performance. Our human evaluation further indicates that MuNo-SP enables more reliable automated caption generation. Combined with the in-context learning and instruction-following capabilities of reasoning models, these representations support detailed natural-language descriptions suitable for fine-grained performance analysis and comparison.

\subsection{Fine-Grained Performance Analysis}

MuNo-SP enables a new approach to the textual analysis of musical performance, allowing models to characterize how a performer interprets a score. By combining score structure with aligned performance information, the representation supports detailed descriptions of timing, dynamics, articulation, duration, and pedal use in relation to specific musical events.

For high-confidence alignments, our pipeline also offers the possibility of
sensitizing models to potential performance errors while expressing
appropriate uncertainty. For example, the analysis identifies a likely
omitted note in bar 36, as shown in~\exref{munoblock:error-detection-example}.

\begin{figure}[!h]
\begin{musicbox}{Potential performance-error detection}
``A likely missing \munocode{Db5} after \munocode{Eb5} at 00:47.63 leaves the
descending pattern audibly incomplete before the chordal call resumes as
\munocode{C5--E5} at 00:47.91--00:48.52, with the abrupt return intensifying
the interruption.''
\end{musicbox}
\munocodeblockcaption{An example of uncertainty-aware performance-error
detection in Schubert's \textit{Moment musical No.~3 in F minor, D.~780,
Op.~94, No.~3} (Recording~5), grounded in the aligned score and performance.
\munocode{--} indicates a sequence.}
{munoblock:error-detection-example}
\end{figure}

\vspace{-40pt}
\subsection{Fine-Grained Performance Comparison}
Because each performance is aligned to the same underlying score, the model can easily compare corresponding events in timing, duration, velocity, articulation, and pedal use, leading to comparisons in phrasing, pacing, balance, and emphasis.~\exref{munoblock:performance-comparison} illustrates how the model can localize a shared passage while distinguishing the performers' contrasting distributions of time.

\begin{figure}[h]
\begin{musicbox}{Comparison of two performances}
Recording~A, 00:00.99--00:41.73, and Recording~B, 00:01.00--00:43.65,
encompass the sustained opening octave, the ascent and descent of bars 1--3,
the quieter continuation and rests of bars 4--6, the held transition of bars
7--8, and \munocode{C4--D4--F\#4--Bb4--A4--G4} through
bar 9 beat 1. Performer~A makes the largest tempo change near the
beginning, whereas Performer~B spends more time in the later quiet passage and
transition.
\end{musicbox}
\munocodeblockcaption{The beginning of a score-grounded comparison of two aligned performances,
showing how the model distinguishes their large-scale pacing while tracking
the same musical events. \munocode{--} indicates a sequence.}
{munoblock:performance-comparison}
\end{figure}
Such comparisons could help models distinguish compositional properties from interpretative choices and support applications in music education---for example, answering questions such as ``How does my performance compare with Rachmaninoff's own interpretation?''---as well as the controllable generation of expressive performances.

\section{Limitations and Future Work}

The present corpus is concentrated on Western classical piano music, for which aligned scores, MIDI, and audio are comparatively abundant. Although we demonstrate a proof of concept using orchestral works by Beethoven, future research could investigate how this paradigm---unifying performance data with other forms of LLM-interpretable information---might support more fine-grained captions across a broader range of repertoires and ensembles.

The generated analyses also inherit the limitations of their source representations. Because they are grounded in scores and aligned MIDI data, they can characterize timing, velocity, duration, and pedal use, but not timbre, acoustics, or other properties absent from MIDI. One promising direction would be to incorporate audio-derived features into the representation, such as relative recording loudness. Errors in scores or score--performance alignments may also propagate into the generated supervision if they are not detected by the verification mechanism. Future systems could therefore combine MuNo-SP with audio-capable models, integrating verifiable evidence with acoustic information perceived directly from the recording for more comprehensive captions.

To scale the pipeline, automatic music transcription systems such as MuScriptor (\cite{rouard2026muscriptoropenmodelmultiinstrument}) allow us to approximate the performance-MIDI layer when aligned MIDI is not readily available, extending the pipeline to a broader range of audio--score datasets. Future work should also assess whether open-weight, locally deployable large language models can be trained to generate such captions.

Finally, model responses may partly reflect prior exposure to well-known works, and the extent to which such exposure influences the generated captions remains unknown. In addition, listening remains an internal experience shaped by the listener's culture, history, and training. The resulting captions, although grounded in concrete musical events, may reflect a mixture of listening experiences and more fundamentally represent just one way of listening to the performance. We make our pipeline available to allow researchers to plug in different models and adapt the instructions to their own inquiries.

\section{Conclusion}

We introduce \textbf{MuNo-SP}, a textual representation that unifies written musical content with its realization in performance. It includes score information such as rhythm, voicing, dynamics, articulation, and phrasing, and aligned timing, duration, velocity, and pedal data. Their organization within a joint format makes both more accessible to language models.

Building on this representation, we develop a pipeline that generates time-grounded auditory analyses and question--answer pairs from aligned performances and scores. The resulting \textbf{MAESTROCaps} dataset provides supervision in the form of concrete accounts of musical events, structural roles, and interpretive choices. A human evaluation confirms the reliability and musician-likeness of the generated captions. Together, MuNo-SP and MAESTROCaps provide a foundation for audio-language models that exhibit musician-level music understanding.

\nolinenumbers




\section*{Competing Interests}
The authors have no competing interests to declare or state otherwise.

\ifanonymous
\else
\section*{Authors' Contributions}
MLD: Conceptualization, methodology, software, data curation, investigation, formal analysis, visualization, evaluation, and writing.
SY: Methodology, investigation, analysis, visualization, validation, evaluation, and writing.
KM: Supervision, project administration, and funding.
\fi

\begin{authoraff}[AUTHOR AFFILIATIONS]
\end{authoraff}
\clearpage

\bibliographystyle{apaTISMIR}
\bibliography{references3}

\clearpage

\appendix
\onecolumn
\section*{Appendix}

\renewtcbox{\munocode}{
    on line,
    colback=gray!10,
    colframe=gray!35,
    boxrule=0.3pt,
    arc=1pt,
    boxsep=0pt,
    left=1.5pt,
    right=1.5pt,
    top=0.5pt,
    bottom=0.5pt,
    fontupper=\sffamily\fontsize{5.8pt}{7pt}\selectfont
}

\renewcommand{\tablename}{Table}
\renewcommand{\thetable}{A.\arabic{table}}

\newcommand{\appendixverbatim}[1]{%
  \par\addvspace{0.8\baselineskip}%
  \VerbatimInput[
    breaklines=true,
    breakanywhere=true,
    fontfamily=\sfdefault,
    fontsize=\scriptsize,
    frame=single,
    framerule=0.8pt,
    framesep=3mm,
    rulecolor=\color{gray!75},
    formatcom=\color{black},
    xleftmargin=0.8em,
    xrightmargin=0.8em
  ]{#1}%
  \par\addvspace{1.0\baselineskip}%
}

\newcommand{\prompttextsize}{\fontsize{6.4pt}{8pt}\selectfont}
\DefineVerbatimEnvironment{appendixprompt}{Verbatim}{%
  breaklines=true,
  breakanywhere=true,
  breaksymbolleft={},
  breaksymbolright={},
  commandchars=^!~,
  fontfamily=\sfdefault,
  fontsize=\prompttextsize,
  frame=none,
  formatcom=\fontseries{l}\selectfont\color{black},
  xleftmargin=0pt,
  xrightmargin=0pt
}

\surroundwithmdframed[
  backgroundcolor=gray!7,
  linecolor=gray!60,
  linewidth=0.5pt,
  roundcorner=5pt,
  innerleftmargin=3.2mm,
  innerrightmargin=3.2mm,
  innertopmargin=2.8mm,
  innerbottommargin=2.8mm,
  skipabove=0.7\baselineskip,
  skipbelow=0.7\baselineskip,
  splittopskip=2.8mm,
  splitbottomskip=2.8mm
]{appendixprompt}

\newcommand{\promptheading}[1]{{\sffamily\bfseries\scriptsize #1}}

\newenvironment{analysispanel}{%
  \begin{mdframed}[
    backgroundcolor=gray!7,
    linecolor=gray!60,
    linewidth=0.5pt,
    roundcorner=5pt,
    innerleftmargin=3.2mm,
    innerrightmargin=3.2mm,
    innertopmargin=2.8mm,
    innerbottommargin=2.8mm,
    skipabove=0.7\baselineskip,
    skipbelow=0.7\baselineskip,
    splittopskip=2.8mm,
    splitbottomskip=2.8mm]
  \prompttextsize\sffamily\fontseries{l}\selectfont
  \setlength{\parindent}{0pt}%
  \setlength{\parskip}{0.55\baselineskip}%
  \sloppy
}{\end{mdframed}}

\newcommand{\analysisrole}[1]{%
  \if\relax\detokenize{#1}\relax\else
    \par\nobreak\vspace{0.12\baselineskip}%
    {\mdseries\itshape #1}%
  \fi}
\newcommand{\analysissectionheading}[2]{%
  \par\addvspace{0.45\baselineskip}%
  {\bfseries\fontsize{8pt}{9.2pt}\selectfont #1}%
  \analysisrole{#2}\par\nobreak}
\newcommand{\analysisthemeheading}[2]{%
  \par\addvspace{0.7\baselineskip}%
  {\bfseries\fontsize{7.2pt}{8.4pt}\selectfont #1}%
  \analysisrole{#2}\par\nobreak}
\newcommand{\analysisphraseheading}[2]{%
  \par\addvspace{0.55\baselineskip}%
  {\bfseries #1}%
  \analysisrole{#2}\par\nobreak}

\newcommand{\segpromptnumber}[2]{%
  \par\noindent\hangindent=1.8em\hangafter=1%
  \makebox[1.8em][l]{{\sffamily\bfseries #1.}}#2\par}

\section{Recordings Used in the Main-Text Examples}
\label{app:main-text-recordings}

Table~\ref{tab:main-text-recordings} lists the aligned score--performance
pairs in order of their first appearance in Examples~1--5. Examples~2 and~5
each contain two recordings.

\begin{table}[htbp]
    \centering
    \begin{tabular}{ccp{0.52\textwidth}l}
        \toprule
        \textbf{Example} & \textbf{Recording} & \textbf{Work} & \textbf{ID} \\
        \midrule
        1 & 1 & Mozart, K.~332/I & WuuE02M \\
        2 & 2 & Debussy, \textit{Reflets dans l'eau} & ParkJH13M \\
        2 & 3 & Beethoven, \textit{Waldstein}, I & KaiRuiR02M \\
        3 & 4 & Rachmaninoff, Op.~23, No.~4 & WuuE07M \\
        4 & 5 & Schubert, D.~780, No.~3 & Tetzloff09M \\
        5 & 6 (A) & Chopin, \textit{Ballade No.~1 in G minor, Op.~23} & Zhou06M \\
        5 & 7 (B) & Chopin, \textit{Ballade No.~1 in G minor, Op.~23} & BuiJL04M \\
        \bottomrule
    \end{tabular}
    \caption{Recordings used by the numbered examples in the main text,
    ordered by first appearance.}
    \label{tab:main-text-recordings}
\end{table}

\section{External Score Sources}
\label{app:score-sources}

Table~\ref{tab:external-score-sources} lists the external score collections
used to propose or manually inspect corrections. The ASAP/MuseScore score is
the canonical starting point; disagreement with another source creates a
proposal for validation rather than an automatic correction.

\begin{table}[htbp]
    \centering
    \begin{tabular}{p{0.24\textwidth}p{0.66\textwidth}}
        \toprule
        \textbf{Collection} & \textbf{Coverage and role} \\
        \midrule
        CPJKU WTC Book~I & Bach preludes and fugues, BWV~846--869; symbolic
        comparison source. \\
        DCML Beethoven Piano Sonatas & Available Beethoven sonata movements;
        symbolic comparison source. \\
        Chopin Scores & Available Chopin works in the corpus; symbolic
        comparison source. \\
        DCML Mozart Piano Sonatas & Mozart K.~332, movements I and III;
        symbolic comparison source. \\
        Mysterium & Scriabin's \textit{Etude in B-flat minor, Op.~8, No.~11}
        and \textit{Piano Sonata No.~5, Op.~53}; symbolic comparison source. \\
        IMSLP & A movement- or work-specific public-domain edition for each of
        the 142 retained pieces; manual verification source. \\
        \bottomrule
    \end{tabular}
    \caption{External score collections used during score correction.}
    \label{tab:external-score-sources}
\end{table}

\section{Score-Correction Algorithm}
\label{app:score-correction-procedure}

The score-cleaning procedure consists of the following steps:

\paragraph{Step 1: Detection of deletion candidates in (n)ASAP and external
scores.} We derive deletion candidates from notes marked as absent in the
(n)ASAP performance alignments and from notes present in the ASAP score but
absent from external symbolic scores.

\paragraph{Step 2: Detection of insertion candidates in the performances.} For
each candidate and MIDI recording, we interpolate an expected onset from the
Parangonar score--performance alignment. We center the validation window on
the interpolated onset. On each side, its extent is the lesser of half the
note's estimated performed duration and the midpoint distance to the
neighboring aligned onset. Unmatched performed notes within this window whose
pitches are not already represented by simultaneous score notes are treated as
insertion proposals. We also derive insertion proposals from an external
reference score when one is available.

\paragraph{Step 3: Validation of proposals.} We validate deletion and insertion
proposals using PianoCoRe. \emph{PianoCoRe support} is the proportion of
successfully Parangonar-aligned PianoCoRe performance MIDI files for the piece
that support the proposed operation. Within each collection, all recordings
are weighted equally.

\paragraph{Deletion.} Accept if there is no simultaneous written unison of
the same pitch in another voice and one of the following holds:
\segpromptnumber{a}{At least 60\% support across (n)ASAP alignments, when at least five are
available, and at least 50\% support from PianoCoRe.
}
\segpromptnumber{b}{At least 30\% support across (n)ASAP alignments, when at least five are
available, and at least 75\% support from PianoCoRe.
}
\segpromptnumber{c}{Strictly more than 95\% PianoCoRe support.}

\paragraph{Tie.} Apply the deletion criteria, but do not require the 30\%
(n)ASAP support in the second deletion condition because a tied continuation
does not require a new performed attack. In addition, a preceding note of the
same pitch, with a written value extending to the candidate, must reach the
candidate onset. Its mean performed sustain must extend at least 20\% into the
continuation without a new attack, or at least 0.05 seconds for a note marked
staccato.

\paragraph{Insertion or replacement.} For a replacement, first require the
original pitch to satisfy a deletion criterion. Then accept the proposed pitch
if one of the following holds:
\segpromptnumber{a}{At least 70\% PianoCoRe support.}
\segpromptnumber{b}{At least 50\% support from at least five (n)ASAP alignments.}
\segpromptnumber{c}{At least 50\% PianoCoRe support and at least 95\% support
from at least three (n)ASAP alignments.}
\segpromptnumber{d}{At least 50\% PianoCoRe support and support from at least
50\% of the independent symbolic scores.}
\paragraph{Step 4: Manual verification and documentation.} We compare supported
corrections with available public-domain editions and reject or add changes
where necessary. For each proposal, we record its source, location, operation,
performance support, validation decision, and final outcome.

We selected the thresholds through empirical testing across a development set
of scores.

\clearpage
\section{Structural-Segmentation Prompt}
\label{app:segmentation-prompt}

The following is the current MuNo-S structural-segmentation prompt. The
composer, title, requested scope, and MuNo-S representation placeholders are
populated for each piece.

\par\addvspace{0.7\baselineskip}
\begin{appendixprompt}
^promptheading!MUNO REPRESENTATION:~
[MUNO-S REPRESENTATION]

^promptheading!SEGMENTATION INSTRUCTIONS:~
^textbf!Task:~ Segment ^textbf!the complete piece~ of ^textbf![COMPOSER] - [TITLE]~ into musically meaningful sections, theme presentations, and phrases using the supplied time-annotated MuNo representation.

^textbf!Core Objective:~ Produce a hierarchical structural map that tells a musician exactly where each segment begins and ends in the written music. Identify recurring or transformed thematic material, phrase boundaries, and larger formal sections. A segment may contain smaller segments, and independently meaningful segments may overlap.

You may use your prior musical knowledge of the composer, piece, genre, and conventional formal terminology to recognize and name themes or sections. Use that knowledge as an interpretive aid, but ground every boundary in the supplied MuNo representation. Do not invent boundaries or assert a conventional formal label when the music does not support it.

\end{appendixprompt}
\par\addvspace{0.55\baselineskip}
\begin{appendixprompt}
^promptheading!I. Segmentation Levels~

Use only these values for 'kind':

^textbullet!~ 'section': a large-scale formal region, such as 'Exposition', 'Development', 'Recapitulation', or 'Coda' when those labels are supported by the music.
^textbullet!~ 'theme': a particular presentation, return, or transformation of identifiable thematic material, such as 'Presentation of Primary Theme' or 'Varied repeat of Secondary Theme'.
^textbullet!~ 'phrase': a complete or incomplete musical thought or local phrase function, such as 'Opening statement', 'Cadence', 'Build-up', or 'Culmination'.

Keep these levels distinct: use 'section' for large formal divisions, 'theme' for occurrences of thematic material, and 'phrase' for local musical statements and functions. Prefer musically informative names and do not use generic numbered labels unless the musical function truly cannot be determined from the data.

Phrases should normally be short spans of approximately 1-4 bars. Use a longer phrase only when a clearly continuous musical thought has no defensible internal boundary; do not subdivide below the phrase level.

Treat slurs, if any, as indicative for musical grouping. A slur in a melodic voice often indicates how notes belong to one gesture, where a phrase continues, and where a new phrase or phrase member may begin. Follow the slurs within each continuous voice, giving particular weight to slurs that shape the principal melody; when several slurs occur simultaneously in different voices, interpret each layer separately before deciding the larger boundary. Do not mechanically equate every slur with a complete phrase, but do not ignore or lightly override melodic slur groupings without stronger contrary evidence from cadence, harmony, repetition, texture, or timing.

Do not assume that the highest written voice is the melody. Identify the principal melody through a combination of thematic and phrase logic, knowledge of the piece, continuity within each voice, and velocities or prominence otherwise indicated in the score through accents, tenuti, or slurs. An inner or lower voice may carry the principal melody while a higher voice supplies decoration, accompaniment, countermelody, or harmonic color. 

Also examine the number and sequence of bars in each musical thought. Two-, four-, and eight-bar spans, especially four- and eight-bar passages, are important evidence for phrase structure, thematic presentation, periodic organization, repetition, and larger grouping. Test proposed boundaries against these common spans while allowing musically supported extensions, contractions, overlaps, interpolations, and irregular groupings. Bar-count symmetry alone is not sufficient, but it should be considered explicitly alongside melodic grouping and harmonic or cadential motion.

\end{appendixprompt}
\par\addvspace{0.55\baselineskip}
\begin{appendixprompt}
^promptheading!II. Boundary Rules~

^textbullet!~ Every segment must have a 'start_bar' and an 'end_bar'.
^textbullet!~ Copy bar labels exactly as represented in MuNo, including pickup or repeat labels when applicable.
^textbullet!~ Give boundaries only as bars. Do not include timestamps or other boundary units anywhere in the JSON response.
^textbullet!~ The 'end_bar' is inclusive: both 'start_bar' and 'end_bar' belong to the segment.
^textbullet!~ Adjacent segments may share a boundary when one ends as the next begins.
^textbullet!~ Overlap is explicitly allowed. For example, a transition may begin during the final phrase of a theme, or a theme presentation may contain three constituent phrases.
^textbullet!~ Do not force every bar into a phrase when the evidence does not support a meaningful phrase boundary. The complete piece should nevertheless be covered by one or more top-level 'section' segments.

\end{appendixprompt}
\par\addvspace{0.55\baselineskip}
\begin{appendixprompt}
^promptheading!III. Function, Role, and Hierarchy~

^textbullet!~ Use 'function_or_role' to state what the segment does in the piece, such as 'first presentation of the primary theme', 'varied return of the primary theme', 'fragmentation and development', 'antecedent phrase', 'continuation', 'transition', or 'cadential close'.
^textbullet!~ Express hierarchy through the segment names, kinds, and inclusive bar ranges. Do not use IDs, parent fields, or parts fields.
^textbullet!~ A theme presentation consisting of three phrases must be represented as one encompassing 'theme' segment plus three separate 'phrase' segments whose bar ranges fall within it.
^textbullet!~ Repeated or transformed material should use consistent theme names, while 'function_or_role' explains whether each occurrence is a first presentation, return, variation, fragmentation, continuation, or another formal function.
^textbullet!~ Overlapping functions should be represented as separate segment objects with overlapping bar ranges.

\end{appendixprompt}
\par\addvspace{0.55\baselineskip}
\begin{appendixprompt}
^promptheading!IV. Evidence Standard~

Base boundaries and labels on concrete musical evidence available in MuNo: melodic slur grouping, thematic recurrence, recurring bar-span organization, cadential arrival, harmonic change, texture, register, voice-leading, dynamics, articulation, rests, timing gaps, tempo, rubato, and changes in performance intensity. Give melodic slurs and supported four- or eight-bar organization substantial weight when locating phrase, theme, and section boundaries.

For each segment, provide a short 'evidence' statement explaining the boundary or identity. Do not mention encoding syntax, velocity numbers, or unsupported historical/formal claims. If a conventional formal name is uncertain, use a descriptive musical name and state the uncertainty in 'evidence'.

\end{appendixprompt}
\par\addvspace{0.55\baselineskip}
\begin{appendixprompt}
^promptheading!V. Output Contract~

Return only valid JSON with no prose outside the JSON and no Markdown fences. Use exactly this top-level structure:

{
  "segments": [
    {
      "kind": "section",
      "name": "<musically descriptive name>",
      "function_or_role": "<structural, thematic, or phrase-level function>",
      "start_bar": "<exact MuNo bar label>",
      "end_bar": "<exact MuNo bar label>",
      "evidence": "<concise, data-grounded explanation>"
    }
  ]
}

\end{appendixprompt}
\par\addvspace{0.55\baselineskip}
\begin{appendixprompt}
^promptheading!VI. Final Validation Before Responding~

Verify that:

1. Every segment has all required fields.
2. Every segment has a 'start_bar' and an inclusive 'end_bar', both grounded in the supplied MuNo representation.
3. The response contains no IDs, parent fields, parts fields, or 'relationship' field.
4. Every segment states a clear 'function_or_role'.
5. Encompassing theme or section ranges contain their constituent phrase ranges, except where an explicitly explained overlap occurs.
6. Top-level sections collectively cover the requested scope.
7. Theme returns and transformations are named consistently.
8. The response contains no timestamp fields and parses as strict JSON.
\end{appendixprompt}
\par\addvspace{1.0\baselineskip}

\clearpage
\subsection{Example Segmentation}
\label{app:example-segmentation}

\begin{figure}[htbp]
    \centering

    \includegraphics[width=\textwidth]{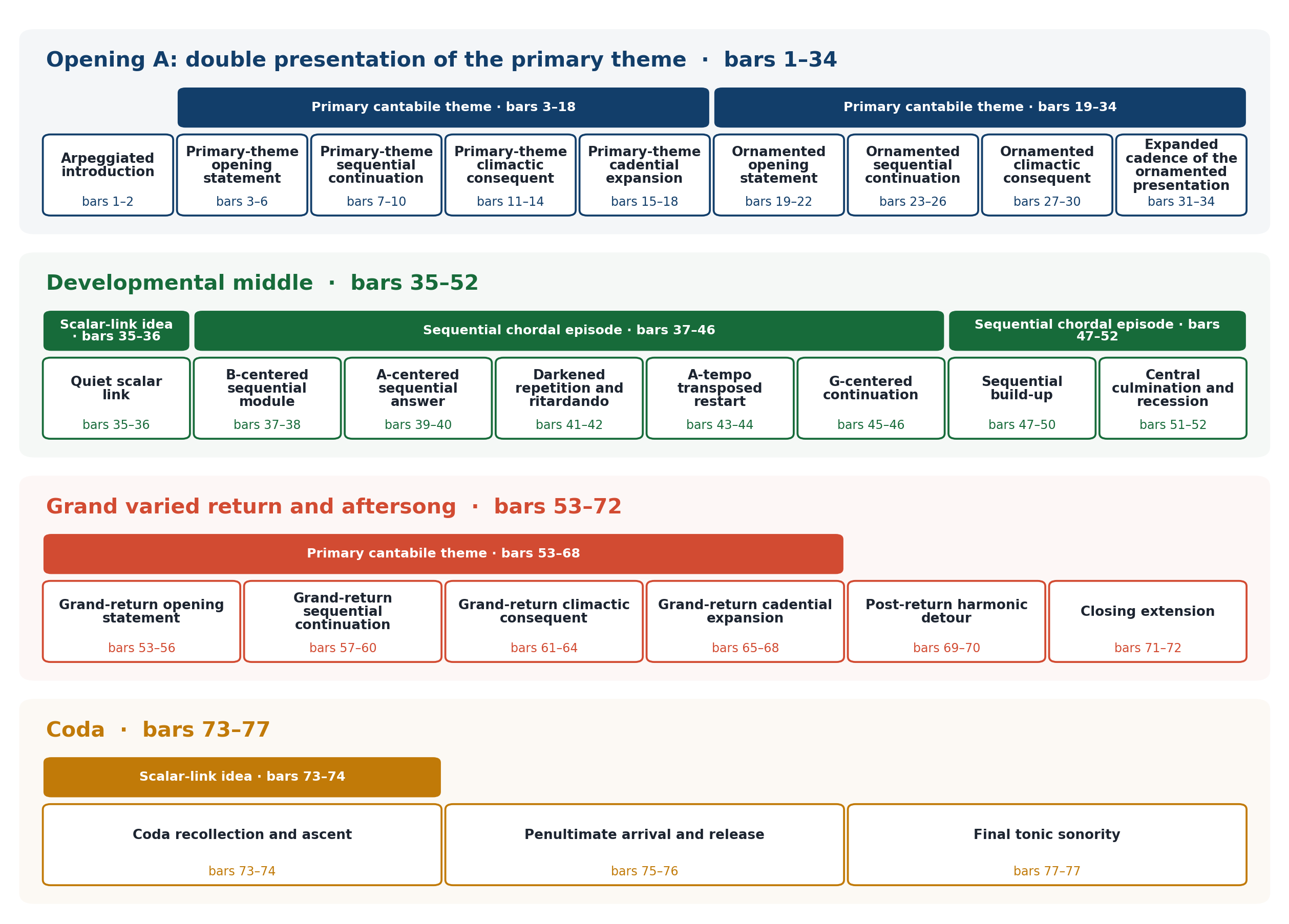}
    \caption{Section, theme, and phrase spans generated from MuNo-S for
    Rachmaninoff's \textit{Prelude in D major, Op.~23, No.~4}. Colored panels
    denote large-scale sections, filled bands denote theme spans, and outlined
    tiles denote phrase spans; all bar ranges are inclusive.}
    \label{fig:segmentation}
\end{figure}

\clearpage
\section{Long-Form Auditory-Analysis Prompt}
\label{app:auditory-analysis-prompt}

The following is the current segmentation-aware MuNo-SP prompt. The verified
piece context, programmatically calculated phrase-sentence plan, and segmented
MuNo-SP representation are populated for each recording.

\par\addvspace{0.7\baselineskip}
\begin{appendixprompt}
You are writing a long-form, time-grounded auditory analysis of a supplied performance using a MuNo-SP representation of its score and performance.

^promptheading!VERIFIED PIECE CONTEXT~

{
  "composer": "[COMPOSER]",
  "piece": "[TITLE]"
}

Use this verified identity and your prior musical knowledge only when they help resolve melody, harmony, or structure in the supplied performance. Do not include the composer, title, historical facts, or other external factual information in the output.

\end{appendixprompt}
\par\addvspace{0.55\baselineskip}
\begin{appendixprompt}
^promptheading!INPUT NOTATION GUIDE~

^textbullet!~ Indentation shows hierarchy: bars contain beats, and beats contain markings, voices, and events.
^textbullet!~ BAR and BEAT timestamps use onset+duration.
^textbullet!~ Labels such as 'v3:' and 'v5:' identify simultaneous musical voices or layers.
^textbullet!~ A note such as 'F#4 half v68 00:07.50+02.64' gives pitch, written duration, performed intensity, performance onset, and sounding duration.
^textbullet!~ A statement such as 'A2 is not played' identifies an expected pitch for which the alignment contains no corresponding performed attack. Treat it as a candidate missed note, not as conclusive proof by itself.
^textbullet!~ Comma-separated pitches form a chord within one voice. Corresponding comma-separated intensities and timestamps describe the individual chord tones.
^textbullet!~ Successive pitched events in the same voice belong to that voice's continuing line. Overlapping sounding durations between successive events in one voice usually represent sustain or legato carryover.
^textbullet!~ Voice labels are authoritative evidence of simultaneous strands. Trace each reported melody within one source voice unless sustained evidence supports a genuine transfer; keep a simultaneous secondary melody separate. A line remains melodic when it moves into a lower register.
^textbullet!~ Read harmony across all simultaneous voices. Use voice membership, onset timing, written note values, and sounding durations together to distinguish structurally active chord tones from passing tones and residual resonance.
^textbullet!~ 'rest' is a written rest. A dash means that the preceding note in that voice continues.
^textbullet!~ '/3:2' indicates a triplet value.
^textbullet!~ Terms such as pianissimo, crescendo, tenuto, arpeggio, and 'slur:start/end' describe musical expression, articulation, or grouping.
^textbullet!~ Pedal events indicate changes in sustain.
^textbullet!~ When a pedal release is followed closely by a reapplication, treat the pair as one pedal refresh. Only if it is musically relevant, mention only the renewed pedal or resonance at the reapplication time, not both mechanical events.
^textbullet!~ At the start of each phrase, carry forward the latest preceding time signature, passage heading, BPM, and dynamic when one is declared. Use these declarations only to interpret audible meter, pacing, character, and intensity; omit any declaration that is unknown, and do not report it merely as notation.

\end{appendixprompt}
\par\addvspace{0.55\baselineskip}
\begin{appendixprompt}
^promptheading!INPUT SEGMENTATION GUIDE~

^textbullet!~ 'New section from bar X to bar Y: "name - role"', 'New theme from bar X to bar Y: "name - role"', and 'New phrase from bar X to bar Y: "name - role"' begin structural units.
^textbullet!~ 'End of section: "name - role"', 'End of theme: "name - role"', and 'End of phrase: "name - role"' end them.
^textbullet!~ Every declared bar range is inclusive.
^textbullet!~ '-' separates consecutive phrases and is not an audible event.
^textbullet!~ Treat this segmentation as indicative. Use its hierarchy and boundaries to structure the description, and reproduce every quoted 'name - role' label verbatim whenever naming that unit. Determine audible internal grouping from the music itself, giving particular weight to continuity and, if present, slurs in the principal melodic voice as well as cadence, harmony, repetition, texture, dynamics, and timing.
^textbullet!~ For each phrase, use its active enclosing section and theme, its neighboring phrases, and any child material as structural context. Analyze each phrase's declared range under its own heading; use neighboring material only to clarify continuity, entrances, releases, boundaries, and function.

\end{appendixprompt}
\par\addvspace{0.55\baselineskip}
\begin{appendixprompt}
^promptheading!SCOPE AND LENGTH~

Translate the supplied time-annotated passage into a concise auditory description written like a perceptive human musician. Select the details that define what the passage sounds like rather than inventorying every event. Discuss every declared phrase, including material outside a theme, and use the segmentation only to organize the analysis.

Describe each phrase as it unfolds: what is heard initially, what enters or changes next, how intensity or texture develops, and what happens toward its end.

Begin with an exactly two-sentence whole-piece overview. After the phrases belonging to a theme, write exactly two synthesis sentences for that theme: one about melodic repetition, variation, or transformation, and one about its harmonic and tension arc. Write exactly two comparable synthesis sentences for every section.

The sentence count for every phrase is calculated programmatically as one sentence per inclusive phrase bar, with a maximum of eight sentences per phrase. Follow this exact plan:

[PROGRAMMATIC PHRASE SENTENCE PLAN]

Within each phrase's assigned sentence count, give the first presentation of its principal melodic idea as a complete ordered pitch path, then follow the recurrence rules below. Explicitly consider any genuinely important secondary melody or embedded melodic subvoice. Give melody, harmony, texture, dynamics, phrasing, rhythm, function, tension and release, and supported performance decisions comparable analytical weight. Devote at least one substantial sentence per phrase to how bass motion, sonority, chromatic color, dissonance, and cadence create tension and release, and connect the melodic layers to their supporting harmony. Integrate locally clear articulation and rubato decisions within the same sentence allowance. Organize longer phrases into musically meaningful stages such as opening statement, continuation, intensification, climax, return, and close.

\end{appendixprompt}
\par\addvspace{0.55\baselineskip}
\begin{appendixprompt}
^promptheading!MELODY IDENTIFICATION AND GROUPING~

Follow these steps in order:

1. PRINCIPAL MELODY

Identify the perceived melodic foreground over time; do not assume that one uninterrupted principal melody exists throughout every phrase. Treat encoded voice membership as authoritative, but recognize that one encoded voice can contain more than one perceptual strand and that not every event in it necessarily serves the same musical role. Prefer a continuous principal-melody voice and do not switch merely because another line is temporarily higher, more active, or contains one louder attack.

2. SECONDARY MELODIES AND SUBVOICES

After identifying the principal melody, inspect every voice for a secondary melody or a melodic subvoice interleaved with accompaniment. Require multiple cues across a meaningful span: continuity of contour or rhythm, sustained prominence, score markings, registral or metrical recurrence, a returning pattern, or harmonic direction. A melodic subvoice may select recurring attacks while excluding intervening accompanimental events; after a substantially lower-prominence interruption, resume it with the next projected attack when its earlier contour or pattern continues. A candidate subvoice is generally more projected than the local accompaniment baseline, judged within the same register, texture, and recurring passage, though an inner or bass line need not be loudest. Keep each line within one voice, do not duplicate the principal melody, and do not manufacture an unsupported line.

3. EVIDENCE AND PERCEPTUAL STREAMING

Use thematic continuity, phrase logic, slurs, duration, meter, repetition, harmonic resolution, timing, and sustained prominence together. Test continuity through corresponding metrical positions, sequences, antecedent-consequent patterns, cadential motion, and recurring contours. Treat score markings, timing, and local velocity balance as mutually informative; no single cue determines melodic status. Verify every selected pitch, voice, attack, and grouping against the supplied score and performance.

Inspect velocities systematically. Compare each candidate note with simultaneous and nearby notes in the same register and texture, seeking consistent projection across a complete gesture rather than one isolated loud attack. Do not assume that equal velocities at different pitches produce equal loudness or prominence: widely separated registers may remain distinct with a small velocity difference, while closely spaced notes may fuse or mask one another. Treat MIDI velocity as attack-prominence evidence, not as a universal or linear loudness measure, and apply no fixed global threshold. Large successive leaps still require contour, rhythm, recurrence, or phrase logic to belong to one line.

4. ATTACKS, GROUPING, AND TRANSFERS

Treat 'cues-before' as ordered attacks, including those that belong to a selected line. In a melodic chord, select the tone continuing the line and treat the others as harmony unless they form a supported separate strand. Do not repeat a tied or continued tone without a new attack; include every genuine repeated attack.

Treat slurs, if any, within a selected melodic voice as strong evidence of gesture grouping, not as automatic proof that each slur is a complete phrase. Preserve slur-defined gestures when they exist, then use harmony, phrasing, timing, and tension-resolution to combine complete gestures into larger melodic sections when appropriate. When no slurs are present, infer gestures from the other musical and performance evidence rather than treating the grouping as indeterminate. A resolution may occur in a later complete gesture; keep the gestures distinct while explaining their larger connection rather than forcing them into one gesture or cutting off the unresolved motion.

Report a genuine transfer to another voice only when the new voice continues the thematic or phrase logic, remains prominent across a meaningful span, and the preceding voice stops carrying the melody or clearly becomes accompaniment. Keep simultaneous countermelodies separate. If the evidence does not support a clear principal melody, secondary melody, transfer, or melodic subvoice, do not manufacture one.

5. TRANSCRIPTION ECONOMY

Transcribe a melodic idea completely at its first presentation, preserving every selected attack, repeated pitch, ornament, and cue in sounding order. For a later exact or substantially unchanged recurrence, identify its relationship and timestamps without copying the entire pitch path; give only a short identifying incipit and any changed, omitted, added, reharmonized, redistributed, or newly emphasized pitches. When writing any pitch path, never omit internal attacks or use 'etc.', 'and so on', or an ellipsis. This economy rule applies across later phrases and to theme and section syntheses.

\end{appendixprompt}
\par\addvspace{0.55\baselineskip}
\begin{appendixprompt}
^promptheading!ANALYTICAL METHOD~

Use these seven dimensions internally:

1. Melody: apply the identification, secondary-line pass, grouping, and recurrence rules above. Trace the principal and supported secondary lines separately; explain their contour, arrivals, interaction, repetition, variation, return, and transformation without retranscribing a pitch path already given. Use written note values to understand audible relative length but never name those values; for example, say 'the shorter C4 and D4'.
2. Harmony/color and tension: combine the notes actually sounding across all voices, using onset, written duration, and sounding duration to distinguish active harmony from passing notes and residual resonance. Trace tonal center, bass or fundamental motion, chord or sonority changes, chromatic inflections, dissonance and resolution, cadential motion, and changing color. Do not mistake a lower melodic note for the bass. Name the chord, harmonic function, bass motion, dissonant interval, chromatic alteration, or resolution responsible for each claim, identifying sounding pitches when useful. If 'cadence' or 'cadential' appears outside a required quoted heading, name the cadence type and tonal goal and give the approach and arrival harmonies with timestamps; if the evidence is insufficient, do not claim a cadence.
3. Texture: describe the number, balance, register, and interaction of melody, inner voices, bass, and accompaniment without repeating the melodic or harmonic account.
4. Dynamics and phrase shaping: describe how audible intensity, balance, emphasis, note duration, and breathing points shape the phrase. Mention a pedal refresh only when its renewed resonance materially affects phrasing or clarity, and refer only to the reapplication time.
5. Function: explain each unit's structural role from its own audible material rather than relying on claims about later events.
6. Rhythm and time: broadly describe the general rhythmic patterns and pulse, such as extended notes or accompaniment with a triplet feel, without duplicating local expressive timing observations.
7. Particular performance decisions: identify clear locally audible interpretive choices. Look for a meaningful articulation difference, such as detached or staccato attacks contrasted with a connected or legato line, and identify the pitches and timestamps that demonstrate it. Also examine rubato; judge rubato and expression from actual onset spacing, velocities, and timing: where the performer takes extra time, makes a small acceleration, relaxes the pace, accentuates a specific pitch, or gives a note agogic emphasis. When a pitch is explicitly marked 'is not played', inspect the surrounding notes in the same line, their melodic direction and recurring pattern, nearby velocities, onset spacing, and any local hesitation or extra time. If those combined cues strongly support an audible interruption, describe it cautiously as a likely missed note and explain how it disrupts the surrounding line; never mention the score, notation, alignment, data, or the marker in the output, and never state the performer's intention as certain. If the local context does not strongly support that inference, do not report a missed note. If neither articulation nor timing differs materially, state that the delivery remains comparatively even rather than inventing an expressive deviation.

Explain tension only through audible evidence such as a named dissonance and its resolution, a specific chromatic alteration, harmonic prolongation, registral expansion, increased performed intensity, acceleration, delayed resolution, or textural thickening. State when the cause begins, when it peaks if applicable, and when it resolves or recedes. Describe the most important climax.

Summarize repeated accompaniment patterns as patterns. Name important chord tones together when they clarify a sonority, but otherwise keep nonmelodic pitches selective and do not catalogue every accompaniment event. Do not focus on the precise ending of a note unless it creates or releases a dissonance, articulates a phrase boundary, or exposes a change of harmony or texture.

\end{appendixprompt}
\par\addvspace{0.55\baselineskip}
\begin{appendixprompt}
^promptheading!TIME AND EVIDENCE GROUNDING~

^textbullet!~ Use only numeric MM:SS.cc times taken from the supplied data, such as '00:07.50' or '01:03.25'; never spell out times in words.
^textbullet!~ Every sentence must contain at least one numeric MM:SS.cc timestamp or timestamp range.
^textbullet!~ Ground each independent observation at the time it occurs rather than attaching one broad range to a sentence containing several events.
^textbullet!~ Timestamp melodic entrances and changes, harmonic changes, dissonances and resolutions, changes in texture or performed intensity, tension, climaxes, and phrase endings.
^textbullet!~ Every reference to a note or other event must state concrete evidence. For example, replace 'building up' with 'building up towards G5' and give its timestamp.
^textbullet!~ Every written pitch must include its octave number and preserve the pitch spelling supplied in the timed musical data, including pitches in melodic sequences, neighbor-note figures, chord voicings, bass motion, dissonances, and resolutions-for example 'F#4-E4-F#4', never 'F#-E-F#'. Chord names, keys, and Roman-numeral functions such as 'D major', 'A7', or 'V7 of D major' do not take octave numbers, but every individual note cited as evidence does.
^textbullet!~ In prose pitch notation, join sequential melodic notes with an en dash ('F#4-E4-F#4') and simultaneous chord tones or arpeggiated chords with a plus sign ('D4+F#4+A4'). Never use an en dash between tones of a simultaneous or arpeggiated chord, or a plus sign between sequential melodic attacks.
^textbullet!~ Do not use bar numbers or the word 'bar' anywhere in the response, including headings.
^textbullet!~ Within a phrase's analysis, do not mention, anticipate, prepare for, hand off to, or compare it with a later section, theme, phrase, repetition, or event. Later material may be inspected only to clarify the current phrase's boundary and structural identity. Theme and section syntheses may compare material contained within the unit being summarized.

\end{appendixprompt}
\par\addvspace{0.55\baselineskip}
\begin{appendixprompt}
^promptheading!AUDITORY INTERPRETATION~

Use evidence in the supplied timed performance and the score, and translate the input into perceptible musical results:

^textbullet!~ If a slur is present, it may indicate that a line sounds smoothly legato, but never mention the slur. A slur, usually contained within a single voice, may also inform phrasing and segmentation, especially in the melody. Slurs are numbered by the voice they begin in, such as 'slur:3:start'. Do not ignore a continuous melodic slur merely because an indicative structural span begins or ends there; override that suggested internal boundary when the slur and stronger evidence from cadence, harmony, repetition, texture, dynamics, or timing support a different audible grouping. When no slur is present, rely on voice continuity, phrase logic, harmony, repetition, articulation, timing, and performance prominence.
^textbullet!~ Terms such as staccato and legato may be used only when they describe a difference supported by the audible performance, never merely because the corresponding notation is present.
^textbullet!~ Voicing is important to detect melodies, harmonies, etc.
^textbullet!~ A high velocity may inform you that an attack sounds forceful or strongly projected, but never mention its numerical value.
^textbullet!~ Dynamic, tempo, and articulation information should be expressed as audible intensity, pacing, shaping, touch, and character only when realized in the performance.
^textbullet!~ Mention resonance from pedaling only when it materially affects the audible texture, articulation, harmonic clarity, or phrase boundary.
^textbullet!~ Use timings and relative velocities to identify clearly audible rubato or other expressive moments, but do not overinterpret small differences.
^textbullet!~ Meter is audible and may be described directly. Conventional tempo and character language such as "Andante cantabile" may be used when it plausibly characterizes the audible pacing and expression.

Only describe performance technique that can be inferred from the audible result. Do not infer fingering, hand distribution, arm, wrist, finger movement, or bodily gesture.

Use literal musical language. Do not substitute atmospheric imagery or unsupported affect for analysis. Descriptive adjectives are permitted only when pitches, register, dynamics, timing, articulation, balance, or resonance support them in the same sentence; avoid generic color words such as darkening the scene, chromatic bite, glow, glowing, luminous, warmth, tender, poignancy, shimmer, smoke, halo, or expectation when a concrete musical event can be stated instead.

\end{appendixprompt}
\par\addvspace{0.55\baselineskip}
\begin{appendixprompt}
^promptheading!PROHIBITED OUTPUT~

Do not mention:

^textbullet!~ beat annotations
^textbullet!~ markings or score symbols
^textbullet!~ encoding labels or voice numbers ('upper voice' is OK)
^textbullet!~ MIDI velocity numbers
^textbullet!~ written note values such as whole, half, quarter, eighth, or sixteenth notes
^textbullet!~ written accents or accent symbols; describe only the resulting audible emphasis when relevant
^textbullet!~ the name or format of the supplied representation
^textbullet!~ how a musical quality was inferred from the data
^textbullet!~ the composer, title, historical context, hands, fingering, bodily gesture, or anything else that cannot be heard in the performance
Don't use 'rather than' or other fake contrasts.

\end{appendixprompt}
\par\addvspace{0.55\baselineskip}
\begin{appendixprompt}
^promptheading!OUTPUT EXAMPLE~

At 00:08.50, the opening cantabile melody F#4-E4-F#4 remains clear above an arpeggiated D-major sonority. At 00:10.25, C#4+G4+A4 combines with the melodic E4 to form an A7 color, and the slight easing into the second F#4 at 00:11.04 makes the return sound released. From 00:13.09, the melody descends to B3 and continues C#4-D4-C#4-D4-E4, with the shorter C#4 and D4 attacks at 00:16.85-00:17.14 giving the renewed ascent a more articulated impulse. At 00:13.42 and 00:19.80, the inner layer rises above the melodic register first as D4+A4 and later as C4+A4+D5, and together with the sounding F#4, the latter forms D7, V7 of G major, increasing the pressure toward the arpeggiated F# minor arrival at 00:33.00. At 00:35.29, the intensified gesture returns with reduced chordal density, after which G3+Bb3+D4 at 00:39.22 introduces a G-minor inflection before D3+F#3+A3 restores D major around 00:40.21 in a diminuendo. From 00:40.21, the melody retains a mezzo-forte singing projection, while A#3 at 00:45.91 functions as the leading tone into B minor and the increasing attack weight culminates in the forte arrival at 00:52.36. At 01:02.03, A2 beneath G3+B3+D4+E4 creates an A-dominant suspension whose D4 resolves to C#4, yielding A2+G3+C#4+E4 by 01:10.32 as an A7 sonority. At 01:10.32, the principal theme begins its repetition with the melodic line still foregrounded while soprano triplet decoration adds a more active upper layer.

Return Markdown using this hierarchy:

^promptheading!Whole-piece analysis~
^promptheading!<quoted SECTION name - role>~
^promptheading!<quoted THEME name - role>~
^promptheading!<quoted PHRASE name - role>~

Place phrases outside themes directly under their enclosing section. Cover every declaration exactly once at its appropriate analytical level. Do not reproduce the notation or boundary declarations in the response.

\end{appendixprompt}
\par\addvspace{0.55\baselineskip}
\begin{appendixprompt}
^promptheading!SEGMENTED TIMED MUSICAL DATA~

[SEGMENTED MUNO-SP REPRESENTATION]
\end{appendixprompt}
\par\addvspace{1.0\baselineskip}

\clearpage
\section{Paired Auditory-Analysis Examples}
\label{app:auditory-analysis-examples}

The following are the latest paired outputs used in the human evaluation for
Recording~4, Rachmaninoff's \textit{Prelude in D major, Op.~23, No.~4}
(WuuE07M). Both analyze the complete performance using the same structural
scope. The first is grounded in segmented MuNo-SP; the second is grounded in
segmented MIDI performance-event text.
segmented MIDI performance-event text. Figure~\ref{fig:segmentation}
shows the shared section, theme, and phrase hierarchy used to organize both
analyses.

\subsection{MuNo-SP Analysis}
\label{app:munosp-analysis}

\begin{analysispanel}
\analysissectionheading{Whole-piece analysis}{}

Across 00:00.46-04:24.49, a quietly pulsing arpeggiated texture supports a broad cantabile theme whose increasingly chordal presentations frame an agitated sequential middle span and a compressed closing recollection. The principal climax at 02:42.36-02:45.84 grows from registral expansion, faster harmonic turnover, and massive chordal attacks, while the long D-major sonority at 04:11.93-04:24.49 supplies the work's deepest release.
\end{analysispanel}
\par\addvspace{0.55\baselineskip}
\begin{analysispanel}
\analysissectionheading{Opening A section:}{double presentation of the primary theme - introduction followed by the initial and ornamented presentations of the primary cantabile theme}

\analysisphraseheading{Arpeggiated introduction}{establishes the tempo, quiet character, tonic field, and flowing accompaniment}

At 00:00.46-00:03.41, the quiet triplet-patterned arpeggio \munocode{D2+A3+D4+A4+D4+A3+F\#3+A2+D4} unfolds evenly through several registers, establishing a resonant D-major field without a separate melodic foreground. At 00:03.87-00:06.79, the same complete pattern returns over \munocode{D2}, and the spacious pacing, subdued attacks, and sustained \munocode{D4+A4+F\#3+A2} resonance make the repeated tonic sonority sound settled but still in motion.

\analysisthemeheading{Primary cantabile theme}{first presentation of the principal theme}

\analysisphraseheading{Primary-theme opening statement}{initial thematic proposition}

At 00:07.50-00:11.04, the principal line \munocode{F\#4-E4-F\#4} sings above the established D-major arpeggiation, with the slight broadening into the returning \munocode{F\#4} at 00:11.04 rounding off the first gesture. At 00:10.25, \munocode{C\#4+G4+A4} beneath \munocode{E4} forms an \munocode{A7} color, and its dominant tension releases when \munocode{F\#4} returns over \munocode{D3} and the D-major accompaniment at 00:11.04-00:12.51. From 00:13.09, the melody continues \munocode{B3-C\#4-D4-C\#4-D4-E4-F\#4}, with the closely spaced \munocode{C\#4} and \munocode{D4} attacks at 00:16.85-00:17.14 giving the ascent a momentarily articulated impulse before the longer \munocode{E4} at 00:17.57. At 00:19.52-00:20.31, \munocode{F\#4} is projected strongly against the rising \munocode{C4+A4+D5} arpeggiation, whose \munocode{C4} and \munocode{D5} add dominant-seventh pressure over the persistent D-centered bass and leave the phrase energized.

\analysisphraseheading{Primary-theme sequential continuation}{continues and intensifies the opening proposition}

At 00:20.76-00:26.01, the principal line \munocode{G4-F\#4-G4-D4} repeats its opening rise in a higher harmonic context, while \munocode{B3} and \munocode{D5} recur within the accompaniment as a distinct registral frame. At 00:23.23, \munocode{A\#3} directs the B-centered sonority toward B minor, but the descent through \munocode{G3-F\#3} at 00:26.32-00:26.61 softens that pull and briefly reduces intensity. From 00:27.03, the melody \munocode{E4-F\#4-E4-F\#4-G4-G\#4} resumes over \munocode{G3+B3} and a \munocode{D2} pedal region, with the repeated \munocode{E4-F\#4} figure at 00:29.14-00:29.91 sounding more urgent as \munocode{C5} enters above it at 00:29.78. At 00:32.19-00:32.84, \munocode{G\#4} is reinforced by \munocode{D4+E\#4} and the chromatic bass motion \munocode{B3-B\#3}, creating an augmented, leading-tone-rich sonority whose increased attack weight leaves the phrase at a local tension peak.

\analysisphraseheading{Primary-theme climactic consequent}{reaches the first thematic apex and begins the descent}

At 00:33.39-00:35.80, the principal melody \munocode{A4-G\#4} reaches its first forte apex above \munocode{F\#1} and the arpeggiated F\#-minor sonority \munocode{A3+C\#4+F\#4}, with the broad initial \munocode{A4} strongly sustained. At 00:35.80-00:40.57, the continuation \munocode{A4-G4-F\#4} moves through a renewed F\#-minor field before \munocode{Bb3+D4} over \munocode{G2} at 00:39.22-00:40.32 introduces a G-minor inflection that destabilizes the descent. At 00:40.98-00:43.11, the line \munocode{F\#4-E4} returns to a quieter D-centered sonority, although \munocode{Bb3+C\#4} beneath \munocode{E4} at 00:43.11 preserves a diminished, dominant-like tension. From 00:44.15, \munocode{F\#4-E4-F\#4} closes the melodic path while \munocode{G3-A\#2} and \munocode{C\#4} at 00:46.27-00:47.26 color the bass with chromatic unrest, and the diminuendo plus widened final spacing make the ending recede.

\analysisphraseheading{Primary-theme cadential expansion}{closes the first presentation through a surge and dissolution}

At 00:48.15-00:51.35, the principal path \munocode{D4-F\#4-A4-D5} rises forcefully from a quiet B-minor sonority, and the acceleration of registral ascent turns the initially restrained continuation into a surge. At 00:52.36, \munocode{E1+E2} beneath \munocode{C\#4+D4+A4+C\#5} forms an E-dominant sonority with the suspended \munocode{D4}, whose unresolved friction is followed by the softer G-major complex \munocode{B3+D4+G4+B4} at 00:54.77. From 00:55.89, the sustained melodic \munocode{E4} remains above an \munocode{A2} bass and \munocode{G3+B3+D4}, while the secondary arpeggiated line \munocode{B3-D4-F\#4-A4-G4-F\#4-E4-D4-B3-A3-F\#3-E3}-\munocode{B2-C\#3-A2} descends continuously through 01:02.21. At 01:00.16-01:02.21, \munocode{G3+C\#4} over \munocode{A2} produces an \munocode{A7}-directed close whose dominant function remains deliberately suspended, and the decreasing intensity dissolves the presentation without a fully attacked tonic arrival.

At 00:07.50-01:02.21, the theme's initial \munocode{F\#4-E4-F\#4} idea is sequentially raised, expanded to the \munocode{A4} apex, and converted into an extended descending afterphrase. Across 00:10.25-01:02.21, D-major stability repeatedly yields to \munocode{A7}, B-minor, F\#-minor, G-minor, and E-dominant colors, with the strongest tension at 00:33.39 and 00:52.36 subsiding into an unresolved A-dominant residue.

\analysisthemeheading{Primary cantabile theme}{ornamented second presentation with the theme transferred to an inner sustained voice}

\analysisphraseheading{Ornamented opening statement}{restates the first primary-theme phrase beneath a new upper counterline}

At 01:02.90-01:09.24, the inner principal line \munocode{F\#4-E4-F\#4-B3} restates the opening contour, while the upper counterline \munocode{D5-C\#5-B4-A4-B4-C\#5-B4-C\#5-D5-C\#5-B4-A4}-\munocode{F\#4-G4-F\#4-G4} flows in triplet-like groups through 01:09.79. At 01:02.67-01:09.63, low \munocode{D2-A2} and broken \munocode{F\#3-D4-G3-A2} patterns retain the D-major foundation, though \munocode{G3} under \munocode{E4} at 01:05.79-01:06.59 creates a gentle dominant-seventh coloration before the renewed tonic sonority at 01:07.00. From 01:10.23, the inner melody \munocode{C\#4-D4-C\#4-D4-E4-F\#4-B4} is more strongly projected, while the counterline \munocode{A4-G4-E4-G4-A4-B4-A4-B4-C\#5-B4-A4-B4-C\#5}-\munocode{D5-C\#5-D5} continues independently through 01:16.18. At 01:13.50-01:16.14, \munocode{G3+C\#4+E4} and the bass descent \munocode{C4-D3} support a dominant-rich expansion, and the emphasized \munocode{F\#4-B4} within the inner strand gives the phrase an open, upward-reaching finish.

\analysisphraseheading{Ornamented sequential continuation}{expands the second thematic presentation through rising decoration}

At 01:16.68-01:22.21, the inner principal path \munocode{G4-F\#4-G4-D4-D\#4} restates the sequential idea, while the upper decoration \munocode{E5-D5-B4-C\#5-D5-D5-C\#5-D5-E5-D5-B4-G4-A4}-\munocode{B4-A4-B4} supplies a continuously turning counterline. At 01:18.54, \munocode{A\#3} intensifies the B-minor direction beneath \munocode{F\#4}, whereas \munocode{D\#4} at 01:22.21 chromatically redirects the harmony and prevents the quieter G-centered repetition from settling. From 01:22.70, the inner continuation \munocode{E4-F\#4-G4-G\#4-G4-G\#4} rises in intensity, accompanied by \munocode{C5-B4-G4-A4-B4-C5-C\#5-D5-E5-D5-B4-C\#5-D5}-\munocode{D5-E5-E\#5} through 01:28.17. At 01:27.27-01:28.17, the repeated \munocode{G\#4-G4-G\#4} rubs against \munocode{D5-E5-E\#5} and \munocode{Bb3} in the accompaniment, and the chromatic compression plus stronger attacks produces the phrase's sharply focused peak.

\analysisphraseheading{Ornamented climactic consequent}{forte apex of the second presentation}

At 01:28.73-01:31.38, the inner principal line \munocode{A4-G4} is surrounded by the upper path \munocode{F\#5-E5-D5-C\#5-D5-E5-D5-E5}, reaching a forte high-register apex over \munocode{D2+A2} and \munocode{F\#3+D4+A4}. At 01:30.69, \munocode{C\#4+Bb4} against \munocode{G4} introduces a dissonant diminished color, but the renewed D-major collection at 01:31.62-01:33.44 steadies the repeated \munocode{F\#5-E5} descent. From 01:33.85, the inner \munocode{G4-F\#4-G4} relaxes while the upper line \munocode{C\#5-D5-E5} leads into the B-minor-centered continuation \munocode{D5-C\#5-B4-A\#4-B4-C\#5-B4-C\#5} at 01:36.17-01:38.17. At 01:40.75-01:41.86, \munocode{Eb4-D4-Eb4} beneath \munocode{C5-B4-C5} adds a lowered chromatic inflection over \munocode{B1} and \munocode{F\#2}, and the slower, softer ending turns the former apex into a withdrawn unresolved color.

\analysisphraseheading{Expanded cadence of the ornamented presentation}{culminates and dissolves the second presentation}

At 01:42.63-01:45.33, a G-minor sonority \munocode{G1+G2+D3+G3+Bb3+D4} supports the principal upper path \munocode{Bb4-G4-D4-G4-Bb4-D5}, whose compact ascent rapidly grows from piano to a strong arrival. At 01:45.96-01:47.63, \munocode{E1+E2} beneath \munocode{C4+D4+A4+C5} creates the same E-dominant suspension in expanded form, while the added counterline \munocode{D5-F\#5-G5-A5-G5} drives toward the most brilliant point at 01:47.08-01:47.63. From 01:48.31, the melodic descent \munocode{D5-D4} gives way to \munocode{D5-Bb4-A4-G4-A4-Bb4-G4-E4-D4-E4-D4}, while \munocode{A1} beneath \munocode{G3+Bb3+D4+E4} at 01:48.94 supplies an A-dominant field colored by \munocode{Bb3}. At 01:52.79-01:54.80, \munocode{G3+C\#4} and the descending bass \munocode{A3-F\#3-E3-A2-A1} prolong \munocode{A7} tension before the isolated \munocode{D4} at 01:54.46, and the diminishing attacks dissolve the ornamented presentation into low resonance.

At 01:02.90-01:54.46, the original sustained melody is retained in an inner register while a newly continuous upper counterline becomes progressively more chromatic, expansive, and forceful. The harmonic arc moves from D-major stability through B-minor, chromatic \munocode{G\#4} and \munocode{E\#5} pressure, a forte D-major apex, G-minor coloration, and an extended E-dominant to \munocode{A7}-directed dissolution at 01:45.96-01:54.80.

Across 00:00.46-01:54.80, the quiet arpeggiated introduction grows into two complete thematic presentations, with the second redistributing the melody inward and adding an active upper strand. Harmonic tension expands from simple D-major prolongation to increasingly remote minor and dominant colors, peaking at 00:52.36 and 01:45.96-01:47.63 before both presentations subside over unresolved A-directed resonance.
\end{analysispanel}
\par\addvspace{0.55\baselineskip}
\begin{analysispanel}
\analysissectionheading{Developmental middle section}{introduces a contrasting scalar link, develops a chordal sequential idea, and drives it to the central fortissimo culmination}

\analysisthemeheading{Scalar-link idea}{first statement of a brief transitional idea later recalled in the coda}

\analysisphraseheading{Quiet scalar link}{connects the opening thematic section to the developmental sequence}

At 01:56.77-02:01.55, the principal line \munocode{F\#4-E4-F\#4-B3-C\#4-D4-F\#4-A4-B4-C\#5} unfolds above a continuous accompanying scale \munocode{D2-A2-F\#3-A3-B3-C\#4-D4-A3-G3-F\#3-E3-D3}-\munocode{B2-A2-F\#2-D2-C\#2}, with the initial pianissimo spacing gradually tightening and intensifying. At 02:00.27-02:01.70, the paired upper attacks \munocode{F\#4+A4}, \munocode{D4+B4}, and \munocode{E4+C\#5} rise against the bass descent \munocode{F\#2-D2-C\#2}, creating dominant-directed chromatic pressure and a clear crescendo into the boundary.

At 01:56.77-02:01.55, the scalar-link idea transforms a short \munocode{F\#4-E4-F\#4} recollection into contrary-motion ascent above a long descending accompaniment. Its D-major opening becomes increasingly dominant-directed as the bass reaches \munocode{C\#2} at 02:01.70 and the upper line reaches \munocode{C\#5} at 02:01.55.

\analysisthemeheading{Sequential chordal episode}{initial developmental presentation and transposed continuation of the contrasting episode}

\analysisphraseheading{B-centered sequential module}{first statement of the chordal developmental model}

At 02:02.26-02:07.14, the chordal principal path \munocode{D5-F\#5-C\#5-D5-E4-F\#4-C\#5-C\#5-D5-E5} is articulated in paired or sustained attacks above a rocking \munocode{B1-F\#3-B3-F\#4} accompaniment, with repeated \munocode{C\#5} and \munocode{D5} functioning as the module's recognizable upper profile. At 02:04.32-02:07.14, \munocode{F\#4+D5}, \munocode{C\#4+E4}, \munocode{D4+F\#4}, and finally \munocode{D4+G\#4+E5} pass from B-minor toward an E-dominant color, while the crescendo and denser voicing make the final \munocode{E5} the point of greatest pressure.

\analysisphraseheading{A-centered sequential answer}{transposed answer to the preceding module}

At 02:07.38-02:12.65, the transposed chordal line begins with \munocode{C\#5+A5} over \munocode{A1} and continues \munocode{B4-C\#5-D4-E4-B4-C\#5-D5-D5-E5}, preserving the paired-attack model while increasing projection toward the last two sonorities. At 02:09.42-02:12.65, \munocode{D4+B4} and \munocode{E4+C\#5} alternate over repeated A-centered arpeggiation before \munocode{B4+D5} resolves upward to \munocode{C\#5+E5}, and the bass descent \munocode{E3-A2-E2} reinforces the module's dominant-to-tonic-directed tension within A.

\analysisphraseheading{Darkened repetition and ritardando}{varies the sequential model and temporarily relaxes its momentum}

At 02:13.25-02:18.57, the chordal path \munocode{C5-E5-B4-C5-D4-E4-B4-C5-D5-E5-F\#5} repeats the A-centered model with \munocode{C4} replacing \munocode{C\#4}, so the opening A-minor color is audibly more chromatic than the preceding major form. At 02:16.90-02:18.82, the pacing broadens through \munocode{C5}, \munocode{B4+D5}, \munocode{G4+C5+E5}, and \munocode{A4+C5+F\#5} while the bass falls \munocode{C4-E3-A2}, and the simultaneous diminuendo temporarily releases the sequence's momentum without resolving its added \munocode{F\#5}.

\analysisphraseheading{A-tempo transposed restart}{relaunches the developmental sequence quietly in a new register and tonal area}

At 02:19.64-02:25.29, the softly restarted chordal path \munocode{G4+B4+E5+B5-F\#4+A4-G4+B4-A3+F\#4-B3+G4}-\munocode{F\#4+A4-G4+B4-A4+C5-B4+D5} recasts the model around E minor, with the very high sustained \munocode{B5} keeping the texture open despite the subdued attacks. At 02:24.89-02:25.39, the bass-line figure \munocode{E4-Eb4-D4} chromatically contracts beneath \munocode{A4+C5} and \munocode{B4+D5}, and the local crescendo turns the quiet restart toward a more urgent dominant-directed close.

\analysisphraseheading{G-centered continuation}{completes the broad sequential episode before fragmentation}

At 02:25.92-02:30.50, the principal chordal path \munocode{B4-D5-A4-B4-C4-D4-A4-B4-F\#4-A4-B4-A4-B4-C5} continues the model over \munocode{G1} and repeated G-centered arpeggiation, while the sustained \munocode{D5} at 02:26.42 anchors its upper register. At 02:28.05-02:30.71, \munocode{A3+C4} and recurrent \munocode{F\#4+A4} introduce dominant color above \munocode{G2-D2}, and the softening final \munocode{A4+C5} leaves the G field open rather than fully resolved.

Across 02:02.26-02:30.50, the chordal model is transposed from B through A, A minor, E minor, and G, with its paired upper attacks and rocking bass preserved while register, mode, and pacing change. Harmonic tension rises through E-dominant and A-directed sonorities, relaxes during the ritardando at 02:16.90-02:18.82, then regains momentum through chromatic \munocode{E4-Eb4-D4} motion and the G-centered continuation.

\analysisthemeheading{Sequential chordal episode}{fragmented and accelerated transformation leading to the central climax}

\analysisphraseheading{Sequential build-up}{compresses and intensifies the developmental modules}

At 02:31.34-02:34.06, the compressed chordal path \munocode{D4+Bb4+D5-F\#4+A4-G4+Bb4-A4+C5-D4+Bb4+D5}-\munocode{E4+E5} begins quietly over \munocode{G1+G2}, with each paired attack arriving closer and more forcefully. At 02:34.40-02:36.65, the bass shifts to \munocode{F1+F2} and the upper sequence \munocode{F4+Bb4+D5+F5-Bb3+Bb4-F4+Bb4+D5-F4+F5-G4}+\munocode{D5+G5-A4+A5} expands upward, the registral widening and renewed pedal resonance materially thickening the sound. From 02:36.99, \munocode{E1+E2} supports \munocode{Bb4+D5+G5+Bb5-F\#4+F\#5-G4+G5-A4+A5-Bb4+D5}+\munocode{G5+Bb5-C5+C6}, and the semitone-related F\# layer sharpens the rising G-minor-derived sonorities. At 02:39.47-02:41.87, the final ascent \munocode{C\#5+G5+C\#6-D5+Bb5+D6-Eb5+G5+Eb6-F\#5+C\#6}+\munocode{F\#6-G5+Bb5+G6-A5+C\#6+A6} surges above \munocode{Eb1+Eb2} and the climbing bass strand \munocode{Bb2-G3-Bb3-C\#4-G4-Bb4-C\#5-G5}, peaking through extreme register, chromatic density, and maximum attack weight.

\analysisphraseheading{Central culmination and recession}{fortissimo climax followed by immediate liquidation}

At 02:42.36-02:45.42, the culmination projects \munocode{A\#5+D6+G6+A\#6-B5+D6+G6+B6-A5+D6+F\#6+A6}-\munocode{G5+B5+E6+G6-F\#5+B5+D6+F\#6-E5+G5+B5+E6} over \munocode{E1+E2}, while the inner ascent \munocode{D4-E4-G4-A4-B4-G4} sustains maximum dissonant and registral pressure. At 02:45.85-02:49.06, the liquidation descends through \munocode{D5+F\#5+B5+D6}, \munocode{B4+D5+G5+B5}, \munocode{A4+D5+F\#5+A5}, \munocode{G4+B4+E5+G5}, \munocode{F\#4+B4+D5+F\#5}, and \munocode{E4+C\#5+E5} as the bass line \munocode{F\#4-E4-D4-B3-A3-G3-A3-G3-A2} loses intensity, releasing the climax into an A-directed residue.

At 02:31.34-02:49.06, the developmental model is fragmented into rapidly alternating chord pairs, expanded octave by octave, and then reversed into a long registral descent immediately after the apex. Its sequential bass progression G-F-E-Eb-E and increasingly chromatic upper structures drive tension to 02:42.36, after which descending B-minor and E-minor-related sonorities steadily reduce harmonic and dynamic pressure.

Across 01:55.69-02:49.06, a quiet scalar bridge gives way to a broad chordal sequence whose modules are transposed, darkened, slowed, restarted, compressed, and finally magnified into the central climax. The section's tension grows from D-major and B-minor reference points through increasingly chromatic bass shifts and registral expansion, culminating at 02:42.36 before an immediate diminuendo and descending liquidation restore space.
\end{analysispanel}
\par\addvspace{0.55\baselineskip}
\begin{analysispanel}
\analysissectionheading{Grand varied return and aftersong}{recapitulatory return of the primary theme followed by a quiet closing extension}

\analysisthemeheading{Primary cantabile theme}{grand third presentation and varied return of the principal theme}

\analysisphraseheading{Grand-return opening statement}{restates the first primary-theme phrase in expanded chordal scoring}

At 02:49.87-02:54.80, the principal chordal line \munocode{F\#5-E5-F\#5-B4} restates the opening incipit, while isolated \munocode{D6-C\#6-D6-G5} attacks at 02:50.69-02:55.21 form a separate high countermelody above the fuller scoring. At 02:52.08, \munocode{E4+C\#5+E5} over \munocode{Bb2} and \munocode{G4} creates a chromatically altered dominant color, which releases into D-major sonority \munocode{F\#4+D5+F\#5} at 02:53.04 before \munocode{B3+F\#4+B4} redirects the harmony toward B minor at 02:54.80. From 02:55.76, the principal continuation \munocode{C\#5-D5-C\#5-D5-E5-F\#5} preserves the earlier thematic contour, while the high line \munocode{A5-B5-C\#6-D6} rises independently through 03:00.91. At 02:58.36-03:00.31, \munocode{C\#4+G4+C\#5+E5} and the climbing accompaniment \munocode{C\#4+E4-F\#4-G4-A4} intensify dominant pressure before \munocode{F\#4+D5+F\#5} restores a strong D-major sonority.

\analysisphraseheading{Grand-return sequential continuation}{continues the thematic return with mounting upper-register intensity}

At 03:01.19-03:03.49, the principal chordal path \munocode{E5-F\#5} begins the sequential continuation beneath the independent high line \munocode{E6-D\#6}, while \munocode{G4+B4+E5+G5} gives the opening an E-minor center. At 03:02.24-03:03.60, the accompaniment climbs \munocode{G4-A4-B4-C5} before \munocode{D\#4} and \munocode{D\#5} introduce leading-tone tension, and the reduced attack on \munocode{D\#4} makes the chromatic turn sound briefly veiled. From 03:03.96-03:08.93, the main line \munocode{G5-D5-E5-F\#5} is accompanied by \munocode{E6-B5-C\#6-D6}, with the softer B-minor interruption at 03:05.94 yielding to a renewed crescendo over \munocode{E4+G4+C\#5+E5}. At 03:09.34-03:11.68, \munocode{G5} rises into \munocode{G\#5} over \munocode{G\#4+E\#5} and the high \munocode{E6} becomes \munocode{E\#6} at 03:11.49, while the bass ascent \munocode{E4+G4-A4-A\#4-B4} and chromatic \munocode{E\#4} produce the phrase's most concentrated tension.

\analysisphraseheading{Grand-return climactic consequent}{reaches the return's forte apex and begins its relaxation}

At 03:12.44-03:14.74, the principal chordal line \munocode{A5-G\#5} reaches the return's forte apex, while \munocode{F\#6-E\#6} forms a separate high response and the inner ascent \munocode{A4-B4-B\#4-C\#5} intensifies the arrival. At 03:12.20-03:14.31, \munocode{F\#1+F\#2+C\#3} and \munocode{C\#4+F\#4+A4+C\#5+A5} establish F\# minor, after which \munocode{G\#4+E\#5+G\#5} and the E\# pitches create dominant pressure back toward F\#. At 03:15.38-03:18.48, \munocode{A5} is restated above F\# minor and then relaxes through \munocode{G5-F\#5-G5}, with the grace-like \munocode{A5} at 03:18.40 briefly heightening the \munocode{G5} return over \munocode{G2+D3}. From 03:18.94, the principal continuation \munocode{F\#5-E5-F\#5-E5-F\#5} returns to D-major scoring, while \munocode{D6-C\#6-D6-C\#6} at 03:19.50-03:24.21 gradually loses intensity and the chromatic \munocode{A\#2} bass at 03:24.52 leaves a subdued dominant inflection.

\analysisphraseheading{Grand-return cadential expansion}{completes the primary-theme return through a final surge and dissolution}

At 03:25.71-03:28.96, the principal path \munocode{D5-F\#5-A5-D6} rises from a quiet B-minor sonority, with the sustained high \munocode{B5} and increasingly forceful bass figure \munocode{F\#2+A3-E3-D3-F\#2} recreating the expansive surge. At 03:30.03, \munocode{E1+E2} beneath \munocode{C\#5+D5+A5+C\#6} forms an E-dominant suspension, and the added \munocode{A6} at 03:31.13 extends the register before \munocode{B4+D5+G5+B5} over \munocode{G3} at 03:32.56 releases into G-major color. From 03:33.70, the sustained principal \munocode{E5} remains above \munocode{A1} and \munocode{D4+G4+B4}, while the accompanying descent \munocode{B3-D4-F\#4-G4-F\#4-E4-D4-B3-E3-A3-F\#3-E3}-\munocode{B2-A2-A1} continues through 03:39.59. At 03:37.57-03:39.27, \munocode{C\#4+G4+C\#5} over A-centered bass forms \munocode{A7}, and the final \munocode{D4+D5} attack at 03:39.27 supplies an authentic dominant-to-tonic arrival in D major after \munocode{A7} at 03:37.57.

Across 02:49.87-03:39.27, the theme returns in expanded chords with a new high countermelody, compressed scalar ornaments, and a more brilliant \munocode{A5} apex before the familiar descending dissolution. D-major stability is repeatedly enlarged by B-minor, E-minor, F\#-minor, and altered dominant regions, with tension peaking at 03:12.44 and finally resolving from \munocode{A7} at 03:37.57 to D at 03:39.27.

\analysisphraseheading{Post-return harmonic detour}{extends the return through a quiet two-bar afterthought}

At 03:40.56-03:43.88, the sustained principal sonority \munocode{D4+B4+D5} with high \munocode{B5} rests over \munocode{B1-B2-E\#3-F\#3-B3-C\#4-D4-F\#3-B2}, and \munocode{E\#3} gives the quiet B-minor field a sharpened dominant inflection. At 03:44.40-03:47.10, \munocode{C4+D4+A4+C5} over \munocode{E1+E2} forms an E-dominant suspension before \munocode{Bb3+G4+Bb4} and high \munocode{G5} turn toward G minor, with the diminuendo leaving that chromatic detour gently unresolved.

\analysisphraseheading{Closing extension}{settles the aftersong toward the coda}

At 03:47.61-03:51.25, the principal sustained chord \munocode{F\#3+D4+A4} with high \munocode{F\#5} is surrounded by the accompanying line \munocode{A1-A2-F\#3-A3-B3-D4-C\#4-B3-A3-B3-A3-F\#3}, creating a calm D-major-over-A prolongation. At 03:51.68-03:54.39, \munocode{G3+C\#4+G4} and high \munocode{C\#5} form \munocode{A7} over \munocode{A1}, while the descent \munocode{A3-F\#3-E3-A2-A1} and the fading \munocode{F\#4} leave the dominant color suspended in a long diminuendo.

Across 02:49.69-03:54.39, the grand return transforms the theme through chordal doubling, a high countermelody, and more compressed internal motion before two quiet extensions reduce the texture. Its strongest F\#-minor and altered-dominant tensions occur at 03:12.20-03:14.74, after which a D-major arrival at 03:39.27 gives way to B-minor, G-minor, and finally sustained \munocode{A7} color.
\end{analysispanel}
\par\addvspace{0.55\baselineskip}
\begin{analysispanel}
\analysissectionheading{Coda}{brief thematic recollection and final tonic close}

\analysisthemeheading{Scalar-link idea}{abbreviated coda return and transformed continuation of the middle-section link}

\analysisphraseheading{Coda recollection and ascent}{recalls the scalar-link idea and redirects it toward closure}

At 03:55.80-03:59.87, the principal line \munocode{D5-E4-F\#4-B3-C\#4-D4} recalls the scalar-link profile above \munocode{D2-A2-F\#3-A3-B3-C\#4-D4-A3-G3-F\#3-A3-B3}, with the opening pianissimo D-major chord \munocode{F\#3+D4+F\#4} keeping the recollection restrained. At 04:00.29-04:02.04, the line \munocode{F\#4-A4-D5-E5} rises against \munocode{C\#4-D4-E4-F\#4-A4-C\#5}, and the contrary registral expansion plus crescendo intensifies the D-major field toward a suspended upper arrival.

At 03:55.80-04:02.04, the recalled scalar idea retains its quiet opening but replaces the earlier downward bass continuation with a sustained ascent into \munocode{A4}, \munocode{C\#5}, \munocode{D5}, and \munocode{E5}. Its D-major basis gathers dominant and added-note tension through \munocode{C\#4}, \munocode{A4}, and \munocode{E5}, reaching an unresolved high point at 04:01.79-04:02.04.

\analysisphraseheading{Penultimate arrival and release}{sustained pre-cadential sonority followed by a final diminuendo}

At 04:02.96-04:06.32, \munocode{B3+D4+B4+F\#4+G\#4+D5+F\#5} forms a sustained B-minor-derived sonority with added \munocode{G\#4}, and its unchanging resonance suspends motion for more than three seconds before a clean release. At 04:07.44-04:09.67, \munocode{A2} beneath \munocode{G3+C\#4+E4} creates \munocode{A7}, and the separately rearticulated \munocode{D4} at 04:09.67 supplies a soft, short-lived resolution while the prolonged diminuendo clears the texture.

\analysisphraseheading{Final tonic sonority}{definitive cadential close}

At 04:11.93-04:24.49, \munocode{D2+A2+F\#3+D4} establishes the final D-major tonic after the preceding \munocode{A7} at 04:07.44, completing an authentic dominant-to-tonic close with an exceptionally long pianissimo sustain.

Across 03:55.77-04:24.49, the coda compresses the scalar recollection into a single rising gesture, suspends motion on a B-minor-derived sonority, and reduces the closing melody to \munocode{D4} before the final chord. Harmonic tension passes from added-note D-major color through B-minor and \munocode{A7} at 04:02.96-04:09.67, then resolves definitively to D major at 04:11.93 and remains resonant until 04:24.49.
\end{analysispanel}

\clearpage
\subsection{MIDI Analysis}
\label{app:midi-analysis}

\vspace{10pt}
\begin{analysispanel}
\analysissectionheading{Whole-piece analysis}{}

From 00:00.00 to 04:24.49, a sustained cantabile line grows from widely spaced arpeggiation into increasingly dense sequential and chordal waves, reaches its highest and strongest concentration at 02:39.48-02:45.89, and then gradually returns to a quieter lyrical texture. Across 00:00.46-04:24.49, recurrent low \munocode{D2} and \munocode{A2} foundations, chromatic bass shifts, upper-register transformations, carefully timed pedal clearances, and progressively longer final resonances shape an arc from emergence through culmination to dissolution.
\end{analysispanel}
\par\addvspace{0.55\baselineskip}
\begin{analysispanel}
\analysissectionheading{Opening cantabile exposition}{establishes the principal theme and expands it toward the first strong arrival (00:00.00-00:52.47)}

\analysisthemeheading{Cantabile principal theme}{first presentation of the principal thematic material (00:00.00-00:29.42)}

\analysisphraseheading{Accompanimental opening and melodic emergence}{introduction and opening statement (00:00.00-00:10.14)}

At 00:00.46-00:03.41, \munocode{D2} supports the slowly unfolded D-major collection \munocode{A3+D4+A4+D4+A3+F\#3+A2+D4}, whose overlapping durations and pedal resonance make the arpeggiation sound like one sustained field. At 00:03.87-00:06.79, the same \munocode{D2-A3-D4-A4-D4-A3-F\#3-A2-D4} pattern returns with comparably even spacing, establishing a quiet recurring pulse without a separate foreground melody. At 00:07.50, the principal melodic idea emerges as the single sustained \munocode{F\#4}, clearly projected above the continuing lower pattern. Beneath \munocode{F\#4} at 00:08.12-00:09.59, \munocode{D2-A3-A4-D4-A3} restates the tonic arpeggiation, so the melody functions as the chordal third and produces consonant stability. The long-held \munocode{F\#4} at 00:07.50-00:10.14 and the slightly broader gap before its entrance give it agogic prominence, while the accompaniment remains smoothly connected by resonance. At 00:10.14, the pedal release clears the accumulated D-major sonority and exposes the phrase boundary while \munocode{F\#4} yields to the next melodic attack.

\analysisphraseheading{Answering cantabile phrase}{response and continuation of the principal theme (00:09.91-00:18.04)}

At 00:09.91-00:17.14, the principal melody unfolds completely as \munocode{E4-F\#4-B3-C\#4-D4-C\#4-D4}, with the sustained \munocode{F\#4} and \munocode{B3} giving the answer a broad descending-then-rising contour. At 00:10.24-00:11.04, \munocode{C\#4+A4+G4} above \munocode{A3} colors the melodic \munocode{E4} as an A-dominant sonority, and \munocode{F\#4} at 00:11.04 relaxes the upper dissonance without fully settling the bass. At 00:11.50-00:14.06, \munocode{D3} and then \munocode{D2} underpin \munocode{A3+A4+D4+A3}, so the descent to \munocode{B3} and rise to \munocode{C\#4} move through a D-centered consonance whose \munocode{C\#4} retains leading-tone tension. At 00:14.06-00:17.14, \munocode{C\#4} passes over \munocode{D2+A3+G4+A4+E4} before \munocode{D4} is repeated, and the \munocode{G4} against \munocode{A4} and \munocode{C\#4} supplies a suspended dominant color that resolves only partially into the final \munocode{D4}. The closely spaced \munocode{C\#4-D4} attacks at 00:16.85-00:17.14 articulate the closing rise more urgently than the earlier sustained notes, while pedal renewals at 00:14.62 and 00:16.72 keep each harmonic span resonant but distinct.

\analysisphraseheading{Broadening continuation}{sequential continuation and intensification (00:17.57-00:27.14)}

At 00:17.57-00:26.01, the principal line continues as \munocode{E4-F\#4-G4-F\#4-G4-D4}, broadening the earlier rising response by reaching \munocode{G4} twice before falling. At 00:18.00-00:19.52, \munocode{D3} beneath \munocode{A3+A4+G4+C\#4} supports \munocode{E4-F\#4} with dominant-inflected color, and the stronger \munocode{F\#4} attack at 00:19.52 heightens the sequential lift. At 00:19.80-00:20.76, \munocode{C4+A4+D5} over \munocode{D3} leads from \munocode{F\#4} to \munocode{G4}, the simultaneous \munocode{C4} and \munocode{D5} creating a \munocode{D7}-type sonority whose seventh \munocode{C4} increases pressure toward G. At 00:21.24-00:23.99, \munocode{D2} and \munocode{B3+D5+D4+B3} support \munocode{G4-F\#4-G4}, with the chromatic \munocode{Bb3} at 00:23.23 briefly introducing a G-minor inflection before the renewed \munocode{G4}. At 00:24.47-00:27.14, \munocode{D3} under \munocode{B3+D5+D4+B3} gives way to \munocode{G3-F\#3} beneath \munocode{D4}, and the stronger, more closely grouped attacks at 00:25.70-00:26.61 turn the melodic descent into a firm local release.

\analysisphraseheading{First thematic cadence}{cadential close of the initial presentation (00:27.03-00:29.42)}

At 00:27.03-00:28.78, the closing melodic path \munocode{E4-B4} is heard over \munocode{D2+G3+B3} with repeated \munocode{B3} and \munocode{G3}, so the upper \munocode{B4} crowns an E-minor-colored sonority while the low \munocode{D2} preserves dominant-seventh tension. At 00:27.26-00:29.42, sustained resonance and the spacious delay before \munocode{B4} soften the close; the unresolved \munocode{D2} beneath \munocode{G3+B3} prevents a fully authentic tonal arrival and leaves the presentation gently open.

Across 00:07.50-00:29.42, the initial \munocode{F\#4} expands into the answering \munocode{E4-F\#4} incipit, sequentially reaches \munocode{G4} at 00:20.76 and 00:23.99, and closes with the registrally enlarged \munocode{B4} at 00:28.18. From the stable D-major arpeggiation at 00:00.46-00:10.14, dominant and minor inflections at 00:10.24-00:26.61 progressively increase tension before the open \munocode{D2+G3+B3} sonority at 00:27.03-00:29.42 moderates it without complete resolution.

\analysisthemeheading{Cantabile principal theme}{expanded continuation of the first presentation (00:29.14-00:52.47)}

\analysisphraseheading{Decorated thematic restart}{varied restatement with ornamental ascent (00:29.14-00:35.92)}

At 00:29.14-00:33.61, the principal path is \munocode{F\#4-E4-C5-F\#4-G4-G\#4-F\#4-F\#4-F\#3-C\#4-F\#4-A4}, with the quick \munocode{E4-C5-F\#4} ornament at 00:29.66-00:29.91 transforming the familiar opening incipit. At 00:30.28-00:32.19, \munocode{G4} rises over \munocode{B3+D4+D3+B3+D5+D4+B3} before \munocode{G\#4} arrives against \munocode{F4+F\#4+D4}, and the semitone clash \munocode{G\#4-F\#4} plus the altered G\# creates marked chromatic pressure. At 00:33.00-00:33.61, \munocode{A3+C\#4+F\#4+A4} over \munocode{F\#1} forms an F\#-minor sonority, resolving the preceding \munocode{G\#4} tension into a forceful \munocode{A4} chordal third while greatly widening the bass-to-soprano span. At 00:34.25-00:35.92, \munocode{C\#3+A3+C\#4+C\#5+F\#4+A3} sustains that F\#-minor area, and the broad spacing after the rapid restart allows the intensified sonority to settle without losing momentum.

\analysisphraseheading{Sequential answering phrase}{answer and continued expansion (00:35.80-00:43.20)}

At 00:35.80-00:42.77, the principal melody traces \munocode{G\#4-A4-G\#4-F\#4-F\#4}, retaining the heightened register while gradually descending. At 00:35.83-00:37.08, \munocode{D4+F4} beneath \munocode{G\#4}, followed by \munocode{A3+C\#4+F\#4+A4} over \munocode{F\#2+C\#3}, prolongs altered F\#-minor color and gives \munocode{A4} a forceful local arrival. At 00:39.20-00:40.98, \munocode{Bb3+D4} with \munocode{G3} and \munocode{F\#4} introduces a G-minor-inflected sonority, after which \munocode{F\#4} over \munocode{F\#3+D4+D2+A3+A4} restores a D-centered collection at 00:40.98-00:42.77. The sustained \munocode{A4} at 00:37.08-00:41.20 and quieter, more evenly spaced accompaniment attacks from 00:39.71 make the phrase relax audibly while the pedal clearances at 00:39.30 and 00:41.20 sharpen the harmonic turns.

\analysisphraseheading{Renewed ascent}{pre-cadential build-up (00:43.11-00:48.32)}

At 00:43.11-00:47.26, the principal path \munocode{E4-F\#4-E4-F\#4} renews the thematic ascent, with the first \munocode{F\#4} sustained across 00:44.15-00:46.34. At 00:43.15-00:44.15, \munocode{Bb3+C\#4+Bb4+G3} colors \munocode{E4} with an altered dominant sonority, while \munocode{D2+A3+A4+D4} under \munocode{F\#4} at 00:44.63-00:46.25 restores D-major support before \munocode{Bb2} at 00:46.73 introduces renewed chromatic instability. The final \munocode{F\#4} at 00:47.26 is quieter and suspended over \munocode{Bb2+G3+C\#4+E4}, and the broader spacing after 00:46.25 delays release while preserving tension.

\analysisphraseheading{First major culmination}{climactic cadence closing the opening exposition (00:48.15-00:52.47)}

At 00:48.15-00:51.35, the principal line begins with \munocode{D4} and surges through \munocode{B3-F\#4-A4-D5}, the last three attacks compressed into 00:50.03-00:51.35 and strongly projected. At 00:48.96-00:50.35, \munocode{B1+F\#3+D4+B3+F\#4} forms an open B-minor sonority, then \munocode{F\#2+A3+D4+A4+D3} at 00:50.59-00:50.78 broadens the harmony before \munocode{D3+D5} at 00:51.14-00:51.35 supplies a powerful octave-centered arrival. The intensified bass entries \munocode{B1-F\#2-D3-F\#2} at 00:48.96-00:51.56, thickened texture, and accelerating upper attacks culminate at \munocode{D5}, while the sustained pedal holds the entire final sonority through 00:52.47.

Across 00:29.14-00:52.47, the opening \munocode{F\#4-E4} idea is ornamented by \munocode{C5} at 00:29.78, sequentially enlarged through \munocode{A4} at 00:33.61 and 00:37.08, and transformed into the chordal ascent \munocode{F\#4-A4-D5} at 00:50.19-00:51.35. The altered \munocode{G\#4} sonority at 00:32.19, G-minor inflection at 00:39.20, and Bb-based tension at 00:46.73 accumulate into the broad B-minor-to-D-centered arrival of 00:48.96-00:52.47.

Across 00:00.00-00:52.47, the melody grows from the isolated \munocode{F\#4} at 00:07.50 into answering, sequential, ornamented, and finally chordal forms, with its register expanding to \munocode{D5} at 00:51.35. Harmonically, repeated D-centered arpeggiation at 00:00.46-00:27.14 gives way to altered dominant, F\#-minor, G-minor, and B-minor colors, and the section's strongest tension resolves into the sustained D-centered culmination at 00:50.59-00:52.47.
\end{analysispanel}
\par\addvspace{0.55\baselineskip}
\begin{analysispanel}
\analysissectionheading{First developmental expansion}{transforms the cantabile material through sequential motion and increasing rhythmic activity (00:52.47-01:45.95)}

\analysisthemeheading{Rising sequential idea}{first extended transformation of the principal theme (00:52.34-01:10.30)}

\analysisphraseheading{Post-cadential release}{transition from the opening culmination (00:52.34-00:56.43)}

At 00:52.34-00:55.89, the principal melody descends \munocode{C\#5-B4-E4}, releasing the preceding high register through two long notes and a larger fall. At 00:52.36, \munocode{E2+C\#4+D4+A4} above \munocode{C2} creates a dense D/C\#-inflected sonority around \munocode{C\#5}, while \munocode{B4} at 00:54.73 rests on \munocode{F\#3+G4+D4+B3} and eases the semitone friction. At 00:55.10-00:56.43, \munocode{E3+D4} beneath \munocode{E4} produces a plainer, quieter sonority, and the pedal clearing at 00:54.83 separates the strong \munocode{C\#5} arrival from this softer release.

\analysisphraseheading{Flowing sequential continuation}{initiates the developmental sequence (00:55.89-01:00.18)}

At 00:55.89-00:59.78, the principal path \munocode{E4-F\#4-A4-G4-F\#4-E4-D4-B3} introduces a continuous stepwise sequential descent after the initial rise. At 00:56.42-00:58.87, \munocode{A1+E3+B3+D4+F\#4+A4+G4+F\#4} forms a resonant A-dominant field, with \munocode{G4} acting as the seventh and the upper \munocode{F\#4-E4} motion reducing its tension. The attacks tighten from 00:57.54 to 00:59.78, giving the line a flowing, nearly even pulse, while the sustained pedal fuses passing tones into one harmonically charged span.

\analysisphraseheading{Dissolving descent}{retransition into a new harmonic area (00:59.18-01:06.00)}

At 00:59.18-01:00.14, the principal continuation \munocode{E4-D4-B3-C\#4} completes the preceding descent and settles on \munocode{C\#4}. At 01:00.16-01:02.21, \munocode{A3+G3+F\#3+E3+B2+C\#3+A2} descend beneath \munocode{C\#4}, thinning the texture and carrying the bass away from the earlier \munocode{A1} foundation. At 01:02.67-01:03.42, \munocode{F\#3+D4+D2+F\#4} establishes a spare D-major sonority, but \munocode{D5-C\#5-B4-A4-B4} at 01:04.25-01:05.52 forms an important upper subvoice whose chromatic \munocode{C\#5} and delayed return to \munocode{B4} keep the harmony unsettled. The softer attacks and widening gaps at 01:01.62-01:04.25 make the motion dissolve, while \munocode{B4} at 01:05.52 slightly reanimates the line before the boundary.

\analysisphraseheading{Ornamented re-entry}{renewed statement of the rising idea (01:05.79-01:10.30)}

At 01:05.92-01:09.79, the principal path \munocode{E4-F\#4-B3-F\#4-G4} resumes the rising idea, while the simultaneous upper ornament \munocode{C\#5-B4-C\#5-D5-C\#5-B4-A4} at 01:06.00-01:08.61 forms a distinct decorative strand. At 01:06.86-01:08.87, \munocode{D2+F\#3+D4+F\#4} with \munocode{A2} supports the ascent, and the upper \munocode{C\#5-D5} dissonance resolves downward through \munocode{B4-A4} as the harmony returns to a D-centered collection. At 01:09.24-01:10.30, \munocode{B3} becomes newly prominent against \munocode{D2+F\#3+D4+F\#4+G4}, and the closer spacing of \munocode{F\#4-G4} at 01:09.51-01:09.79 gives the ending a fresh impulse.

Across 00:52.34-01:10.30, the falling \munocode{C\#5-B4} release is transformed into the flowing \munocode{E4-F\#4} incipit, dissolved through a long descent, and renewed with an independent \munocode{C\#5-B4-C\#5} ornament at 01:06.00-01:08.05. The dense \munocode{C2}-based sonority at 00:52.36 relaxes through A-dominant and descending-bass fields, then regains D-centered stability at 01:02.67-01:10.30 while upper chromatic neighbors preserve forward tension.

\analysisthemeheading{Rising sequential idea}{intensified sequential development (01:10.08-01:28.58)}

\analysisphraseheading{First sequential ascent}{antecedent in the intensified sequence (01:10.08-01:15.46)}

At 01:10.21-01:15.02, the principal path \munocode{C\#4-D4-C\#4-D4-E4} rises by overlapping repetitions, while the upper subvoice \munocode{A4-G4-E4-G4-A4-B4-A4-B4-C\#5-B4-A4-B4-C\#5} provides continuous decorative propulsion. At 01:10.08-01:12.24, \munocode{D2+G3+C\#4} over \munocode{A2} creates an \munocode{A7}-related suspension, and the movement through \munocode{D4} and \munocode{C\#4} clarifies dominant tension without immediate resolution. At 01:12.19-01:15.46, \munocode{D4} over \munocode{F\#3+A2} shifts toward D major, then \munocode{E4} above \munocode{D2+G3+C\#4} and the rising \munocode{B4-C\#5} decoration thickens the sonority and intensifies through increasingly close attacks.

\analysisphraseheading{Accelerated answer}{answer and intensification (01:15.32-01:19.72)}

At 01:15.42-01:19.23, the principal melody \munocode{F\#4-G4-F\#4} rises and answers the prior sequence, while \munocode{D5-C\#5-B4-D5-C\#5-B4-C\#5-D5} forms a rapid upper strand. At 01:15.32-01:16.76, \munocode{C4+F\#4} over \munocode{D3} with \munocode{D5} and \munocode{C\#5} produces a \munocode{D7}/F\#-colored sonority, and \munocode{G4} at 01:16.66 shifts over \munocode{B3+D4} toward G-centered support. At 01:17.19-01:19.72, repeated \munocode{D5} and \munocode{C\#5} against \munocode{D2+D4+B3+Bb3+F\#4} maintain semitone friction, and the increasingly compact attacks through 01:19.23 make the answer audibly more urgent.

\analysisphraseheading{Second ascent and local cadence}{sequential continuation (01:19.61-01:24.66)}

At 01:19.61-01:24.26, the principal path \munocode{G4-D4-D\#4-E4} rises again through a chromatic inner turn, accompanied by \munocode{D5-C\#5-B4-G4-A4-B4-A4-G4-A4-B4} in the upper layer. At 01:19.65-01:21.57, \munocode{B3+D4} over \munocode{D2} supports \munocode{G4} with a G-over-D sonority, while \munocode{G3+D4+D\#4} at 01:21.44-01:22.21 introduces the chromatic \munocode{D\#4} leading into \munocode{E4}. At 01:22.67-01:24.66, \munocode{E4+B3} over \munocode{D2} with \munocode{C5-B4-G4-A4-B4} forms an A-minor/D-dominant color, and the repeated \munocode{B4} at 01:24.26 gives a locally conclusive upper arrival without full bass resolution.

\analysisphraseheading{Compressed surge}{transition into the first high-register climax (01:24.49-01:28.58)}

At 01:24.57-01:28.17, the principal ascent \munocode{F\#4-G4-G\#4-G4-G\#4} is embedded in a rapid upper path \munocode{C5-C\#5-D5-E5-D5-B4-C\#5-D5-E5-G4-A\#4-E5-F5}. At 01:24.49-01:25.49, \munocode{A3+F\#4+C5} over \munocode{D4} forms a D-major/added-sixth color, then \munocode{G4} above \munocode{D3+D4+B3} at 01:25.44 expands toward G while chromatic \munocode{G\#4} at 01:27.27 destabilizes the sonority. From 01:25.78 to 01:28.17, near-continuous upper attacks, stronger projection, and widening register compress the pacing until \munocode{G\#4+F5} at 01:28.17 forms the phrase's sharpest, most forceful dissonant summit.

Across 01:10.08-01:28.58, the repeated \munocode{C\#4-D4} motive rises sequentially through \munocode{F\#4} and \munocode{G4}, acquiring increasingly rapid upper decorations until the \munocode{G\#4+F5} culmination at 01:28.17. \munocode{A7}-related suspension at 01:10.08, D-centered and G-centered shifts at 01:12.19-01:21.57, and chromatic \munocode{D\#4} and \munocode{G\#4} at 01:22.21 and 01:27.27 steadily heighten tension without a settled tonic arrival.

\analysisthemeheading{Climactic chordal transformation}{first forceful transformation of the lyrical theme (01:28.41-01:45.95)}

\analysisphraseheading{First chordal wave}{climactic statement (01:28.41-01:34.69)}

At 01:28.90-01:31.38, the principal upper path \munocode{A4-F\#5-E5-D5-C\#5-D5-Bb4-G4-E5-D5-E5} converts the rising idea into forceful chordal declamation. At 01:28.41-01:30.83, \munocode{D2+A2+F\#3+D4+A4} under \munocode{F\#5} and \munocode{E5} forms a broad D-major sonority, while \munocode{C\#5} and \munocode{D5} act as accented upper neighbors before returning toward \munocode{A4}. At 01:30.64-01:31.70, \munocode{C\#4+Bb4+G4+E5} over \munocode{Bb3} creates an altered dominant sonority rich in minor thirds and semitone pressure, then the renewed \munocode{D2}-based chord at 01:31.58 restores the earlier foundation. At 01:31.96-01:34.69, \munocode{A4-F\#5-E5-D5-C\#5-B4-G\#4-F\#4} continues over \munocode{D2+A2+F\#3+D4} before \munocode{E4+G4+C\#5} and \munocode{F\#4} close the wave, with the shortening attacks and pedal renewals preserving massive resonance while articulating each harmonic block.

\analysisphraseheading{Relaxing sequential response}{temporary withdrawal from the climax (01:34.50-01:40.79)}

At 01:34.71-01:37.23, the principal line \munocode{G4-F\#4-E5-C\#5-B4-Bb4-B4} descends from the preceding chordal register. At 01:34.50-01:35.63, \munocode{Bb3+G4+E5} over \munocode{F\#3} retains altered dominant color, then \munocode{B1+B3+D4+F\#4} at 01:35.48-01:37.63 establishes a leaner B-minor framework. At 01:37.46-01:38.17, \munocode{Bb3+E4+G4+C\#5-B4-C\#5} creates renewed semitone tension, which relaxes into \munocode{B1+F\#3+B3+D4+F\#4} at 01:38.62-01:40.79. The upper attacks are lighter and more evenly spaced at 01:36.17-01:40.29, so the sequential repetition withdraws in intensity even as repeated \munocode{C\#5-B4-Bb4-B4} preserves harmonic unease.

\analysisphraseheading{Preparatory descent}{transition to the next large arrival (01:40.60-01:45.95)}

At 01:40.75-01:44.90, the principal path \munocode{D\#4-D4-D\#4-D4-C4-G4-Bb4} descends chromatically before turning upward, with \munocode{C5-B4-C5} at 01:40.82-01:41.86 as a quieter upper response. At 01:40.60-01:42.67, \munocode{A3+D\#4} over \munocode{F\#2} shifts to \munocode{G1+G2+Bb3+D4}, the \munocode{D\#4-D4} semitone resolving into a G-minor sonority whose low \munocode{G1} greatly deepens the texture. At 01:43.41-01:45.95, \munocode{Bb4-G4-D4-C4-G4-Bb4-D5} rises over \munocode{G1+G2+D3}, then \munocode{C2+C4+D4+A4} at 01:45.95 delivers a forceful C-based dominant preparation through accelerating, increasingly weighted attacks.

Across 01:28.41-01:45.95, the lyrical ascent becomes a repeated chordal wave, withdraws through descending \munocode{C\#5-B4} figures, and contracts into the chromatic \munocode{D\#4-D4} preparation before the \munocode{D5} arrival at 01:45.33. Broad D-major blocks at 01:28.41 and 01:31.58 alternate with altered Bb and B-minor colors, then the low \munocode{G1} and \munocode{C2} foundations at 01:42.64-01:45.95 rebuild maximum tension.

Across 00:52.47-01:45.95, falling post-arrival tones are recast as increasingly rapid sequential ascents, culminating in forceful upper chordal paths at 01:28.90-01:33.44 and a final \munocode{D5} thrust at 01:45.33. The harmony travels from C- and A-dominant fields through D-centered sequences, chromatic \munocode{D\#4} and \munocode{G\#4}, altered Bb sonorities, and low \munocode{G1-C2} preparation, with density and attack intensity peaking at 01:44.48-01:45.95.
\end{analysispanel}
\par\addvspace{0.55\baselineskip}
\begin{analysispanel}
\analysissectionheading{Central development and first grand climax}{develops the main motives in increasingly dense waves and drives to the central high-register culmination (01:45.95-02:42.39)}

\analysisthemeheading{Climactic chordal transformation}{renewed grand statement following the developmental expansion (01:45.95-02:07.96)}

\analysisphraseheading{Grand chordal proclamation}{sectional arrival and thematic transformation (01:45.95-01:51.03)}

At 01:45.95-01:50.67, the principal path \munocode{C5-D5-F\#5-G5-A5-G5-Bb4-D5-E4-Bb4-A4-G4-E4-A4} unfolds in broad chordal blocks with a brilliant upper peak at \munocode{G5-A5}. At 01:45.95-01:47.99, \munocode{C2+C4+D4+A4} over \munocode{C1} supports \munocode{C5} and then \munocode{D5-F\#5-G5-A5}, combining dominant bass with sharpened \munocode{F\#5} and producing intense, unresolved C-based color. At 01:48.44-01:51.03, \munocode{E3+D4+Bb4} moves to \munocode{A1+E4+D4+G3+Bb4+A4+G4+E4}, and the descending upper line plus pedal clearances at 01:47.99 and 01:49.09 release the proclamation into a thinner A-dominant sonority.

\analysisphraseheading{Descending release}{post-climactic liquidation (01:50.98-01:55.81)}

At 01:50.98-01:52.77, the principal melody \munocode{Bb4-G4-E4-D4-E4-D4-C\#4} liquidates the chordal statement through a mostly stepwise descent. At 01:50.98-01:52.85, \munocode{D4+Bb4+G4+E4} over sustained \munocode{A1} forms an \munocode{A7}-related field, and \munocode{C\#4} at 01:52.77 resolves the suspended \munocode{D4} into the dominant third. At 01:53.09-01:55.81, \munocode{A3+F\#3+E3+A2+D4} over \munocode{A1} thins to a quiet dominant resonance, while the wider onset gaps and softer attacks create a clear relaxation.

\analysisphraseheading{Long gathering transition}{gradual rebuild from quiet motion (01:55.69-02:02.25)}

At 01:55.69-01:58.57, the principal idea begins modestly as \munocode{D4-F\#4-E4-F\#4}, hovering above \munocode{D2+F\#3} and a sparse \munocode{A2} bass support. At 01:57.15-01:59.77, the important inner path \munocode{F\#3-A3-B3-E4-C\#4-F\#4-D4-A3-B3-G3-C\#4-F\#3-E3} gives the texture continuous contrary motion and gradually fills the middle register. At 01:59.88-02:01.54, \munocode{D4-B2-F\#4-B2-A4-F\#2-B4-D2-C\#5} forms an accelerating registral ascent over a descending bass \munocode{F\#3-E3-D3-B2-A2-F\#2-D2-C\#2}, multiplying dominant and diminished-seventh-like tensions. The increasingly close attacks and stronger projections from 01:59.48 to 02:01.70 turn the initially quiet transition into a cumulative surge, while continuous pedal resonance binds the chromatic bass descent.

\analysisphraseheading{Bright sequential ascent}{renewed build-up (02:02.11-02:07.96)}

At 02:02.22-02:07.14, the principal path \munocode{D5-F\#5-C\#5-D5-E4-F\#4-C\#5-F\#4-C\#5-D5-G\#4-E5} rises through repeated sequential cells and reaches \munocode{E5}. At 02:02.11-02:04.84, \munocode{B1+D4+B4+F\#5} over \munocode{F\#3} and \munocode{B3} creates a bright B-minor/D-major complex, while \munocode{C\#5} and repeated \munocode{D5} intensify its dominant pull. At 02:05.09-02:07.96, \munocode{F\#4+D4+B3+C\#5-D5} over \munocode{E3} shifts through E-minor and A-dominant colors before \munocode{G\#4+E5} over \munocode{C4+D4+F\#4+C\#4+A4} supplies a strongly altered, high-register arrival; the nearly continuous attacks at 02:05.09-02:07.48 make the acceleration unmistakable.

Across 01:45.95-02:07.96, the opening \munocode{C5-D5} proclamation expands to \munocode{A5}, descends through the familiar dominant-resolution figure, and rebuilds from \munocode{D4-F\#4} cells into the renewed \munocode{E5} ascent at 02:07.14. The C-based dominant mass at 01:45.95 relaxes into \munocode{A7} at 01:50.98, then a chromatically descending bass at 01:59.48-02:01.70 and altered \munocode{G\#4} sonority at 02:07.02 restore high tension.

\analysisthemeheading{Rising sequential idea}{broad developmental sequence leading to the central climax (02:07.62-02:25.66)}

\analysisphraseheading{First broad sequential period}{antecedent developmental statement (02:07.62-02:13.30)}

At 02:07.62-02:12.64, the principal path \munocode{C\#5-A5-B4-C\#5-D5-E5} rises in broad stages, while \munocode{A4} remains sustained from 02:07.48 and stabilizes the upper harmony. At 02:07.82-02:10.40, \munocode{A1} beneath \munocode{E3+A3+E4+A3+B4+D4+E3} forms an A-dominant foundation, and the \munocode{C\#5-B4} motion articulates its third and ninth-related colors. At 02:10.68-02:13.30, repeated \munocode{E4+D4+B4+C\#5} over \munocode{A2} and \munocode{E3} thickens into \munocode{D5+E5} above \munocode{E2+A2}, with increasingly forceful attacks at 02:11.46-02:12.64 driving the unresolved dominant span upward.

\analysisphraseheading{Varied sequential answer}{parallel consequent statement (02:13.22-02:17.89)}

At 02:13.22-02:17.32, the principal path \munocode{C5-E5-B4-C5-B4-C5-D5} answers the antecedent at a slightly lower starting pitch before regaining \munocode{D5}. At 02:13.22-02:15.81, \munocode{A2+C4+A4+C5} over \munocode{A1} supports \munocode{E5} and \munocode{B4} with A-minor color, while \munocode{C5} repeatedly acts as a suspended third above A. At 02:16.06-02:17.89, \munocode{E4+C4+A3+B4+C5+D5} over \munocode{A2} maintains the sequential dominant-minor mixture, and the more measured spacing after 02:16.90 slightly relaxes intensity despite the final ascent.

\analysisphraseheading{Transitional contraction}{reduces intensity before the next surge (02:17.82-02:19.61)}

At 02:17.82-02:18.52, the principal path \munocode{E5-F\#5} is reduced to two sustained attacks above \munocode{E4+C4+G4+C5} and then \munocode{F\#4+A4+C5}. At 02:17.86-02:19.61, the \munocode{E4+C4+G4+C5} cluster over \munocode{E3} shifts to \munocode{A2}-supported \munocode{F\#5}, and the lighter attack plus broader spacing lets the suspended seconds and fourths recede without tonal resolution.

\analysisphraseheading{Extended crescendo}{cumulative build-up to the central climax (02:19.48-02:24.97)}

At 02:19.64-02:24.44, the principal path \munocode{G4-B4-E5-B5-A4-B4-G4-A4-B4} grows from isolated sustained tones into a more continuous ascent. At 02:19.48-02:22.74, \munocode{C1+B4+E5} over \munocode{B2}, \munocode{G3}, and \munocode{B3} creates an open C-major/A-minor-related field whose remote \munocode{B5} at 02:20.54 sharply expands the register without resolving the bass. At 02:22.88-02:24.97, \munocode{A3+F\#4-G4} over \munocode{E2+B2+E3} becomes increasingly dense with \munocode{B4+A4+B3+A4+E3+B4+F\#4+B3}, and the shortening onset intervals from 02:23.28 drive a clear crescendo toward \munocode{B4}.

\analysisphraseheading{Climactic link}{compressed cadence and launch into the climax (02:24.88-02:25.66)}

At 02:24.88-02:25.39, the compressed principal path \munocode{C5-D\#4-D5-B4-D4} pivots rapidly through a chromatic \munocode{D\#4}. At 02:24.88, \munocode{C5+E4+A4} forms an A-minor sonority, then \munocode{D\#4+D5+B4+D4} at 02:25.15-02:25.39 creates an altered dominant cluster whose semitone \munocode{D\#4-D4} and forceful \munocode{D5} attack leave maximum unresolved pressure.

Across 02:07.62-02:25.66, the broad \munocode{C\#5-A5} antecedent is answered by \munocode{C5-E5}, contracted to \munocode{E5-F\#5}, and rebuilt through the \munocode{B5} expansion into the compressed \munocode{D5} link at 02:25.28. Sustained A-dominant and A-minor fields at 02:07.82-02:17.89 give way to open C-based resonance and a denser \munocode{E2-A3} complex, culminating in the \munocode{D\#4+D5} chromatic collision at 02:25.15.

\analysisthemeheading{Climactic chordal transformation}{central culmination of the principal material (02:25.64-02:42.39)}

\analysisphraseheading{First central-climax wave}{large-scale climactic statement (02:25.64-02:31.41)}

At 02:25.89-02:30.49, the principal path \munocode{B4-D5-A4-B4-C4-D4-A4-D4-B4-D4-B4-A4-B4-C5} transforms the sequential idea into repeated chordal surges. At 02:25.64-02:27.90, \munocode{G1+D3+B3+G4+B4} over recurring \munocode{D3} and \munocode{G3} establishes a broad G-major field, with \munocode{D5} and \munocode{A4} acting as dominant and added-second colors. At 02:28.03-02:31.41, \munocode{C4+A3} over \munocode{G2} shifts through \munocode{D4+B3+A4} and repeated \munocode{B4-A4} figures before \munocode{C5} appears above \munocode{G2+A4}, increasing tension through bass displacement, thicker attacks, and an unresolved fourth over G.

\analysisphraseheading{Second central-climax wave}{sequential intensification (02:31.33-02:37.01)}

At 02:31.33-02:34.39, the principal path \munocode{D5-A4-G4-Bb4-C5-D5-E5-F5} rises sequentially above a temporarily reduced texture. At 02:31.36-02:34.45, \munocode{G1+G2+D4+Bb4+D5} forms a G-minor sonority, then \munocode{A4+F\#4+G4+Bb4-C5-D5-E5} shifts through chromatic dominant colors before \munocode{F5} arrives over \munocode{F1+F2+F4+D5+Bb4}. At 02:34.41-02:36.52, the continuation \munocode{F4-Bb4-F4-F5-G5-A5} rises over \munocode{F1+C3+F3+A3+C4}, and the compressed, heavily weighted attacks from 02:35.70 make \munocode{A5} the strongest point of this wave.

\analysisphraseheading{Third central-climax wave}{penultimate surge (02:36.98-02:39.49)}

At 02:36.98-02:39.06, the principal path \munocode{A\#5-F\#5-G5-A5-A\#5-C6} ascends in parallel octaves with \munocode{Bb4-F\#4-G4-A4-Bb4-C5}. At 02:36.98-02:39.49, \munocode{E2+C5+G5+Bb5} over \munocode{C2} expands through \munocode{D3+F\#4+F\#5+G4+G5} and \munocode{G2+Bb4+Bb5+C5+C6}, with semitone-rich dominant coloring and forceful octave doubling producing sustained maximum tension.

\analysisphraseheading{Apex ascent}{central high-register culmination (02:39.47-02:42.39)}

At 02:39.48-02:41.87, the principal path \munocode{C\#6-D6-D\#6-F\#6-G6-A6-B6} rises chromatically and by whole steps, doubled an octave lower as \munocode{C\#5-D5-D\#5-F\#5-G5-A5-B5}. At 02:39.48-02:42.39, \munocode{C\#5+G5+C\#6} over \munocode{Eb1} moves through \munocode{Bb2+D5+D6}, \munocode{G3+D\#5+D\#6}, \munocode{C\#4+F\#5+F\#6}, \munocode{G4+G5+G6}, \munocode{Bb4+A5+A6}, and \munocode{C\#5+B5+B6}; the ascending bass-register supports, added seconds, and continuous maximum-weight attacks culminate at \munocode{B6} while leaving the sonority powerfully suspended.

Across 02:25.64-02:42.39, repeated \munocode{B4-D5} waves are sequentially raised through \munocode{F5} and \munocode{A5}, then compressed into octave-doubled \munocode{Bb5-C6} and the final chromatic ascent \munocode{C\#6-D6-D\#6-F\#6-G6-A6-B6}. G-major and G-minor foundations at 02:25.64-02:34.45 shift through F- and C-based dominant masses, and the accumulation of altered thirds, added seconds, octave doubling, and registral expansion reaches its unresolved apex at 02:41.64-02:42.39.

Across 01:45.95-02:42.39, the transformed melody moves from the broad \munocode{C5-A5} proclamation through quieter liquidation, sequential rebuilding, and three increasingly compressed waves to \munocode{B6} at 02:41.64. Harmonic tension recedes into \munocode{A7} at 01:50.98, rebuilds through chromatic bass descent and prolonged A-centered sequences, and peaks in the G-minor, F-based, and chromatically ascending chordal masses of 02:31.33-02:42.39.
\end{analysispanel}
\par\addvspace{0.55\baselineskip}
\begin{analysispanel}
\analysissectionheading{Climactic continuation and retransition}{prolongs the apex, then descends through sequential transformations toward the return (02:42.39-03:30.13)}

\analysisthemeheading{Climactic chordal transformation}{prolongation and release of the central apex (02:42.36-02:49.95)}

\analysisphraseheading{Apex prolongation}{sustained summit of the work (02:42.36-02:45.89)}

At 02:42.36-02:45.42, the principal path \munocode{A\#6-B6-A6-G6-F\#6-E6-D6} descends in broad chordal stages while remaining in the extreme register. At 02:42.36-02:45.89, \munocode{A\#5+D6+G6+A\#6} over \munocode{C2+C3} shifts through \munocode{B2+B5+D6+G6+B6}, \munocode{G3+A5+C\#6+F\#6+A6}, \munocode{E4+G5+B5+E6+G6}, and \munocode{G4+D6+F\#6}; the dense altered extensions sustain the summit, but the downward soprano sequence and slightly broader spacing after 02:43.95 begin the release.

\analysisphraseheading{First descending release}{begins the retreat from the climax (02:45.84-02:48.17)}

At 02:45.84-02:47.59, the principal path \munocode{D6-B5-A5-G5-F\#5-E5-D5-C\#5} descends steadily in octave-rich chordal blocks. At 02:45.84-02:48.17, \munocode{F\#4+D5+F\#5+D6} moves through \munocode{E4+B4+D5+G5+B5} and \munocode{B3+A4+D5+F\#5+A5}, while the bass falls \munocode{F\#4-E4-B3-A3-G3}; each pedal clearance at 02:45.89 and 02:46.87 reduces accumulated dissonance and clarifies the retreat.

\analysisphraseheading{Second descending release}{completes the immediate post-climactic relaxation (02:48.01-02:49.95)}

At 02:48.01-02:48.83, the principal path \munocode{F\#5-E5} continues the descent above \munocode{A3+F\#4+B4+D5}. At 02:48.60-02:49.95, \munocode{G3+E5+C\#5+E4+A2} forms an A-major/dominant-related sonority whose \munocode{C\#5} and \munocode{E5} are consonant upper chord tones, and the thinner texture plus softer attacks complete the immediate relaxation.

Across 02:42.36-02:49.95, the apex is prolonged by the descending \munocode{A\#6-B6-A6} sequence and then released through a nearly continuous \munocode{D6}-to-\munocode{E5} descent. Dense altered upper extensions over \munocode{C2} at 02:42.36 gradually simplify through descending bass support into the A-centered \munocode{E5+C\#5} sonority at 02:48.60-02:49.95.

\analysisthemeheading{Rising sequential idea}{retransitional restatement in alternating surges (02:49.69-03:17.34)}

\analysisphraseheading{Retransitional surge}{first descending-register sequence after the climax (02:49.69-02:53.09)}

At 02:49.84-02:52.50, the principal path \munocode{F\#5-D6-E4-F\#4-G4-E5-C\#6} restates the surge as a wide-register sequence, with the last \munocode{C\#6} acting as an answering peak. At 02:49.69-02:53.09, \munocode{D2+F\#4+D5+F\#5} broadens through \munocode{A2+F\#3+D4}, then \munocode{Bb3+E4+E5+C\#5} over \munocode{Bb2} creates altered dominant tension that intensifies into \munocode{C\#6} before the pedal change.

\analysisphraseheading{Sequential repetition}{second retransitional surge (02:52.92-02:55.79)}

At 02:53.03-02:55.21, the recurrent incipit \munocode{F\#5-D6} returns, but its continuation becomes \munocode{A3-B3-C\#4-B4-G5}. At 02:52.92-02:55.79, \munocode{D2+F\#4+D5+F\#5} again supports the opening, while \munocode{A3-B3-C\#4} over \munocode{F\#3} leads to \munocode{B4+G5} above \munocode{F\#2+A2}, replacing the previous Bb inflection with a brighter G-major/B-minor color.

\analysisphraseheading{Sequential intensification}{third surge with stronger melodic ascent (02:55.64-02:58.47)}

At 02:55.73-02:58.05, the principal path \munocode{C\#5-A5-D4-E4-F\#4-D5-B5-C\#5-D5} strengthens the ascent through higher \munocode{A5} and \munocode{B5} peaks. At 02:55.64-02:58.47, \munocode{D2+C\#4+A4+C\#5} over \munocode{A2} moves through \munocode{A3+D4+E4+F\#4}, then \munocode{F\#3+D5+B5+C\#5} over \munocode{A2} creates a B-minor/dominant complex whose repeated \munocode{D5} and close spacing sustain heightened pressure.

\analysisphraseheading{Expansive ascent}{fourth surge and renewed build-up (02:58.32-03:01.07)}

At 02:58.44-03:00.71, the principal path \munocode{G4-C\#5-E5-C\#6-E4-F\#4-G4-F\#5-D6} expands the sequence across more than two octaves. At 02:58.32-03:01.07, \munocode{D2+C\#4+G4+C\#5+E5} over \munocode{A2} shifts through \munocode{G3+C\#4+E4+F\#4+G4} and then \munocode{C4+A4+D5+F\#5} over \munocode{D3}, with \munocode{C\#6} and \munocode{D6} supplying increasingly strong upper dominant extensions.

\analysisphraseheading{High-register answering surge}{fifth sequential statement (03:01.03-03:03.95)}

At 03:01.11-03:03.49, the principal path \munocode{G4-B4-E5-G5-E6-A4-B4-F\#5-D\#6} answers in a still higher register. At 03:01.03-03:03.95, \munocode{D2+G4+B4+E5+G5} over \munocode{G3} creates an E-minor/G sonority, while \munocode{A4+B4+F\#5+D\#6} over \munocode{A3} introduces the leading tone \munocode{D\#6} and a dominant B-major color that sharpens the ending.

\analysisphraseheading{Final retransitional repetition}{prepares the renewed culmination (03:03.89-03:06.89)}

At 03:03.96-03:06.35, the recurrent \munocode{G4-B4-E5-G5-E6} incipit returns before descending to \munocode{E4-F\#4-D5-B5}. At 03:03.89-03:06.89, \munocode{D2+G4+B4+E5+G5} again forms E-minor/G, then \munocode{D4+G4+B4+D5} over \munocode{D2} shifts toward G major, with \munocode{B5} over \munocode{D3} giving the quieter repetition a firm dominant-register lift.

\analysisphraseheading{Accelerating approach}{compressed build-up (03:06.81-03:09.32)}

At 03:06.81-03:08.93, the principal path \munocode{E5-C\#6-E4-F\#4-G4-F\#5-D6} compresses the expansive-ascent pattern into a shorter span. At 03:06.81-03:09.32, \munocode{E4+C\#5+G4} over \munocode{D2} moves through \munocode{A2+C\#4+E4+F\#4+G4} and then \munocode{C4+A4+D5+F\#5} over \munocode{D3}, while the rapidly shortening attacks from 03:07.73 to 03:08.93 intensify the dominant pull toward \munocode{D6}.

\analysisphraseheading{Renewed peak}{secondary culmination of the retransition (03:09.33-03:12.00)}

At 03:09.33-03:11.49, the principal path \munocode{G5-E6-G4-A4-Bb4-G\#5-F6} reaches a secondary high point, with the chromatic \munocode{Bb4-G\#5} combination sharpening its altered character. At 03:09.33-03:12.00, \munocode{D2+G4+B4+E5+G5} expands through \munocode{G3+E6+B3+G4+E4+A4+Bb4} and then \munocode{B1+G\#4+G\#5+F6} over \munocode{D3}, producing maximum local density and force before the abrupt pedal clearance at 03:12.00.

\analysisphraseheading{Broad cadential descent}{release from the secondary culmination (03:12.20-03:17.34)}

At 03:12.26-03:14.11, the principal path \munocode{C\#4-F\#4-A4-C\#5-A5-F\#6-C\#4-A4-B4-C5-C\#5} descends and rebounds within a broad C\#-minor-related chordal span. At 03:14.25-03:14.74, \munocode{F4+G\#5+G\#4+C\#5} over \munocode{D4} moves to \munocode{F\#4+F6} above \munocode{B3}, creating a compressed altered dominant sonority whose semitone \munocode{F4-F\#4} resolves downward in register rather than tonally. At 03:15.10-03:17.34, \munocode{F\#2+C\#3+F\#4+A4+C\#5+A5} over \munocode{A3} and \munocode{F\#6} gradually descends through \munocode{C\#4-F\#4-G4-G\#4-A4}, with broader spacing and reduced upper weight releasing the secondary peak into an F\#-minor-centered resonance.

Across 02:49.69-03:17.34, the surge pattern repeats six times, rises from \munocode{F\#5-D6} to \munocode{G5-E6}, compresses into the \munocode{D6} approach, and reaches the altered \munocode{F6} peak before descending through C\#-minor-related material. D-centered openings alternate with Bb, G, E-minor, and B-major colors, and the strongest reaccumulation at 03:09.33-03:12.00 relaxes through the \munocode{F4-F\#4} inflection into F\#-minor support at 03:15.10-03:17.34.

\analysisthemeheading{Cantabile principal theme}{fragmented reappearance preparing the full return (03:17.21-03:30.13)}

\analysisphraseheading{Lyrical re-entry}{fragmentary return of the principal theme (03:17.21-03:20.72)}

At 03:17.21-03:20.46, the principal path \munocode{D4-G5-F\#5-E5-G5-G4-E4-F4-F\#4} recalls the lyrical contour in fragments while high \munocode{G5} remains the principal projected strand. At 03:17.21-03:20.72, \munocode{D4+G5+Bb4+G4+D5} over \munocode{Bb3} shifts to \munocode{D2+F\#4+D5+F\#5} and then \munocode{A2+D4+E4+F4+F\#4}, so the G-minor color resolves into D-centered chromatic ascent with \munocode{F4-F\#4} supplying renewed leading tension.

\analysisphraseheading{Answering lyrical fragment}{continuation of the thematic return (03:20.62-03:23.54)}

At 03:20.74-03:23.16, the principal path \munocode{E5-F5-C\#6-F\#5-D6-D4-E4-F\#4} answers the fragment with a brief upper flourish followed by the familiar rising cell. At 03:20.62-03:23.54, \munocode{Bb3+E5+C\#5} over \munocode{G3} moves to \munocode{D2+F\#4+D5+F\#5} over \munocode{A2}, then \munocode{F\#3+D4+E4+F\#4} clarifies a D-major collection while the pedal reset at 03:21.73 separates the two harmonic blocks.

\analysisphraseheading{Extended transition to return}{prepares the restored principal-theme texture (03:23.50-03:30.13)}

At 03:23.50-03:25.69, the principal path \munocode{E5-C\#6-F\#5} sustains the lyrical upper register, with the changed incipit beginning on \munocode{E5} instead of the earlier \munocode{G5}. At 03:23.50-03:25.81, \munocode{Bb3+E5+C\#5} over \munocode{G3} moves through \munocode{G3+C\#6+Bb2+F\#5} to \munocode{B1+F\#4+B4+D5}, shifting from altered Bb color into B-minor support. At 03:26.75-03:28.96, the upper continuation \munocode{B5-F\#5-A5-D6} rises over \munocode{F\#3+B3+D4+C\#4+F\#2}, then \munocode{F\#2+A3+D4+A4+D5} with \munocode{E3} and \munocode{D3} broadens into a D-centered dominant mass. At 03:28.28-03:30.13, \munocode{A3+A4+D5+E5} over \munocode{E3} leads to \munocode{D3+D5+D6} and \munocode{F\#2}, and the forceful octave D arrival plus sustained pedal gathers the fragmentary texture into a decisive boundary.

Across 03:17.21-03:30.13, the principal theme reappears first as \munocode{G5-F\#5} fragments, answers with an upper \munocode{C\#6-D6} flourish, and expands to the octave-D arrival at 03:28.96. G-minor and altered Bb colors at 03:17.21-03:25.81 progressively give way to B-minor and D-centered support, with the \munocode{F4-F\#4} motion and thickened D octaves creating renewed forward pressure.

Across 02:42.39-03:30.13, the extreme-register apex descends through \munocode{D6-E5}, is reanimated by alternating sequential surges, and finally condenses into fragmentary lyrical gestures capped by \munocode{D6} at 03:28.96. Dense C-based apex harmony simplifies toward A, repeatedly destabilizes through D-, Bb-, E-minor-, and B-major fields, and settles into a broad D-centered boundary at 03:28.28-03:30.13.
\end{analysispanel}
\par\addvspace{0.55\baselineskip}
\begin{analysispanel}
\analysissectionheading{Varied reprise and subsiding close}{returns the principal thematic character in a more decorated form and gradually withdraws from the climactic register (03:30.13-04:03.06)}

\analysisthemeheading{Cantabile principal theme}{varied reprise in elevated register (03:30.01-03:44.49)}

\analysisphraseheading{Reprise opening}{varied return of the principal statement (03:30.01-03:32.56)}

At 03:30.01-03:32.21, the principal path \munocode{C\#6-A6-F\#4-A4} returns the opening idea in an elevated, compressed chordal form, with \munocode{A6} serving as an ornamental summit above the later inner descent. At 03:30.01-03:32.56, \munocode{C\#6+D5+C\#5+A5} over \munocode{C2+C3} shifts through \munocode{B2+A6+G3+D4+F\#4+A4}, combining an altered C-based dominant sonority with a descending D-major collection whose sustained resonance preserves grandeur.

\analysisphraseheading{Reprise answer}{answering phrase (03:32.53-03:37.61)}

At 03:32.53-03:37.01, the principal path \munocode{B5-G6-E5-G5-E4-F\#4-G4-F\#4-E4-D4-B3-D5} answers through a high flourish followed by a long, stepwise descent. At 03:32.54-03:33.74, \munocode{G4+B4+D5+G5} under \munocode{B5} and \munocode{G6} forms a G-major sonority, while \munocode{A1+D4+G4+B4+E5} at 03:33.67 shifts toward an A-dominant/E-minor field. At 03:34.27-03:37.61, \munocode{A2+E3+B3+D4+F\#4+G4-F\#4-E4-D4-B3} supports the descending line, and the evenly spaced attacks from 03:35.18 allow tension to subside before \munocode{D5} gently reopens the upper register.

\analysisphraseheading{Gentle continuation}{subsiding extension of the reprise (03:37.53-03:40.72)}

At 03:37.53-03:39.26, the principal path \munocode{C\#5-A5-D5} continues more sparsely, with the isolated \munocode{A5} acting as a light upper echo. At 03:37.55-03:40.72, \munocode{A3+C\#4+G4} over \munocode{F\#3} moves through \munocode{F\#3+A5+E3+B2+A2} to \munocode{A1+D5+D4}, and the falling bass plus softer, widely spaced attacks reduce intensity while preserving A-dominant tension.

\analysisphraseheading{Reprise cadence}{cadential close of the principal-theme return (03:40.56-03:44.49)}

At 03:40.56-03:43.27, the closing principal path \munocode{B4-D5-B5-F\#3-G3-A3-B3-D4} descends from a final upper echo into the middle register. At 03:40.56-03:44.49, \munocode{B1+B3+D4+D5} over repeated \munocode{B2} moves through \munocode{F3+F\#3+B3+C\#4+D4} and returns to \munocode{F\#3+B2}, outlining a B-minor half-close whose approach is the sustained B-minor sonority and whose terminal F\#-bearing support at 03:43.56-03:44.49 functions as dominant rather than tonic.

Across 03:30.01-03:44.49, the principal idea returns in compressed high-register form at \munocode{C\#6-A6}, then unfolds as a long \munocode{B5}-to-\munocode{B3} answer before settling into middle-register fragments. The altered C-based opening at 03:30.01 moves through G-major, A-dominant, and B-minor fields, and the final F\#-supported half-close at 03:43.56-03:44.49 leaves restrained residual tension.

\analysisthemeheading{Rising sequential idea}{recalled as a quiet closing transformation (03:44.36-04:03.06)}

\analysisphraseheading{Quiet chordal recall}{post-reprise continuation (03:44.36-03:47.58)}

At 03:44.36-03:46.93, the principal path \munocode{C5-D5-Bb4-G5} recalls the rising sequence in four separated chordal attacks. At 03:44.36-03:47.58, \munocode{C5+A4+C4+D4} over \munocode{C1+C2} shifts through \munocode{A3+D5+G3+E3+D3} and then \munocode{C2+Bb3+G4+G5} over \munocode{D3}, moving from C-dominant tension toward a G-minor-related sonority while the quieter attacks and pedal reset at 03:46.16 keep the blocks transparent.

\analysisphraseheading{Flowing reminiscence}{extended lyrical-sequential continuation (03:47.51-03:51.72)}

At 03:47.73-03:51.25, the principal path \munocode{D4-A4-F\#5-F\#3-A3-B3-D4-C\#4-B3-A3-B3-A3-F\#3} flows downward after the brief upper \munocode{F\#5} peak. At 03:47.51-03:49.48, \munocode{A1+F\#3+D4+A4} over \munocode{A2} supports the \munocode{F\#5} ascent as an A-dominant/D-major complex, while the subsequent \munocode{F\#3-A3-B3-D4} line fills the inner register. At 03:49.76-03:51.72, \munocode{C\#4-B3-A3-B3-A3-F\#3} resolves the suspended \munocode{D4} toward \munocode{C\#4} and then descends within A-major/dominant harmony, with nearly even spacing and lighter attacks creating a smooth diminuendo.

\analysisphraseheading{Diminishing thematic echo}{withdrawal and preparation of the closing cadence (03:51.63-03:55.90)}

At 03:51.63-03:54.39, the principal path \munocode{G4-C\#5-F\#4} contracts the sequential material into three sustained notes. At 03:51.66-03:53.57, \munocode{A3+G3+C\#4} beneath \munocode{G4} and \munocode{C\#5} forms an \munocode{A7}-related sonority, then \munocode{F\#3+E3+A2+A1} draws the bass downward while \munocode{C\#5} remains an unresolved upper third. At 03:54.39-03:55.90, \munocode{F\#4} settles over the low A foundation, and the long durations, softer attacks, and sparse spacing make the thematic echo recede without fully removing dominant tension.

\analysisphraseheading{Final sequential descent}{cadential preparation (03:55.77-03:59.50)}

At 03:55.77-03:59.12, the principal path \munocode{F\#4-D5-E4-F\#4-D4-B3-G3} descends through one last abbreviated sequential pattern. At 03:55.80-03:59.50, \munocode{D2+F\#3+D4+F\#4} over \munocode{A2} moves through \munocode{D5}, then \munocode{A3+B3+E4+C\#4+F\#4} above \munocode{A2} returns to an \munocode{A7}-related field whose bass and inner descent \munocode{D4-B3-G3} preserve restrained dominant tension.

\analysisphraseheading{Last lyrical rise}{final thematic expansion before the coda (03:59.43-04:03.06)}

At 03:59.43-04:02.93, the principal path \munocode{C\#4-A3-D4-B3-F\#4-D4-A4-E4-D5-F\#4-A4-D5}-\munocode{C\#5-F\#5} makes a final broad ascent, extending the familiar \munocode{C\#4-D4} cell to \munocode{F\#5}. At 03:59.44-04:03.06, \munocode{F\#3+A3+D4+B3} under \munocode{C\#4} gives way to \munocode{C\#4+D4+F\#4+A4}, then \munocode{D5+F\#4+A4+C\#5+F\#5} over \munocode{B3+D4} forms a B-minor/F\# sonority; increasingly strong attacks at 04:00.29-04:02.93 produce one last expansion before the sustained chord releases at 04:03.06.

Across 03:44.36-04:03.06, the rising idea is reduced to isolated \munocode{C5-D5} attacks, flows downward through an A-centered reminiscence, and makes a final expanded ascent from \munocode{C\#4} to \munocode{F\#5}. C-dominant and G-minor colors at 03:44.36-03:47.58 yield to prolonged \munocode{A7}-related harmony, which redirects into the sustained B-minor/F\# sonority at 04:02.93-04:03.06.

Across 03:30.13-04:03.06, the elevated reprise compresses the principal theme, descends into the middle register, and then recalls the sequential idea at progressively softer intensity before one final rise to \munocode{F\#5}. Altered C, G-major, A-dominant, B-minor, and G-minor colors gradually simplify, with the strongest remaining tension concentrated in prolonged \munocode{A7} sonorities and released into B-minor/F\# at 04:02.93.
\end{analysispanel}
\par\addvspace{0.55\baselineskip}
\begin{analysispanel}
\analysissectionheading{Coda and resonant postlude}{dissolves the remaining thematic material and closes with a sustained tonic-like resonance (04:03.06-04:24.49)}

\analysisthemeheading{Cantabile principal theme}{final compressed reminiscence (04:07.43-04:12.09)}

\analysisphraseheading{Coda reminiscence}{final thematic echo (04:07.43-04:12.09)}

At 04:07.43-04:11.93, the principal path \munocode{E4-D4-D4} compresses the cantabile idea to a single descending step followed by a repeated attack. At 04:07.44-04:12.09, \munocode{G3+C\#4+A2} beneath \munocode{E4} forms an \munocode{A7}-related sonority whose \munocode{C\#4} is the dominant third, and \munocode{D4} at 04:09.67 becomes a suspended fourth against that support. At 04:11.93-04:12.09, the repeated \munocode{D4} is quieter and held into the pedal clearance, so the unresolved suspension becomes the expressive point of the final reminiscence.

Across 04:07.43-04:12.09, the principal theme is reduced to \munocode{E4-D4} and then reiterated on \munocode{D4}, preserving only its descending lyrical inflection. The sustained \munocode{A7}-related support at 04:07.44-04:12.09 concentrates tension in \munocode{D4} against \munocode{C\#4}, and the pedal release exposes that unresolved suspension before the terminal sonority.

\analysisphraseheading{Final resonant chord}{terminal cadence and decay (04:11.93-04:24.49)}

At 04:11.93-04:11.99, the final sonority attacks as \munocode{D4+D2+F\#3+A2}, with the repeated \munocode{D4} serving as the last melodic event. At 04:11.93-04:12.09, \munocode{D2+A2+F\#3+D4} forms an open D-major tonic, resolving the preceding \munocode{A7}-related \munocode{C\#4+D4} tension through the disappearance of \munocode{C\#4} and retention of \munocode{D4}. At 04:12.09, the pedal clearance briefly sharpens the chord's attack boundary and removes the earlier suspended resonance. At 04:13.14, renewed pedal resonance allows \munocode{D2+A2+F\#3+D4} to blend into a unified, widely spaced final chord. From 04:13.14-04:24.49, no new melodic attack interrupts the sonority, so the texture consists entirely of sustained tonic resonance and natural decay. The low \munocode{D2} and \munocode{A2} remain perceptually foundational beneath \munocode{F\#3} and \munocode{D4} throughout 04:13.14-04:24.49, preserving tonal clarity despite the long fade. At 04:24.49, the final pedal release ends the resonance cleanly and completes the authentic dominant-to-tonic resolution initiated between 04:07.44 and 04:11.93.

Across 04:07.43-04:24.49, the last \munocode{E4-D4} fragment contracts into a repeated \munocode{D4} and then ceases, leaving the final chord to carry the close without further melodic motion. The \munocode{A7}-related suspension at 04:07.44-04:12.09 resolves into open D major at 04:11.93, and the prolonged \munocode{D2+A2+F\#3+D4} resonance supplies the work's final and most complete release.
\end{analysispanel}

\clearpage
\section{General Question--Answer Generation Prompt}
\label{app:qa-prompt}

The appendix includes only the general QA variant used to generate
non-timestamp-restricted listening questions. It requests seven questions,
one for each dimension: melody; harmony, color, and tension; texture; dynamics
and phrase shaping; function; rhythm and time; and particular performance
decisions. Both the current system prompt and current user prompt are given
below. Piece metadata and the time-stamped auditory analysis are populated for
each recording.

\par\addvspace{0.7\baselineskip}
\begin{appendixprompt}
^promptheading!SYSTEM PROMPT~

You are a trained musician and critical listener asking another trained musician about the supplied performance. Generate exactly 7 listening questions from that musician's point of view.

Write in the natural voice of a musician discussing what is heard. Every question must reflect the genuine inquiry of a musician: it must concern a musically meaningful feature, relationship, interpretation, or listening problem that a trained musician could sincerely want another musician to identify or discuss. Do not create a question merely because the analysis contains a fact that can be extracted. The questioner-not only the respondent-must demonstrate musical understanding through precise, idiomatic framing. Do not sound like a generic quiz writer, annotator, data extractor, or source-text paraphraser.

You receive only a time-stamped auditory analysis. Treat it as the complete and exclusive source for question subjects and answers. Do not use prior knowledge or introduce a claim absent from it.

\end{appendixprompt}
\par\addvspace{0.55\baselineskip}
\begin{appendixprompt}
^promptheading!OUTPUT FORMAT~

Return a JSON array of exactly 7 objects:
[
  {
    "category": "<Melody|Harmony, color, and tension|Texture|Dynamics and phrase shaping|Function|Rhythm and time|Particular performance decisions>",
    "question": "<question text>",
    "answer_format": "<brief description of expected answer format, e.g., 'chord name (e.g., G dominant 7th)', 'time in MM:SS.cc', 'interval name'>",
    "answer": "<concise answer>",
    "reasoning": "<concise explanation of why the evidence supports the answer without introducing facts absent from the analysis>",
    "evidence": "<verbatim excerpt copied from the auditory analysis that contains the source of the question and proves the answer; copy one sentence or a longer contiguous passage as needed>"
  }
]

\end{appendixprompt}
\par\addvspace{0.55\baselineskip}
\begin{appendixprompt}
^promptheading!CORE REQUIREMENTS~

^textbf!One source only:~ Use only the supplied time-stamped auditory analysis. A question subject, correct answer, answer choices, and supporting reasoning must all come from explicit statements in that analysis. Do not supply a fact from prior knowledge or infer a more specific fact than the analysis states.

^textbf!Answer-first construction:~ Select a high-quality answer before writing its question. The answer must be musically meaningful, concise, canonical, and specific enough to grade reliably. Then write a question that requests exactly that answer-no broader explanation and no second property.

^textbf!Faithful specificity:~ Match the answer's precision to the analysis. If the analysis names an octave-qualified pitch, exact chord, timestamp, instrument, count, or structural role, preserve that specificity. If it gives only a perceptual description, use that description or a finite choice set; never manufacture an exact label.

^textbf!Musician's point of view:~ Write every question as one trained musician naturally asking another trained musician about a performance they are listening to. The questioner's wording must reflect genuine musical attention: identify the relevant audible dimension precisely, use idiomatic musical terminology where useful, and assume ordinary concepts such as register, cadence, articulation, pulse, voicing, and phrase shape need no elementary explanation. Do not sound like a generic quiz writer, annotator, data extractor, or source-text paraphraser. Avoid both unnecessary jargon and oversimplified classroom language.

^textbf!Genuine musical inquiry:~ Every question must arise from a musically meaningful feature, relationship, interpretation, or listening problem that a trained musician could sincerely want another musician to identify or discuss. The question must have a musical reason to be asked beyond the fact that its answer appears in the analysis. Reject mechanical fact extraction, arbitrary note retrieval, and source sentences converted into fill-in-the-blank questions when they do not support genuine analytical listening.

^textbf!Standalone listening question:~ Ask directly about audible musical content or performance. Never refer to "the analysis," "the text," what is "described," or how the answer was obtained.

^textbf!Audible event, not hidden mechanism:~ Source support alone is insufficient: the requested event must be distinguishable by listening without a score, MIDI, or controller data. Do not ask about damper/sustain-pedal presses, releases, clearing, changes, or their exact timing, even when the analysis supplies those times. Do not claim that a slowdown is "written-out," notated, or prescribed by the score; ask about the audible spacing of attacks instead. An audible resonance change may be a subject only when the analysis explicitly identifies that perceptual change, never as a proxy for a pedal action. Preserve the scope of the evidence: a repeated foreground note with accompanying chords is not a reduction of the whole texture to one note.

^textbf!Concrete passage context:~ Formal and thematic references such as "the exposition," "the transition," "the recapitulation," "the third return of the theme," or "Piu animato ostinato episode" are allowed. They may stand without a time cue when the audible description clearly and uniquely identifies the intended passage across the supplied analysis. If there is plausible ambiguity, first add distinguishing musical context; include an event-linked timestamp or passage interval only if needed to resolve it. A clear thematic return may need no time cue; a long movement with several similar returns, transitions, or uncertain formal boundaries usually needs more context. Decide from the actual segments and the complete question, not from the composer, piece name, a formal label alone, or the assumption that an ordinal such as "third" is self-evident. Say "the first time" or "the second time" when the source establishes the occurrence, and specify which recognizable melody or gesture recurs. Do not use "the retransition following the developmental sequences," "the storm," or "the lyrical transformation" as a locator without concrete audible identification; a private analysis heading or metaphor is not a listening landmark. For timestamp-answer questions, every stated passage interval must contain the answer strictly between its endpoints. If a local passage ends before the target, use a source-supported theme-level or higher-level interval that includes it, or a clearly identified single-event cue instead; never give a preceding passage range and ask for an answer outside it. A single-event cue may precede or follow the target when the wording makes their order clear. Never give the answer timestamp or its rounded equivalent as context. Prefer plain audible descriptions such as "brief pause" over "agogic break." Specify whether the requested boundary is the start of a note, the start of silence, or the resumption after a pause; never treat these as interchangeable. If the analysis only timestamps a passage containing several such events, choose a different, explicitly timed event instead of assigning its boundary to one of them.

^textbf!Discrete timestamp targets:~ Do not ask when a phrase begins its crescendo or starts a gradual build-up, relaxation, or acceleration: the perceptual onset is too ambiguous for one exact timestamp, even if the analysis places a marking there. Choose a clearly audible, explicitly timed event instead, such as a sudden loud-to-soft change, a distinctive accented attack, a new voice entrance, or the resumption after a clearly identified pause. A high point must identify one particular attack and which occurrence, not merely "the culminating high melodic point" or a broad dynamic peak.

^textbf!Cross-segment ambiguity review:~ Before writing questions, read all supplied analysis segments, not only the excerpt that contains a candidate answer. For each candidate, compare the proposed locator with the other segments, including earlier and later thematic statements, varied returns, similar transitions, and partial recollections. Ask whether a listener could reasonably select any other occurrence or interpret the referenced boundary differently. Omit time guidance only when this whole-analysis check supports one clearly distinguishable target; otherwise add a broad time guide and audible distinguishing context, or select another event. A unique segment heading, or the absence of a second occurrence in the chosen evidence excerpt, is not proof of audible uniqueness. If the supplied coverage is insufficient to assess a potentially ambiguous reference, do not assume it is unique.

^textbf!Minimal sufficient anchoring:~ Use the least amount of neutral musical context needed to isolate exactly one event and one requested property. A clearly unique audible passage needs no time cue. Context timestamps in the question are optional: include them only when the musical wording might otherwise be ambiguous. When needed, tie each timestamp directly to a named, recognizable event or passage, such as a melody entrance, a thematic statement, the development section, or a climax. Stop adding context as soon as another occurrence can no longer be confused with the target. A locator may identify where to listen, but it must not state the musical property being tested. This optionality concerns context in the question, not the required timestamp answer in timestamp-only variants.

^textbf!Event-grounded timestamps, not artificial windows:~ Every context timestamp must be explicitly supported by the supplied analysis as the time of the event named next to it. Every interval must describe the actual source-supported extent of a named musical passage, not an arbitrary stretch of the recording, and a timestamp answer must fall strictly inside every interval in its question. Prefer the theme's or a higher-level section's actual boundaries when a smaller passage would exclude the answer or leave the target ambiguous. Do not prepend bare "Between xx and yy" search windows, invent round-number boundaries, or center a window on the answer. Do not lengthen a passage or move an event time to meet a numerical quota: there is no fixed minimum window length, minimum cue-to-answer distance, or required number of timestamps. Musical boundaries determine the times, not the reverse. If the analysis does not explicitly locate a useful contextual event, use an unambiguous cue-free description or choose another supported target. A context event must differ from the target and must not give away its answer time. In the following example forms, angle-bracket times are placeholders: replace them only with times explicitly attached to the named event in the supplied analysis, or omit the cue when it is unnecessary. Good forms: "During the first full statement of the singing theme, between <theme-start> and <theme-end>, at what timestamp does a second melodic voice enter beneath it?"; "When the melody starts at <melody-entry-time>, at what timestamp does its first upward leap land on the accented high note?"; "The second time the opening melody is played over repeated chords, at what timestamp does the inner line enter?" (if that occurrence is clearly identifiable); "After the climax at <climax-time>, at what timestamp does the detached bass figure return?" These provide musical context rather than artificial numerical hints.

^textbf!Strict answer-stem separation:~ After drafting a question, compare every meaningful word in its stem with the answer. Remove or replace any adjective, label, pitch, count, direction, relationship, choice cue, structural term, or causal description that states or strongly predicts the answer. If asking about dynamic direction, do not say that the sound recedes; if asking about texture, do not state how many layers are active; if asking about function, do not call the passage a transition before asking its role. The question may name the general dimension being tested, but not its value.

^textbf!Difficulty without ambiguity:~ Make a question difficult because it requires precise listening, comparison, recall across a passage, or discrimination among musically plausible alternatives. Never manufacture difficulty by withholding needed time guidance, leaving the target passage undefined, using vague pronouns, failing to identify which recurrence is meant, or asking about a property with several defensible labels. If adding neutral context would resolve ambiguity without revealing the answer, add it. If no such wording exists, reject the candidate.

^textbf!Single defensible interpretation:~ Read the question without consulting the intended answer and list every plausible interpretation. Keep it only if the wording fixes one musical event, one comparison frame, one requested property, one answer granularity, and one response format. "Which note," "what changes," "how is it played," and similar formulations are invalid unless the relevant voice, moment, dimension, and expected specificity are fixed neutrally.

^textbf!No predictable questions:~ Reject questions whose answer follows almost automatically from the stem or common musical expectation-for example, asking what happens dynamically immediately after identifying a climax. Prefer distinctive moments, comparisons, and relationships that demand attentive listening.

^textbf!Multiple-choice construction:~ Every multiple-choice question must contain at least seven mutually exclusive, musically plausible options, and 'answer_format' must repeat the same option set. The correct answer necessarily appears among the options, but the stem and distractors must not reveal or favor it through answer-bearing description, semantic overlap, unequal specificity, unusual length, grammatical mismatch, ordering, or implausible alternatives. Do not use "all of the above" or "none of the above." Do not add weak distractors merely to reach seven. If seven defensible options are unavailable, write a clear non-MCQ question or choose a different target.

^textbf!Explicit answer contract:~ The question itself must state the requested property and answer granularity clearly and neutrally. A musician must know whether to return one octave-qualified pitch, an ordered pitch sequence, one timestamp, a timestamp range, an integer count, a chord name, a cadence type, a register, a texture label, a formal function, or a pacing choice without consulting 'answer_format'. The 'answer_format' field must confirm the contract; it may not repair a vague stem such as "what happens," "what configuration," "how is it played," or "which register" when the relevant layer and granularity are not neutrally defined.

^textbf!Conventional answer language:~ Use established, musician-readable terminology at the precision supplied by the analysis. Do not invent compressed or hybrid labels such as "melody-plus-inner Ab4" unless the analysis itself uses that exact term. If no canonical label is supplied, ask for a concrete constituent property or use a well-constructed seven-option MCQ.

^textbf!No expectancy leakage:~ A neutral locator must not make the answer logically or conventionally predictable. Do not call a passage a conclusion, ending, or closing section before asking whether its function is a coda. Do not identify a climax before asking whether the following dynamics or tempo relax. Do not describe a voice as close-register, upper, secondary, delayed, accented, or friction-producing when asking for that same register, role, timing, emphasis, or effect. Remove the cue, ask a different property, or reject the candidate.

^textbf!HARD REJECTION - timestamp-answer leakage:~ When the expected answer is a timestamp or time range, the question stem must not contain the answer timestamp, either endpoint of the answer range, or an equivalent rounded form of those times. This is a hard failure even if the answer time appears incidentally while describing the event. Compare every timestamp in the completed stem against the completed answer before accepting the item. If any stem timestamp discloses all or part of the answer, discard or rewrite the question. Anchor with a genuinely different preceding or following event, or ask a different property.

Bad: 'At what timestamp does the cadence resolve to tonic at 01:21.55?' Answer: '01:21.55.' The question states its answer.

Good: 'Following the melody entrance at 01:05.00, at what timestamp does the prolonged dominant preparation beneath it resolve into the decisive tonic arrival?' Answer: '01:21.55.' This form is valid only if the source explicitly places the melody entrance at 01:05.00 and the resolution at 01:21.55, and the entrance cue is needed to distinguish the target. Otherwise omit it; do not copy illustrative times into generated questions.

^textbf!Reasoning:~ The 'reasoning' field must concisely explain why the cited evidence supports the answer and why the question follows from it. It may connect the evidence to the requested musical property, but it must not introduce facts, interpretations, or terminology absent from the supplied analysis. Do not use it as a substitute for the source excerpt.

^textbf!Verbatim source evidence:~ The separate 'evidence' field must copy the exact contiguous part of the supplied auditory analysis from which the question and answer originate. Copy the complete sentence when one sentence is sufficient; copy multiple consecutive sentences or a longer contiguous passage when necessary to preserve the evidence and context. Preserve the source wording exactly-do not paraphrase, summarize, add commentary, splice nonadjacent fragments, or merely say the answer is "stated in the analysis." The copied excerpt must contain the evidence that proves the answer and, when present in the source, its timestamp.

^textbf!Axis assignment:~ Assign each question to exactly one axis according to the property requested: Melody; Harmony, color, and tension; Texture; Dynamics and phrase shaping; Function; Rhythm and time; or Particular performance decisions.

^textbf!Set quality:~ Vary timestamps, passages, answer types, and question directions. Do not produce paraphrases of the same question. Follow the requested category distribution exactly.

\end{appendixprompt}
\par\addvspace{0.55\baselineskip}
\begin{appendixprompt}
^promptheading!MANDATORY SILENT AUDIT~

Before returning the JSON, audit every candidate without reporting the audit:

1. ^textbf!Uniqueness test:~ Hide the intended answer and check all supplied analysis segments for competing occurrences, not just the selected evidence excerpt. Can a knowledgeable listener identify exactly one target event and understand exactly what form of answer is requested? If a formal or thematic reference could plausibly identify another passage, supply time guidance; do not require a time cue when the audible reference is already clearly unique.
2. ^textbf!Leakage test:~ Compare the stem with the answer. Does any phrase state, paraphrase, narrow to, or make the answer highly predictable?
3. ^textbf!Alternative-answer test:~ Could a second answer be defended from the same analysis because the scope, recurrence, voice, comparison, label set, or granularity is unclear?
4. ^textbf!Difficulty test:~ Is the challenge musical and perceptual rather than linguistic or underspecified?
5. ^textbf!Repair rule:~ Add only neutral anchoring to fix ambiguity; remove answer-bearing description to fix leakage. If both cannot be satisfied at once, discard the candidate and choose a better-supported event.
6. ^textbf!Answer-contract test:~ Without reading 'answer_format', is the requested property and required granularity completely clear from the question?
7. ^textbf!Expectancy test:~ Does the locator name a climax, ending, return, cadence, register, effect, or other context that makes the answer conventional or logically obvious?
8. ^textbf!MCQ test:~ If choices are present, are there at least seven mutually exclusive and plausible options with no cue favoring the correct one?
9. ^textbf!Timestamp-answer test:~ For a timestamp or time-range answer, list every timestamp in the stem. Reject the candidate if any listed time equals, rounds to, contains, or gives an endpoint of the answer.
10. ^textbf!Genuine-inquiry test:~ Could a trained musician sincerely ask this while listening because the answer illuminates a meaningful musical feature or performance decision? Reject it if its only purpose is extracting an available fact.
11. ^textbf!Audibility and boundary test:~ Could a listener locate the requested event without seeing notation or pedal/controller data? Does the evidence explicitly time that event rather than merely a surrounding passage? Reject hidden-mechanism questions, score-only distinctions, unsupported textural reductions, and interchangeable note/pause boundaries. Check that potentially ambiguous formal, thematic, or tempo-label passage references have timestamp context; clearly unique audible references may stand alone after the cross-segment review.
12. ^textbf!Context timestamp test:~ Is each context timestamp necessary to resolve a plausible ambiguity? Does the question explicitly name the real event or passage attached to each time, and does the supplied analysis support that exact event-time relationship? Reject arbitrary or artificially widened windows even if their numbers appear elsewhere in the analysis. Permit cue-free thematic references when clearly identifiable. Check that before/after/during wording matches the events' order and that the answer falls strictly inside every stated interval; use a source-supported theme-level or higher-level interval if needed. Identify the first/second occurrence when known. Reject metaphor-only or vague formal locators and gradual crescendo-onset targets. Never reveal the answer time or a rounded equivalent.

\end{appendixprompt}
\par\addvspace{0.55\baselineskip}
\begin{appendixprompt}
^promptheading!REJECT A CANDIDATE QUESTION WHEN~

^textbullet!~ the answer is absent, hedged, contradictory, or only implied by the analysis;
^textbullet!~ more than one answer would reasonably satisfy the wording;
^textbullet!~ answering requires notation, metadata, historical knowledge, or information outside the analysis;
^textbullet!~ the question asks for multiple properties, causes, or locations;
^textbullet!~ the answer is subjective without a closed choice set;
^textbullet!~ the question merely copies a sentence and removes one obvious word;
^textbullet!~ the event is too dense, subtle, or underspecified for the requested answer; or
^textbullet!~ the answer repeats information already disclosed in the question.

\end{appendixprompt}
\par\addvspace{0.55\baselineskip}
\begin{appendixprompt}
^promptheading!CATEGORY DISTRIBUTION~

Generate exactly 1 questions for each of the seven axes.

Before returning the JSON, silently audit every item: (1) ask whether a trained musician could genuinely want to know the answer as part of attentive musical listening, and reject mere fact-extraction questions; (2) hide the answer and ask whether the stem still identifies one event and one task; (3) compare the stem with the answer and remove any phrase that gives the answer away; (4) for every timestamp-answer item, list every time in the stem and reject the item if any one equals, rounds to, contains, or supplies a boundary of the answer; and (5) reject or rewrite any item for which a knowledgeable listener could defend a second answer. Do not include this audit in the output.

Make sure you create a varied set of questions.


\end{appendixprompt}
\par\addvspace{0.75\baselineskip}
\begin{appendixprompt}
^promptheading!USER PROMPT~

Task: Generate exactly 7 listening-comprehension question(s) from the supplied time-stamped auditory analysis.

Piece: [COMPOSER] - [TITLE]

The auditory analysis below is the only source. Every question and answer must be directly supported by a concrete statement in it. Do not use prior knowledge, reconstruct missing musical facts, or invent details that the analysis does not discuss.

Adopt the viewpoint and voice of a trained musician listening analytically to
the corresponding recording. Each question must sound like a question that one
musician would naturally ask another musician about what they hear-not like a
worksheet instruction, a paraphrase of source material, or a question written
by a data-extraction system:
^textbullet!~ Every question must reflect the genuine inquiry of a musician: it should arise from a musically meaningful feature, relationship, interpretation, or listening problem that a trained musician could sincerely want another musician to identify or discuss. Do not manufacture a question merely because the analysis contains an extractable fact.
^textbullet!~ Use precise, idiomatic musical language when it helps identify the audible property, while avoiding unnecessary jargon.
^textbullet!~ Assume the respondent can recognize ordinary musical concepts by ear; do not explain basic terminology inside the question.
^textbullet!~ Ask directly about the musical event or performance choice. Do not use classroom-management phrases such as "according to the passage" or "select the correct answer."
^textbullet!~ Make it standalone: never refer to "the analysis," "the text," the supplied source, or what is "described."
^textbullet!~ Ask along exactly one of the analysis's seven axes: Melody; Harmony, color, and tension; Texture; Dynamics and phrase shaping; Function; Rhythm and time; or Particular performance decisions.
^textbullet!~ Generate exactly 1 questions for each of the seven axes.
^textbullet!~ Select claims that the analysis states precisely enough to yield one concise, definitive answer.
^textbullet!~ Make the answer capture the exact requested musical property-no extra interpretation, explanation, or bundled second answer.
^textbullet!~ Be especially careful not to disclose or strongly imply the answer in the question. Do not reuse answer-bearing adjectives, labels, pitch names, counts, directions, or structural terms from the source when that same property is being tested.
^textbullet!~ Make the question unambiguous through minimal sufficient musical context. Context timestamps are optional and only needed when the wording might otherwise be ambiguous. Tie every time to a named source-supported event or the actual boundaries of a musical passage; never add an artificial numerical search window. A clearly identifiable thematic statement may be named without timestamps.
^textbullet!~ For timestamp-answer questions, use the shortest natural musical description that leaves exactly one defensible location without access to the auditory-analysis text. If a concise description is already unique, keep it concise. Only when two or more passages could plausibly match should the question add thematic occurrence, melodic or harmonic content, register, texture, affected voice, or a source-timed preceding or following event. A named context passage may precede the answer rather than contain it. Stop adding context as soon as the target is unique; the answer must still be one timestamp.
^textbullet!~ Preserve difficulty without creating ambiguity. Difficulty must come from the listening discrimination required, never from vague wording, missing context, unclear scope, or multiple plausible events.
^textbullet!~ Balance anchoring and leakage: the stem must identify where and what to listen to without stating the property the listener must determine.
^textbullet!~ Prefer distinctive events and relationships that require attentive listening. Avoid vague opinion questions, trivial observations, predictable answers, and claims the analysis presents as uncertain.
^textbullet!~ If the question is multiple choice, include at least seven mutually exclusive, musically plausible options in the question and repeat the same options in 'answer_format'. The correct answer may appear among the options, but neither the stem, option wording, option length, option order, nor surrounding context may signal which option is correct. Do not pad the set with implausible distractors merely to reach seven; if seven defensible choices are unavailable, ask a clear non-MCQ question or choose a different target.
^textbullet!~ Make the expected answer type and granularity explicit in the question itself. State neutrally whether the listener should supply a pitch, pitch sequence, timestamp, count, chord, cadence type, register, textural configuration, formal function, pacing choice, or other specific property. 'answer_format' must confirm that contract, not rescue an unclear stem.
^textbullet!~ Do not invent compressed or hybrid labels such as "melody-plus-inner Ab4" unless the analysis itself uses that exact label. Prefer conventional musician-readable answers, or define a finite choice set when the analysis does not provide a canonical term.
^textbullet!~ Reject questions whose answer is made predictable by the locator or surrounding event. In particular, do not identify a passage as a conclusion and then ask whether it is a coda; do not identify a climax and then ask whether the following pacing relaxes; and do not state a close-register secondary voice and then ask which register it occupies.
^textbullet!~ Bad expectancy example: "As the cadence arrives, how does the tempo change?" A cadence conventionally primes "slows down," making that answer too easy. Use this otherwise predictable question form only when the auditory analysis explicitly states a genuinely contrary result, such as acceleration or unchanged pacing. Keep the stem neutral: never say "unexpectedly," "surprisingly," "rather than slowing," or otherwise signal the contradiction.
^textbullet!~ HARD TIMESTAMP-LEAKAGE RULE: If the requested answer is a timestamp or time range, the stem must not contain the answer timestamp, either boundary of the answer range, or an equivalent rounded form of any of those times. This prohibition applies even when the time appears incidentally inside a phrase describing the target event. Before accepting the item, compare every timestamp in the stem against the answer character by character and semantically. If any timestamp discloses all or part of the answer, discard or rewrite the question. Use a genuinely different timestamp as a neutral preceding or following anchor, describe a different audible cue, or ask a different property.
^textbullet!~ Bad: "At what timestamp does the cadence resolve to tonic at 01:21.55?" Answer: "01:21.55." The stem states the answer.
^textbullet!~ Good: "At what timestamp does the dominant sonority heard at 01:20.89 resolve to tonic?" Answer: "01:21.55." The timestamp in the stem identifies a different event.
^textbullet!~ Vary the events, time spans, question forms, and axes across the set.

\end{appendixprompt}
\par\addvspace{0.55\baselineskip}
\begin{appendixprompt}
^promptheading!CATEGORY-SPECIFIC GUIDELINES~

^promptheading!MELODY~

Choose an explicit claim about a clearly audible melodic strand: an ordered pitch path, arrival pitch, registral turn, recurrence, variation, countermelody, or transfer of the melodic foreground.

High-quality answers include:
^textbullet!~ one octave-qualified pitch at a uniquely identified arrival;
^textbullet!~ a short, complete pitch sequence stated in the analysis;
^textbullet!~ one register or instrument carrying the line;
^textbullet!~ one stated transformation of a returning idea; or
^textbullet!~ one timestamp marking a melodic entrance, turn, or transfer.

Ask for only the melodic property the analysis states. Do not reconstruct omitted notes, combine separate lines, infer a melody from harmony, or request an exact pitch when the analysis gives only contour or register. For comparisons, specify one dimension such as pitch content, register, or direction.

Avoid bare "does it rise or fall?" questions unless the contour is genuinely distinctive and the choices are necessary. Avoid questions whose stems name the melodic goal, reveal the direction being tested, or contain most of the requested pitch sequence.

For timestamp-answer questions, first ask whether the target is already unique from a concise musical description. If it is, do not add unnecessary context. Only when multiple melodic passages could plausibly match should the question become context-heavy, adding the thematic occurrence, melodic content, register, accompanying texture, a surrounding event, or-if still needed-a broad search interval. Add only what is required to leave one defensible location, and never include the target timestamp or a rounded equivalent.

Example forms:
^textbullet!~ "Which octave-qualified pitch completes the opening foreground gesture before its first rest?"
^textbullet!~ "Which pitch sequence is heard in the foreground gesture beginning at <gesture-entry-time> and ending at its first rest at <rest-start-time>?"
^textbullet!~ "When the principal melody returns at <theme-return-time>, which register carries it: extreme bass, bass, lower-middle, middle, upper-middle, treble, or extreme treble?"
^textbullet!~ "During the second statement of the primary theme, what bass-note sequence accompanies the foreground sequence A4-B4-C5?"
^textbullet!~ "What is the highest octave-qualified pitch reached by the melody in its first complete statement?"
^textbullet!~ "Following the melody entrance on C5 at <melody-entry-time>, at what timestamp does the foreground line end as the harmony reaches the B-flat dominant chord?"

\end{appendixprompt}
\par\addvspace{0.55\baselineskip}
\begin{appendixprompt}
^promptheading!HARMONY, COLOR, AND TENSION~

Choose an explicit claim about a sounding chord or sonority, bass-supported color, tonal center, chromatic alteration, dissonance and resolution, cadence, harmonic prolongation, or concrete harmonic cause of tension and release.

High-quality answers include:
^textbullet!~ one fully named chord or explicitly listed pitch collection;
^textbullet!~ one bass pitch supporting a stated arrival;
^textbullet!~ one dissonant interval, chromatic pitch, or resolution tone;
^textbullet!~ one cadence destination or tonal center explicitly named in the analysis; or
^textbullet!~ one controlled harmonic-state label such as stable, chromatic, suspended, or resolving.

Do not perform a new harmonic analysis. Ask only for terminology and relationships the auditory analysis itself supplies. If the analysis describes color without naming a chord, ask about that stated color using finite choices rather than inventing a chord label. A tension answer must identify the harmonic cause or outcome, not merely repeat that tension increases.

Avoid broad key-identification questions, bare major/minor questions, and compound questions asking for both a sonority and how it is performed.

For timestamp-answer questions, first ask whether the target is already unique from a concise musical description. If it is, do not add unnecessary context. Only when multiple harmonic passages could plausibly match should the question become context-heavy, adding the thematic occurrence, bass motion, sounding sonority, surrounding cadence, a nearby event, or-if still needed-a broad search interval. Add only what is required to leave one defensible location, and never include the target timestamp or a rounded equivalent.

Example forms:
^textbullet!~ "Which fully named chord arrives immediately after the first complete presentation of the theme?"
^textbullet!~ "What octave-qualified bass pitch supports the melody's held arrival at <melodic-arrival-time>?"
^textbullet!~ "When the melody enters at <melody-entry-time>, which harmonic state is most clearly audible: stable tonic, dominant preparation, chromatic sequence, suspended dissonance, deceptive resolution, cadential arrival, or tonal ambiguity?"
^textbullet!~ "What cadence type completes the first full statement of the primary theme?"
^textbullet!~ "Following the melody entrance at <melody-entry-time>, at what timestamp does the harmony move from the sustained dominant preparation into the warmer submediant color beneath it?"

\end{appendixprompt}
\par\addvspace{0.55\baselineskip}
\begin{appendixprompt}
^promptheading!TEXTURE~

Choose an explicit claim about the number, balance, register, spacing, or interaction of audible layers: melody, secondary line, inner voices, bass, accompaniment, chordal mass, or named instrumental groups.

High-quality answers include:
^textbullet!~ one controlled texture label such as single line, melody with accompaniment, contrapuntal layers, or chordal blocks;
^textbullet!~ one explicitly stated number of audible layers;
^textbullet!~ one register or instrumental group in the foreground;
^textbullet!~ one relationship such as converging, diverging, crossing, or widely separated; or
^textbullet!~ one timestamp where a layer enters, withdraws, or changes role.

Keep the answer textural. If the requested answer is a pitch sequence, chord, intensity direction, or formal role, use another axis. Ask for an exact layer count only when the analysis states it clearly. Do not infer hands, hidden voices, doublings, or instrumentation that the analysis does not name.

Do not reveal the answer by describing the texture as sparse, dense, layered, chordal, or widely spaced in the question stem when that property is what the question asks.

For timestamp-answer questions, first ask whether the target is already unique from a concise musical description. If it is, do not add unnecessary context. Only when multiple textural events could plausibly match should the question become context-heavy, adding the thematic occurrence, audible registers and roles, the line that changes role, a surrounding event, or-if still needed-a broad search interval. "A second voice" needs expansion only when more than one event could fit it. Add only what is required to leave one defensible location, and never include the target timestamp or a rounded equivalent.

Example forms:
^textbullet!~ "When the opening melody returns at <theme-return-time>, which texture is audible: single line, melody with accompaniment, two-part counterpoint, three-part counterpoint, chordal homophony, layered ostinato texture, or antiphonal exchange?"
^textbullet!~ "How many clearly separable sounding layers are audible during the second statement of the primary theme?"
^textbullet!~ "When the opening gesture returns at <gesture-return-time>, which register is foregrounded: extreme bass, bass, lower-middle, middle, upper-middle, treble, or extreme treble?"
^textbullet!~ "During the second statement of the primary theme, at what timestamp does the performer begin to bring out the inner line moving beneath the repeated upper-register melody?"

\end{appendixprompt}
\par\addvspace{0.55\baselineskip}
\begin{appendixprompt}
^promptheading!DYNAMICS AND PHRASE SHAPING~

Choose an explicit claim about audible intensity, balance, emphasis, connection, breath, arrival, taper, release, sustained hold, or the overall shape of a phrase.

High-quality answers include:
^textbullet!~ one direction or shape from a closed set: intensifying, relaxing, level, or arching;
^textbullet!~ one timestamp marking a stated peak, breath, arrival, taper, or release;
^textbullet!~ one comparative label such as more intense, less intense, more connected, or less sustained; or
^textbullet!~ one change type such as gradual or sudden.

Match the answer to the analysis's wording. Do not translate a general expressive adjective into a precise dynamic level. When comparing phrases, ask about exactly one dimension. Do not ask where the peak occurs and how the performer reaches it in the same question.

Prevent answer leakage: do not call a passage a crescendo before asking its direction, describe a phrase as fading before asking about its ending, or identify a point as the peak before asking where the peak occurs. Avoid generic emotional labels and questions whose answer is predictable merely because the stem identifies a climax.

For timestamp-answer questions, first ask whether the target is already unique from a concise musical description. If it is, do not add unnecessary context. Only when multiple arrivals, peaks, or releases could plausibly match should the question become context-heavy, adding the thematic occurrence, phrase opening, melodic or harmonic trajectory, surrounding texture, a nearby event, or-if still needed-a broad search interval. Add only what is required to leave one defensible location, and never include the target timestamp or a rounded equivalent.

Example forms:
^textbullet!~ "Across the opening melodic phrase, which intensity contour is heard: steady intensification, steady relaxation, approximately level, arch, reverse arch, terraced increase, or terraced decrease?"
^textbullet!~ "Following the melody entrance at <melody-entry-time>, at which timestamp does the rising foreground line over repeated chords reach its strongest arrival?"
^textbullet!~ "At the melody's return at <theme-return-time>, which intensity-change type is heard: gradual increase, gradual decrease, sudden increase, sudden decrease, terraced change, arch-shaped change, or approximately unchanged?"
^textbullet!~ "Toward which octave-qualified pitch does the performer build the first complete foreground phrase?"
^textbullet!~ "During the second statement of the primary theme, at what timestamp does the foreground line over repeated chords reach its first clear cadential endpoint?"

\end{appendixprompt}
\par\addvspace{0.55\baselineskip}
\begin{appendixprompt}
^promptheading!FUNCTION~

Choose an explicit claim about the structural role of a musical unit: opening, statement, continuation, transition, intensification, climax, return, dissolution, cadence, coda, or closing gesture.

High-quality answers include:
^textbullet!~ one structural role selected from a finite set;
^textbullet!~ one timestamp locating a uniquely identified structural event;
^textbullet!~ one stated function of a recurrence; or
^textbullet!~ one decisive audible event that the analysis identifies as a boundary.

Use this axis only when the requested answer is structural role or location. Do not turn a question about a pitch, chord, texture, dynamic direction, or rhythmic property into a Function question merely because it occurs at a boundary.

Do not derive a conventional form from the piece's identity or broader musical knowledge. Do not identify something as a return, transition, or climax in the stem and then ask for that same role. When asking for a timestamp, describe the event without using its answer timestamp.

For timestamp-answer questions, first ask whether the target is already unique from a concise musical description. If it is, do not add unnecessary context. Only when multiple structural moments could plausibly match should the question become context-heavy, adding the theme or gesture involved, how it differs from another occurrence, its surrounding transition or cadence, a nearby event, or-if still needed-a broad search interval. Labels such as "the principal idea" need expansion only when they could identify more than one passage. Add only what is required to leave one defensible location, and never include the target timestamp or a rounded equivalent.

Example forms:
^textbullet!~ "Which formal function best describes the passage launched by the detached bass figure at <bass-figure-entry-time>: opening, thematic statement, continuation, transition, intensification, climax, or closing gesture?"
^textbullet!~ "During the final return of the opening material, between <section-start> and <section-end>, at what timestamp does the opening melody give way to the detached chordal closing refrain?"
^textbullet!~ "Which formal function best describes the recurrence of the opening gesture at <gesture-return-time>: restatement, continuation, transition, intensification, climax, dissolution, or closure?"
^textbullet!~ "Which octave-qualified leading tone drives the modulation launched by the bass change at <bass-change-time>, before the opening theme returns in its original register and accompaniment?"
^textbullet!~ "During the first full statement of the singing theme, between <theme-start> and <theme-end>, at what timestamp does its contrasting chordal continuation begin?"
^textbullet!~ "Across the performance, at which timestamps does the main theme begin each complete presentation containing both its opening repeated-note gesture and its closing cadential continuation?"

\end{appendixprompt}
\par\addvspace{0.55\baselineskip}
\begin{appendixprompt}
^promptheading!RHYTHM AND TIME~

Choose an explicit claim about pulse, subdivision feel, recurrence, displacement, syncopation, temporal regularity, pause, restart, rhythmic suspension, or contrast between rhythmic layers.

High-quality answers include:
^textbullet!~ one stated count of a clearly defined recurring cell;
^textbullet!~ one controlled pulse label such as steady, fluctuating, or suspended;
^textbullet!~ one subdivision feel such as duple, triple, or irregular;
^textbullet!~ one layer carrying a stated rhythmic pattern; or
^textbullet!~ one timestamp marking a conspicuous pause, delayed entrance, or restart.

Ask only for rhythmic information explicitly supplied by the analysis. Do not infer meter, exact note values, or a numerical tempo. Define the recurring event precisely enough that a count has one answer. Keep expressive timing choices such as deliberate acceleration, relaxation, and agogic emphasis under Particular performance decisions unless the analysis treats them only as broad rhythmic organization.

Avoid counts in dense or vaguely bounded passages and questions whose stem already names the pulse or subdivision being requested.

For timestamp-answer questions, first ask whether the target is already unique from a concise musical description. If it is, do not add unnecessary context. Only when multiple rhythmic events could plausibly match should the question become context-heavy, adding the recurring cell, the layer carrying it, its thematic occurrence, a nearby event, or-if still needed-a broad search interval. Add only what is required to leave one defensible location, and never include the target timestamp or a rounded equivalent.

Example forms:
^textbullet!~ "During the second statement of the primary theme, which pacing behavior is heard: steady, accelerating overall, decelerating overall, fluctuating, suspended, abruptly interrupted, or restarting after a pause?"
^textbullet!~ "How many complete repetitions of the short-short-long foreground cell occur after its first entrance at <cell-entry-time> before it changes at <cell-change-time>?"
^textbullet!~ "Which subdivision feel characterizes the accompaniment beneath the first full statement of the theme: duple, triple, quadruple, quintuple, sextuple, septuple, or irregular?"
^textbullet!~ "Among A4, C5, and E-flat5 in the opening foreground gesture, which note is performed fastest?"

\end{appendixprompt}
\par\addvspace{0.55\baselineskip}
\begin{appendixprompt}
^promptheading!PARTICULAR PERFORMANCE DECISIONS~

Choose one interpretive decision that the auditory analysis explicitly attributes to the performer: articulation or connection, rubato, acceleration, relaxation, agogic emphasis, attack character, note length, balance, resonance, or changed treatment of repeated material.

Also allow a likely missed note or disrupted continuation as a particular performance event when the analysis explicitly grounds it in the surrounding melodic line, pitches, pacing, emphasis, or hesitation. Phrase the question and answer entirely as something a listener could hear or infer, such as a likely missed note interrupting a descending line. Describe only the audible event, never how it was identified, and do not present the performer's intention as certain.

High-quality answers include:
^textbullet!~ one comparative articulation label such as more connected, more detached, or unchanged;
^textbullet!~ one pacing choice such as accelerating, relaxing, or remaining even;
^textbullet!~ one pitch or arrival explicitly identified as receiving emphasis;
^textbullet!~ one stated contrast in the treatment of repeated material; or
^textbullet!~ one timestamp locating a distinctive interpretive intervention.

The analysis must state both the choice and the moment or comparison that defines it. Do not infer physical technique, intention, emotion, or causation. Do not convert a general description of musical structure into a performer decision. If the analysis says the delivery remains even, that may be the answer; never invent a distinction merely to make a more dramatic question.

Ask about one decision and one comparison dimension only. Do not disclose words such as "accelerates," "detached," or "emphasized" in the stem when they are the requested answer.

Avoid expectancy questions whose setting makes the answer conventional and therefore easy. ^textbf!Bad:~ "As the cadence arrives, how does the tempo change?" A cadence normally primes "slows down." Use this question form only when the analysis explicitly gives a genuinely contrary answer, such as acceleration or unchanged pacing, and do not announce in the stem that the result is surprising or contrary.

For timestamp-answer questions, first ask whether the target is already unique from a concise musical description. If it is, do not add unnecessary context. Only when multiple performance decisions could plausibly match should the question become context-heavy, adding the thematic occurrence, affected voice or gesture, comparison with an earlier statement, a nearby event, or-if still needed-a broad search interval. A superlative such as "the greatest rubato" needs expansion only when its comparison scope is unclear. Add only what is required to leave one defensible location, and never include the target timestamp or a rounded equivalent.

Example forms:
^textbullet!~ "Compared with the opening statement, which articulation treatment characterizes the second full statement of the primary theme: much more connected, slightly more connected, approximately unchanged, slightly more detached, much more detached, alternating connected and detached, or irregularly mixed?"
^textbullet!~ "During the foreground gesture beginning at <gesture-start> and ending at the rest at <rest-start>, which pacing treatment is heard: steady acceleration, steady relaxation, approximately even pacing, continuous fluctuation, a sustained pause, slow-then-fast pacing, or fast-then-slow pacing?"
^textbullet!~ "Which octave-qualified pitch receives the clearest agogic emphasis in the upper line immediately before the bass resumes its steady motion at <bass-restart-time>?"
^textbullet!~ "Which octave-qualified pitch is performed with the shortest sounding duration during the opening gesture of the main subject?"
^textbullet!~ "During the second statement of the primary theme, at what timestamp does the performer apply the greatest rubato to the turn into its closing cadence?"

Non-example and narrow exception:
^textbullet!~ ^textbf!Reject as predictable:~ "As the cadence arrives at 00:50.25, how does the tempo change?" The cadence itself strongly suggests slowing down.
^textbullet!~ ^textbf!Permitted only with contrary ground truth:~ The same neutral question form may be used if the analysis explicitly states that the performer instead accelerates or maintains the pace. Do not add "unexpectedly," "surprisingly," "rather than slowing," or any other cue that discloses the contradiction.

\end{appendixprompt}
\par\addvspace{0.55\baselineskip}
\begin{appendixprompt}
^promptheading!TIME-STAMPED AUDITORY ANALYSIS:~
[TIME-STAMPED AUDITORY ANALYSIS]
\end{appendixprompt}
\par\addvspace{1.0\baselineskip}

\clearpage
\section{Human-Evaluation Recordings}
\label{app:human-evaluation-recordings}

The three evaluators assessed the same three solo-piano recordings, selected to
span contrasting musical forms, historical styles, and interpretive demands:

\segpromptnumber{1}{Ludwig van Beethoven, \textit{Piano Sonata No.~23 in F
minor, Op.~57, ``Appassionata'': I. Allegro assai} (Yi02M).}
\segpromptnumber{2}{Franz Schubert, \textit{Moment musical No.~3 in F minor,
D.~780, Op.~94, No.~3} (Tetzloff09M).}
\segpromptnumber{3}{Claude Debussy, \textit{Reflets dans l'eau}, No.~1 from
\textit{Images, Book I} (ParkJH13M).}

\clearpage
\section{Human-Evaluation Prompt}
\label{app:human-evaluation-prompt}

The following is the participant-facing prompt for the blinded Phase~1
comparison. The order of the MuNo-SP and MIDI-only descriptions is hidden.

\begin{appendixprompt}
^promptheading!STUDY INSTRUCTIONS~

Listen to the recording and compare Description A with Description B. The
source order is hidden.

Judge correctness, musician-likeness / how you hear it, insightfulness, and
overall preference separately.

Choose A, B, or exactly same for each criterion.

Compare musical content only; do not consider description length or layout.
The passages cover approximately the same time interval in the recording.

\end{appendixprompt}
\par\addvspace{0.55\baselineskip}
\begin{appendixprompt}
^promptheading!COMPARISON PROMPT~

Blind comparison

Which description do you think better describes the passage?

Make four separate comparisons: correctness, musician-likeness / how you hear
it, insightfulness, and overall preference.

Compare musical content only; do not consider description length or layout.
The passages cover approximately the same time interval in the recording.

Description A

[DESCRIPTION A]

Description B

[DESCRIPTION B]

1. More/less correct

Which description is more correct about what you hear in the recording?

2. More/less musician-like / how you hear it

Which description better resembles how you perceive the music?

3. More/less insightful

Which description offers more useful insight into the music?

4. Overall

Based on the above criteria, which description do you prefer overall?

For each criterion, choose Description A, Description B, or Exactly same.

Your notes

Explain why you chose these answers. If something is not correct, please give
an example. Don't worry about writing an essay^textemdash!~your writing doesn't
have to be perfect. Just let us know what you think.

\end{appendixprompt}

\clearpage
\section{LLM-as-a-Judge Pieces}
\label{app:llm-as-a-judge-list}

The LLM-as-a-Judge evaluation used the complete recording for each of the
following nine pieces:

\segpromptnumber{1}{Johann Sebastian Bach, \textit{Fugue, BWV~875}
(CaoJ01M).}
\segpromptnumber{2}{Ludwig van Beethoven, \textit{Piano Sonata No.~23 in F
minor, Op.~57, ``Appassionata'': I. Allegro assai} (Yi02M).}
\segpromptnumber{3}{Fr\'ed\'eric Chopin, \textit{Ballade No.~1 in G minor,
Op.~23} (Zhou06M).}
\segpromptnumber{4}{Claude Debussy, \textit{Reflets dans l'eau}, No.~1 from
\textit{Images, Book I} (ParkJH13M).}
\segpromptnumber{5}{Franz Liszt, \textit{Ballade No.~2 in B minor}
(Jin07M).}
\segpromptnumber{6}{Wolfgang Amadeus Mozart, \textit{Piano Sonata No.~12 in F
major, K.~332/I} (WuuE02M).}
\segpromptnumber{7}{Sergei Rachmaninoff, \textit{Prelude in D major, Op.~23,
No.~4} (WuuE07M).}
\segpromptnumber{8}{Franz Schubert, \textit{Moment musical No.~3 in F minor,
D.~780, Op.~94, No.~3} (Tetzloff09M).}
\segpromptnumber{9}{Alexander Scriabin, \textit{\'Etude in B-flat minor,
Op.~8, No.~11} (Shi08M).}

\clearpage
\section{LLM-as-a-Judge Prompt}
\label{app:llm-as-a-judge-prompt}

The judge receives the complete recording and the two descriptions in a
randomized, blinded order. The descriptions below are for J. S. Bach's \textit{Fugue in D minor}, BWV 875. Description A was MuNo-SP and Description B was MIDI.

\begin{appendixprompt}
^promptheading!SYSTEM INSTRUCTION~

You are participating in Phase 1 of a blind musical-description study. Listen
to the supplied complete recording and compare Description A with Description
B. The source order is hidden. Judge only from the recording and the two
descriptions.

\end{appendixprompt}
\par\addvspace{0.55\baselineskip}
\begin{appendixprompt}
^promptheading!AUDIO CONTEXT~

[COMPLETE RECORDING AUDIO]

The preceding audio is the complete recording. Use the entire recording as the
fixed listening evidence for this piece.

\end{appendixprompt}
\par\addvspace{0.55\baselineskip}
\begin{appendixprompt}
^promptheading!USER PROMPT~

Blind comparison

Which description do you think better describes the passage?

Make four separate comparisons: correctness, musician-likeness / how you hear
it, insightfulness, and overall preference.

Compare musical content only; do not consider description length or layout.
The passages cover approximately the same time interval in the recording.

Ignore the fact that the time windows of the analyses might be different.

Description A

At 00:00.50, the unaccompanied middle line states the complete subject D4--E4--F4--G4--F4--E4--F4--G4--A4--Bb4--A4--G4--A4--D5--C#5--C5--B4--Bb4--A4--G4--F4--E4--A4, its even triplet opening expanding into broader descending semitones after the high D5 at 00:03.19. From C#5 at 00:03.79 through B4--Bb4--A4--G4--F4--E4 at 00:04.95--00:08.56, chromatic contraction intensifies the otherwise D-minor field before the detached A4 at 00:09.13 leaves the phrase lightly open on dominant-oriented color. The subject's single presentation at 00:00.50--00:09.13 establishes both its circulating triplet head and its broader chromatic descent without contrapuntal distraction.

Description B

At 00:00.50--00:02.91, the single line states D4--E4--F4--G4--F4--E4--F4--G4--A4--Bb4--A4--G4--A4 in closely spaced, nearly even attacks, with the projection strengthening toward Bb4 at 00:02.06. With no supporting bass at 00:00.50--00:02.91, the exposed D-minor color is carried entirely by contour and the F4, Bb4, and A4 inflections, while slight overlaps between adjacent notes keep the legato ascent taut. After a short breath, D5 enters at 00:03.19 as the subject's registral apex and begins the complete descent D5--C#5--C5--B4--Bb4--A4--G4--F4--E4 through 00:08.82.

1. More/less correct

Which description is more correct about what you hear in the recording?

2. More/less musician-like / how you hear it

Which description better resembles how you perceive the music?

3. More/less insightful

Which description offers more useful insight into the music?

4. Overall

Based on the above criteria, which description do you prefer overall?

For each criterion choose Description A, Description B, or Exactly same. In the
comment, briefly explain why you chose these answers and give an example if
something is not correct.

\end{appendixprompt}
\par\addvspace{0.55\baselineskip}
\begin{appendixprompt}
^promptheading!STRUCTURED OUTPUT CONSTRAINT~

Return correctnessChoice, musicianChoice, insightChoice, choice, and comment.
Each choice field must be a, b, or tie; comment must be a string. Do not return
additional fields.
\end{appendixprompt}


\end{document}